\documentclass[preprint,12pt]{elsarticle}

\usepackage[utf8]{inputenc}

\usepackage{dsfont,amssymb,amsmath,graphics,graphicx,fancyhdr,float,verbatim,
ifthen,ifpdf,algorithmic,multirow,array,color, enumerate,theorem}
\usepackage[english]{babel}
\usepackage[perpage]{footmisc}
\usepackage[ruled]{algorithm}
\usepackage[subnum]{cases}

\renewcommand{\parallel}{|\!|}
 
\usepackage[margin=1.15in]{geometry}

\usepackage{natbib} 
\biboptions{authoryear, round}
 \usepackage[
                  breaklinks = true,
                 colorlinks = true,
                 linkcolor = red,
                 urlcolor  = black, 
                 citecolor = blue,
                 anchorcolor = green,
                 ]{hyperref}

\newcommand{\bsr}{\boldsymbol{r}}

\newcommand{\bsx}{\boldsymbol{x}}
\newcommand{\bsX}{\boldsymbol{X}}

\newcommand{\bsZ}{\boldsymbol{Z}}

\newcommand{\bsalpha}{\boldsymbol{\alpha}}
\newcommand{\bsbeta}{\boldsymbol{\beta}}

\newcommand{\bsOmega}{\boldsymbol{\Omega}}

\newcommand{\bstheta}{\boldsymbol{\theta}}

\newcommand{\bsPsi}{\boldsymbol{\Psi}}
\newcommand{\bsvPsi}{\boldsymbol{\varPsi}}

\newcommand{\Indicatrice}{\mathds{1}}

\newcommand{\E}{\mathbb{E}}
\newcommand{\V}{\mathbb{V}}
\newcommand{\Pro}{\mathbb{P}}
\newcommand{\R}{\mathbb{R}}

\newcommand{\ICL}{\text{ICL}}

\journal{Neural Networks}

\begin{document}

\begin{frontmatter}

\title{Robust mixture of experts modeling using the $t$ distribution}

\author[LMNO-Unicaen]{F. Chamroukhi \corref{cor1}} \ead{faicel.chamroukhi@unicaen.fr}
\cortext[cor1]{Corresponding author: Faicel Chamroukhi\\ Universit\'e de Caen-Normandie, LMNO, UMR CNRS 6139 \\  
Campus 2, Bvd Mar\'echal Juin, 14032 Caen Cedex, France\\
Tel: +33(0) 2 31 56 73 67\\Fax:  +33(0) 2 31 56 73 20 }

\address[LMNO-Unicaen]{Normandie Universit\'e, UNICAEN, CNRS, Laboratoire de Math\'ematiques Nicolas Oresme - LMNO, 14000 Caen, France} 

\begin{abstract} 
Mixture of Experts (MoE) is a popular framework for modeling heterogeneity in data for regression, classification, and clustering. For regression and cluster analyses of continuous data, MoE usually use normal experts following  the Gaussian distribution. However, for a set of data containing a group or groups of observations with heavy tails or atypical observations, the use of normal experts is unsuitable and can unduly affect the fit of the MoE model. We introduce a robust MoE modeling using the $t$ distribution. The proposed $t$  MoE (TMoE) deals with these issues regarding heavy-tailed and noisy data.  We develop a dedicated expectation-maximization (EM) algorithm to estimate the parameters of the proposed model by monotonically maximizing the observed data log-likelihood. We describe how the presented model can be used in prediction and in model-based clustering of regression data. The proposed model is validated on numerical experiments carried out on simulated data, which show the effectiveness and the robustness of the proposed model in terms of modeling non-linear regression functions as well as in model-based clustering. Then, it is applied to the real-world data of tone perception for musical data analysis, and the one of temperature anomalies for  the analysis of climate change data. The obtained results show the usefulness of the TMoE model for practical applications.
\end{abstract}


\end{frontmatter}


\section{Introduction}
Mixture of experts (MoE) introduced by \cite{jacobsME} are widely studied in statistics and machine learning. They consist in a fully conditional mixture model where both the mixing proportions, known as the gating functions, and the component densities, known as the experts, are conditional on some input covariates. MoE have been investigated, in their simple form, as well as in their hierarchical form \citep{jordanHME} (e.g Section 5.12 of \cite{McLachlanFMM}) for regression and model-based cluster and discriminant analyses and in different application domains. A complete review of the MoE models can be found in \cite{YukselWG12}. 
For continuous data, which we consider here in the context of non-linear regression and model-based cluster analysis, MoE usually use normal experts, that is, expert components following  the Gaussian distribution. Along this paper, we will call it the  normal mixture of experts, abbreviated NMoE.
It is well-known that the normal distribution is sensitive to outliers, which makes NMoE unsuitable to noisy data. Moreover, for a set of data containing a group or groups of observations with heavy tails, the use of normal experts may be unsuitable and can unduly affect the fit of the MoE model. 
In this paper, we attempt to overcome these limitations in MoE by proposing a more adapted and robust MoE model which can deal with the issues of heavy-tailed and atypical data.

The problem of sensitivity of NMoE to outliers have been considered very recently by \citet{Nguyen2014-MoLE} where the authors proposed  a Laplace mixture of linear experts (LMoLE) for a robust modeling of non-linear regression data. The model parameters are estimated by maximizing the observed-data likelihood via a minorization-maximization (MM) algorithm.  
Here, we propose an alternative MoE model, by relaying on the $t$ distribution.  We call this proposed model the $t$ mixture of experts, abbreviated TMoE.
The $t$ distribution provides indeed a natural robust extension of the normal distribution to model data with possible outliers and tails more heavy compared to the normal distribution. It has been considered to develop the $t$ mixture model proposed by \citet{Mclachlan98robustTmixture} for robust cluster analysis of multivariate data. 
%
We also mention that \cite{Lin07univSkewtMixture} also proposed a mixture of skew $t$ distributions to deal with heavy-tailed and asymmetric distributions.  However, in the skew-$t$ mixture model of \citet{Lin07univSkewtMixture},  the mixing proportions and the components means are constant, that is, they are not predictor-depending. In the proposed TMoE, however, we consider $t$ expert components in which both the mixing proportions and the mixture component means are predictor-depending. More specifically, we use polynomial regressors for the components, as well as multinomial logistic regressors for the mixing proportions.
In the framework of regression analysis, recently, \citet{Bai2012,Ingrassia2012} proposed a robust mixture modeling of regression on univariate data, by using a univariate $t$-mixture model. 
For the general multivariate case using $t$ mixtures, one can refer to for example the two key papers 
\citet{Mclachlan98robustTmixture,Peel2000robusTtmixture}.
The inference in the previously described approaches is performed by maximum likelihood estimation via expectation-maximization (EM)  or extensions \citep{dlr,McLachlanEM2008}, in particular the  expectation conditional maximization (ECM) algorithm \citep{meng_and_rubin_ECM_93}. 
For the Bayesian framework,  \citet{Fruhwirth10BayesSkewMixtures} have considered the Bayesian inference for both the univariate  and the multivariate skew-normal and skew-$t$ mixtures.
For the regression context, the robust modeling of regression data has been studied namely by
\citet{Wei2012,Ingrassia2012}  who considered a $t$-mixture model for regression analysis of univariate data, as well as by \citet{Bai2012} who relied on the M-estimate in mixture of linear regressions. 
%
In the same context of regression, \citet{Song2014} proposed the mixture of Laplace regressions, which has been then extended by \citet{Nguyen2014-MoLE} to the case of mixture of experts, by introducing the Laplace mixture of linear experts (LMoLE).
However, unlike our proposed TMoE model, the regression mixture models of \citet{Wei2012}, \citet{Bai2012}, \citet{Ingrassia2012}, and \citet{Song2014} 
do not consider conditional mixing proportions, that is, mixing proportions depending on some input variables, as in the case of mixture of experts, which we investigate here.  

Here we consider the MoE framework for non-linear regression problems and model-based clustering of regression data, and we attempt to overcome the limitations of the NMoE model for dealing with heavy-tailed data and which may contain outliers.
We investigate the use of the $t$ distribution for the experts, rather than the commonly used normal distribution.
The $t$-mixture of experts model (TMoE) handles the issues regarding namely the sensitivity of the NMoE to outliers.  
%
%
This model is an  extension of the unconditional mixture of  $t$ distributions \citep{Mclachlan98robustTmixture,Wei2012}, to the  mixture of experts (MoE) framework, where the mixture means are regression functions and the mixing proportions are covariate-varying. 
For the models inference, we develop a dedicated expectation-maximization (EM)  algorithm 
to estimate the parameters of the proposed model by monotonically maximizing the observed data log-likelihood.
The EM algorithm is indeed a very popular and successful estimation algorithm for mixture models in general and for mixture of experts in particular.
Indeed, the EM algorithm for MoE has been shown by \citet{NgM-EM-IRLS-04} to be monotonically maximizing the MoE likelihood. The authors have showed that the EM (with IRLS in this case) algorithm has stable convergence and  the log-likelihood is monotonically increasing when a learning rate smaller than one is adopted for the IRLS procedure within the M-step of the EM algorithm.
They have further proposed an expectation conditional maximization (ECM) algorithm to train MoE, which also has desirable numerical properties. 
Beyond the frequentist framework we consider here, We also mention The MoE has also been considered in the Bayesian framework, for example one can cite the Bayesian MoE 
\cite{Waterhouse96bayesianMoE,Waterhouse1997}
and the Bayesian hierarchical MoE \citet{Bishop_BayesianMoE}. 
Beyond the Bayesian parametric framework, the MoE models have also been investigated within the Bayesian non-parametric framework.
We cite for example the  Bayesian non-parametric MoE model 
\citep{Rasmussen01infiniteMoE}
and the Bayesian non-parametric hierarchical MoE approach of \citet{ShiMT05_Hierarchical_GPR} using Gaussian Processes experts for regression.
For further models on mixture of experts for regression, the reader can be referred to for example the book of \citet{ShiGPR_Book2011}. 
In this paper, we investigate semi-parametric models under the maximum likelihood estimation framework.

The remainder of this paper is organized as follows. In Section \ref{sec: MoE} we briefly recall the MoE framework, particularly the NMoE model and its maximum-likelihood estimation via EM. 
Then, in Section \ref{sec: TMoE} we present the TMoE model 
and derive its parameter estimation technique using the EM algorithm in
Section \ref{sec: MLE for the TMoE}. 
Next, in Section \ref{sec: Prediction using the NNMoE} we investigate  the use of the proposed models for fitting non-linear regression functions as well for prediction. 
We also show in Section \ref{sec: MBC using the NNMoE} how the models can be used in a model-based clustering prospective.
In Section \ref{sec: Model selection for the NNMoE}, we discuss the model selection problem for the model. 
In Section \ref{sec: Experimental study}, we perform experiments to assess the proposed models. Finally,  Section \ref{sec: Conclusion} is dedicated ton conclusions  and future work. 

\section{Mixture of experts for continuous data}
\label{sec: MoE}

Mixture of experts \citep{jacobsME,jordanHME} are used in a variety of contexts including regression, classification and clustering. 
Here we consider the MoE framework for fitting (non-linear) regression functions and clustering of univariate continuous data . 
The aim of regression is to explore the relationship of an observed random variable $Y$ given a covariate vector $\bsX \in \R^p$ via conditional density functions for $Y|\bsX = \bsx$ of the form $f (y|\bsx)$, rather than only exploring the unconditional distribution of $Y$. 
Thanks to their great flexibility, mixture models \citep{McLachlanFMM} has took much attention for non-linear regression problems and we  distinguish in particular the classical mixture of regressions model  \citep{Quandt1972, QuandtANDRamsey1978, DeVeaux1989, JonesANDMcLachlan1992, Gaffney99trajectoryclustering, VieleANDTong2002, FariaANDSoromenho2010,HunterANDYoung} 
and mixture of experts for regression analysis \citep{jacobsME,jordanHME,YoungANDHunter}. 
The univariate mixture of regressions model assumes that the observed pairs of data $(\bsx,y)$ where $y \in \R$ is the response for some covariate $\bsx \in \R^p$, are generated from $K$ regression functions and are governed by a hidden categorical random variable $Z$ indicating from which component each observation is generated.
Thus, the mixture of regressions decomposes the nonlinear regression model density $f (y|\bsx)$ into a convex weighted sum of $K$ regression components $f_k(y|\bsx)$ and can be defined as follows: %
\begin{eqnarray}
f(y|\bsx;\bsvPsi) &=& \sum_{k=1}^K \pi_k f_k(y|\bsx; \bsvPsi_k)
\label{eq: mixture of regressions}
\end{eqnarray}where the $\pi_k$'s are defined by $\pi_k = \Pro(Z = k)$ and represent the non-negative mixing proportions that sum to 1, that is, $\pi_k>0 \ \forall k$ and $\sum_{k=1}^K \pi_k = 1$. The model parameter vector is given by $\bsvPsi = (\pi_1,\ldots,\pi_{K-1},\bsvPsi^T_1,\ldots,\bsvPsi^T_K)^T$,  $\bsvPsi_k$ being the parameter vector of the $k$th component of the mixture density.

\subsection{The mixture of experts (MoE) model}
Although similar, the mixture of experts \citep{jacobsME} differ from regression mixture models in many aspects. One of the main differences is that the MoE model consists in a fully conditional mixture while in the regression mixture, only the component densities are conditional on some covariates. Indeed, the mixing proportions are constant for the regression mixture, while in the MoE, they are modeled as a function of some covariates, generally modeled by logistic or a softmax function.  
Mixture of experts (MoE) for regression analysis \citep{jacobsME,jordanHME} extend the model (\ref{eq: mixture of regressions}) by modeling the mixing proportions as function of some covariates $\bsr \in \R^q$. 
The mixing proportions, known as the gating functions in the context of MoE, are modeled by the multinomial logistic (softmax) model and are defined by:
\begin{eqnarray}
\pi_{k}(\bsr;\bsalpha) =\Pro(Z=k|\bsr;\bsalpha) =\frac{\exp{(\bsalpha_k^T\bsr)}}{\sum_{\ell=1}^K\exp{(\bsalpha_{\ell}^T \bsr)}} 
\label{eq: multinomial logistic}
\end{eqnarray}where $\bsr \in \R^q$ is a covariate vector, $\bsalpha_{k}$ is the $q$-dimensional coefficients vector associated with $\bsr$ and $\bsalpha = (\bsalpha^T_1,\ldots,\bsalpha^T_{K-1})^T$ is the parameter vector of the gating network, with $\bsalpha_K$ being the null vector. 
Thus, the MoE model consists in a fully conditional mixture model where both the mixing proportions (the gating functions) and the component densities (the experts) are conditional on predictors (respectively denoted here by $\bsr$ and $\bsx$).  

\subsection{The normal MoE (NMoE) model and its maximum likelihood estimation}
\label{sec: MLE for the NMoE}
In the case of MoE for regression, it is usually assumed that the experts are normal, that is, follow a normal distribution. A $K$-component normal MoE (NMoE)  ($K>1$) has the following formulation:
\begin{eqnarray}
f(y|\bsr,\bsx;\bsvPsi) &=& \sum_{k=1}^K \pi_k(\bsr;\bsalpha) \text{N}\!\left(y; \mu(\bsx;\bsbeta_k), \sigma_k^2\right)
\label{eq: normal MoE}
\end{eqnarray}which involves, in the semi-parametric case, component means defined as parametric (non-)linear regression functions $\mu(\bsx;\bsbeta_k)$. 
 
The NMoE model parameters are estimated by maximizing the observed data log-likelihood by using the EM algorithm \citep{dlr, jacobsME, jordanHME, jordan_and_xu_1995, NgM-EM-IRLS-04, McLachlanEM2008}. 
Suppose we observe an i.i.d sample of $n$ individuals $(y_1,\ldots,y_n)$ with their respective associated covariates $(\bsx_1,\ldots,\bsx_n)$ and $(\bsr_1,\ldots,\bsr_n)$.  Then, under the NMoE model, the observed data log-likelihood for the parameter vector $\bsvPsi$ is given by:
\begin{equation}
\log L(\bsvPsi)  = \sum_{i=1}^n  \log  \sum_{k=1}^K \pi_k(\bsr_i;\bsalpha) \text{N}\!\left(y_i; \mu(\bsx_i;\bsbeta_k), \sigma_k^2\right).
\label{eq: log-lik normal MoE}
\end{equation}
The E-Step at the $m$th iteration of the EM algorithm for the NMoE model requires the calculation of the following posterior probability that the individual $(y_i, \bsx_i,\bsr_i)$ belongs to expert $k$, given a parameter estimation $\bsvPsi^{(m)}$:
\begin{eqnarray}
\tau_{ik}^{(m)} = \Pro(Z_i=k|y_{i},\bsx_i,\bsr_i;\bsvPsi^{(m)})
 =  \frac{\pi_k(\bsr_i;\bsalpha^{(m)}) \text{N}\!\left(y_i;\mu(\bsx_i; \beta^{(m)}_k),{\sigma^2_k}^{(m)}\right)}{f(y_i|\bsr_i,\bsx_i;\bsvPsi^{(m)})}.
\label{eq: posterior prob NMoE}
\end{eqnarray}
Then, the M-step calculates the parameter update $\bsvPsi^{(m+1)}$ by maximizing the well-known $Q$-function (the expected complete-data log-likelihood), that is:
\begin{equation}
\bsvPsi^{(m+1)} =  \arg \max_{\bsvPsi \in \bsOmega} Q(\bsvPsi;\bsvPsi^{(m)})
\label{eq: arg max Q}
\end{equation}where $\bsOmega$ is the parameter space.  
For example, in the case of  normal mixture of linear experts (NMoLE) where each expert's mean has the following linear form: 
\begin{equation}
\mu(\bsx_i;\bsbeta_k) = \bsbeta_k^T \bsx_i, 
\label{eq: linear regression mean}
\end{equation}where $\bsbeta_k \in \R^p$ is the  vector of regression coefficients of expert component $k$, the updates for each of the expert component parameters consist in analytically solving a weighted Gaussian linear regression problem and are given by:
\begin{eqnarray}
\bsbeta_k^{(m+1)}  &=& \Big[\sum_{i=1}^{n}\tau^{(m)}_{ik}  \bsx_i\bsx^T_i \Big]^{-1} \sum_{i=1}^{n} 
 \tau^{(q)}_{ik}  y_i \bsx_i,
\label{eq: beta_k update for NMoE}\\ 
{\sigma^2_{k}}^{(m+1)} &= &
\frac{\sum_{i=1}^n\tau_{ik}^{(m)}\left(y_i - {\bsbeta^T_{k}}^{(m+1)}\bsx_i\right)^2}{\sum_{i=1}^n\tau_{ik}^{(m)}}\cdot
\label{eq: sigma2k update NMoE}
\end{eqnarray}
For the gating network, the parameter update $\bsalpha^{(m+1)}$  cannot however be obtained in a closed form. It can be calculated by Iteratively Reweighted Least Squares (IRLS) \citep{jacobsME,jordanHME,Chen1999,Green1984,chamroukhi_et_al_NN2009}. 

\bigskip
However, the normal distribution, used to model experts in the NMoE model, is not adapted to deal with data with heavy tailed data distribution and  it is also known that the normal distribution is sensitive to outliers. In the proposed model, we propose a robust fitting of the MoE model, which is adapted to data with heavy-tailed distribution and is more robust to outliers, by using the $t$ distribution. This is the $t$ MoE (TMoE) model which we present in the next section. 

\section{The $t$ MoE (TMoE) model}
\label{sec: TMoE}
The proposed $t$ MoE (TMoE) model is based on the $t$ distribution, which is known as a robust generalization of the normal distribution.  The $t$ distribution is recalled in the following section. We also describe  its stochastic and hierarchical representations, which will be used to derive those of the proposed TMoE model.

\subsection{The $t$ distribution}
 
The use of the $t$ distribution in standard mixture models has been shown to be more robust than the normal distribution to handle outliers in the data and accommodate data  with heavy tailed distribution. This has been shown in terms of density modeling and cluster analysis for multivariate data \citep{Mclachlan98robustTmixture,Peel2000robusTtmixture} as well as  for univariate data by using a skewed-$t$ mixture model \citep{Lin07univSkewtMixture}.
The $t$-distribution with location parameter $\mu \in \R$, scale parameter $\sigma^2\in (0,\infty)$ and degrees of freedom $\nu \in (0,\infty)$ has the probability density function
\begin{equation}
  f(y;\mu, \sigma^2, \nu) = \frac{\Gamma(\frac{\nu+1}{2})} {\sqrt{\nu\pi}\,\Gamma(\frac{\nu}{2})} \left(1+\frac{d_y^2}{\nu} \right)^{-\frac{\nu+1}{2}},\!
  \label{eq: t density}
\end{equation}where $d^2_y = \left(\frac{y-\mu}{\sigma}\right)^2$ denotes the squared Mahalanobis distance between $y$ and $\mu$ ($\sigma$ being the scale parameter), and $\Gamma$ is the Gamma function given by $\Gamma(x) = \int_0^\infty x^{t-1} e^{-x}\, d x$.
The $t$ distribution can be characterized as follows. Let $E$ be an univariate random variable with a standard normal distribution with pdf given by $\phi(.)$. Then, let $W$ be a random variable independent of $E$ and following the gamma distribution, that is $W \sim \text{gamma}(\frac{\nu}{2},\frac{\nu}{2})$ where the density function of the gamma distribution is given by $f (u;a,b) = \{b^a u^{a-1}/\Gamma(a)\} \exp(-bu)\Indicatrice_{(0,\infty)}(u); \quad (a,b)>0$ and the indicator function $\Indicatrice_{(0,\infty)}(u)=1$ for $u > 0$ and is zero elsewhere.
Then, a random variable $Y$ having the following representation:
\begin{equation}
Y = \mu + \sigma \frac{E}{\sqrt{W}}
\label{eq: stochastic representation t}
\end{equation}
follows the  $t$ distribution $t_{\nu}(\mu,\sigma^2, \nu)$ with pdf given by (\ref{eq: t density}). 
As given in  \citet{LiuAndRubin95} for the multivariate case, a hierarchical representation of the $t$ distribution in this univariate case can be expressed from the stochastic representation (\ref{eq: stochastic representation t}) as: 
\begin{equation}
\begin{tabular}{lll}
$Y_i|w_i$ & $\sim$ & $\text{N}\!\left(\mu, \frac{\sigma^2}{w_i}\right)$\\
$W_i$ & $\sim$ & $\text{gamma}\left(\frac{\nu}{2},\frac{\nu}{2}\right)$.
\end{tabular} 
\label{eq: hierarchical representation t}
\end{equation} 

 \subsection{The $t$ MoE (TMoE) model}
The proposed   $t$ MoE (TMoE) model extends the $t$ mixture model to the MoE framework. 
The mixture of $t$ distributions have been first proposed by \citet{Mclachlan98robustTmixture,Peel2000robusTtmixture}
 for multivariate data. For the univariate case, a $K$-component $t$ mixture model takes the following form:
\begin{eqnarray}
f(y;\bsvPsi) &=& \sum_{k=1}^K \pi_k ~ t(y; \mu_k, \sigma_k^2, \nu_k)
\label{eq: t mixture}
\end{eqnarray}where each of the mixture components has a $t$ density given by (\ref{eq: t density}).
\cite{Lin07univSkewtMixture} proposed a mixture of skew $t$ distributions to deal with heavy-tailed and asymmetric distributions.  However, in the skew-$t$ mixture model of \citet{Lin07univSkewtMixture},  the mixing proportions and the components means are constant and are not predictor-depending and hence doest not consider the regression problem and is not a mixture of experts model. 
\citet{Wei2012} considered the $t$-mixture model for the regression context on univariate data where the means $\mu_k$ in (\ref{eq: t mixture}) are (linear) regression functions of the form $\mu(\bsx;\bsbeta_k)$. However, this model do not explicitly model the mixing proportions as function the inputs; they are assumed to be constant.

The proposed $t$ MoE (TMoE) is MoE model with $t$-distributed experts and is defined as follows.  
Let $t_{\nu}(\mu,\sigma^2,\nu)$ denotes a $t$ distribution with location parameter $\mu$, scale parameter $\sigma$ and degrees of freedom $\nu$, whose density is given by (\ref{eq: t density}). A $K$-component TMoE model is then defined by:
\begin{eqnarray}
f(y|\bsr,\bsx;\bsvPsi) &=& \sum_{k=1}^K \pi_k(\bsr;\bsalpha) ~ t\left(y; \mu(\bsx;\bsbeta_k), \sigma_k^2, \nu_k\right)
\label{eq: TMoE}
\end{eqnarray}whose parameter vector is given by $\bsvPsi = (\bsalpha^T_1,\ldots,\bsalpha^T_{K-1},\bsvPsi^T_1,\ldots,\bsvPsi^T_K)^T$ where $\bsvPsi_k = (\bsbeta^T_k,\sigma^2_k,\nu_k)^T$ is the parameter vector for the $k$th expert component which has a $t$ distribution. 
When the robustness parameter $\nu_k \rightarrow \infty$ for each $k$, each $t$ expert component approaches a normal expert and thus the TMoE model  (\ref{eq: TMoE}) approaches the NMoE model (\ref{eq: normal MoE}).

In the following section, we present the stochastic and hierarchical characterizations of the proposed TMoE model and then derive the model maximum likelihood inference scheme.
\subsubsection{Stochastic representation of the TMoE}
\label{ssec: stochastic TMoE}
By using the stochastic representation (\ref{eq: stochastic representation t}) of the $t$ distribution, the one for the $t$ MoE (TMoE) is derived  as follows. Let $E$ be a univariate random variable following the standard normal distribution $E \sim \phi(.)$.  
Suppose that, conditional on the hidden variable $Z_i=z_i$, a random variable $W_{i}$ is distributed as $\text{gamma}(\frac{\nu_{z_i}}{2},\frac{\nu_{z_i}}{2})$. Then, given the covariates $(\bsx_i, \bsr_i)$, a random variable $Y_i$ is said to follow the TMoE model (\ref{eq: TMoE}) if it has the following representation:
\begin{equation}
Y_i =  \mu(\bsx_i;\bsbeta_{z_i}) +  \sigma_{z_i}  \frac{E_i}{\sqrt{W_{z_i}}},
\label{eq: stochastic representation TMoE}
\end{equation}where the categorical variable $Z_i$ conditional on the covariate $\bsr_i$ follows the multinomial distribution:  
\begin{equation}
Z_i|\bsr_i \sim \text{Mult}\!\left(1;\pi_{1}(\bsr_i;\bsalpha),\ldots, \pi_{K}(\bsr_i;\bsalpha)\right)
\label{eq: Multinomial}
\end{equation}where each of the probabilities $\pi_{z_i}(\bsr_i;\bsalpha) = \Pro(Z_i=z_i|\bsr_i)$ is given by the multinomial logistic function (\ref{eq: multinomial logistic}). 
In this incomplete data framework,  $z_i$ represents the hidden label of the expert component generating the $i$th  observation.

\subsubsection{Hierarchical representation of the TMoE}
\label{ssec: hierarchical rep TMoE}
 By introducing the binary latent component-indicators $Z_{ik}$ such that $Z_{ik}=1$ iff $Z_i =k$, $Z_i$ being the hidden class label of the $i$th observation,  a hierarchical representation for the TMoE model can be derived from its stochastic representation and is as follows.
From (\ref{eq: hierarchical representation t}), 
(\ref{eq: stochastic representation TMoE}), and (\ref{eq: Multinomial}), following the hierarchical representation of the mixture of multivariate $t$-distributions (see for example  \citet{Mclachlan98robustTmixture}), the hierarchical representation of the TMoE model is written as:
\begin{eqnarray}
Y_i| w_i, Z_{ik}=1, \bsx_i &\sim& \text{N}\!\left(\mu(\bsx_i;\bsbeta_k), \frac{\sigma^2_k}{w_i}\right), \nonumber \\
W_i|Z_{ik}=1&\sim& \text{gamma}\left(\frac{\nu_k}{2},\frac{\nu_k}{2}\right)\label{eq: hierarchical representation TMoE}\\
\bsZ_i|\bsr_i &\sim & \text{Mult}\left(1;\pi_1(\bsr_i;\bsalpha),\ldots,\pi_K(\bsr_i;\bsalpha) \right).\nonumber 
\end{eqnarray}

\subsection{Identifiability of the TMoE model}
\cite{Jiang_and_tanner_NN_99} have established that  ordered, initialized, and irreducible MoEs are identifiable. Ordered implies that there exist a certain ordering relationship on the experts parameters $\bsvPsi_k$ such that $(\bsalpha^T_1, \bsvPsi^T_1)^T \prec  \ldots \prec (\bsalpha^T_K, \bsvPsi^T_K)^T $; initialized implies that $\bsalpha_K$, the parameter vector of the $K$th gating function $\pi_K(\bsr;\bsalpha)$, is the null vector, 
and irreducible implies that $\bsvPsi_{k} \neq \bsvPsi_{k\prime}$ for any $k \neq k\prime$. 
For the proposed TMoE model,  ordered implies that there exist a certain ordering relationship  such that  $(\bsbeta^T_1,\sigma^2_1,\nu_1)^T \prec  \ldots \prec (\bsbeta^T_K,\sigma^2_K,\nu_K)^T $; initialized implies that $\bsalpha_K$ is the null vector, as assumed here in the model, and finally   
 irreducible implies that if $k \neq k\prime$, then one of the following conditions holds: 
 $\bsbeta_k\neq \bsbeta_{k\prime}$,
 $\sigma_k\neq \sigma_{k\prime}$,
or
 $\nu_k\neq \nu_{k\prime}$. 
Then, we can establish the identifiability of  ordered and initialized irreducible TMoE models 
by applying Lemma 2 of \cite{Jiang_and_tanner_NN_99}, which requires the validation of the following nondegeneracy condition. The set $\{t(y; \mu(\bsx;\bsbeta_1), \sigma_1^2,\nu_1),\ldots,t(y; \mu(\bsx;\bsbeta_{3K}), \sigma_{3K}^2, \nu_{3K})\}$ contains $3K$ linearly independent functions of $y$, for any $3K$ distinct triplet $(\mu(\bsx;\bsbeta_k), \sigma_k^2, \nu_k)$ for $k=1,\ldots,3K$. 
%
Thus,  via Lemma 2 of \cite{Jiang_and_tanner_NN_99} we have any ordered and initialized irreducible TMoE is identifiable.
%

\section{Maximum likelihood estimation of the TMoE model}
\label{sec: MLE for the TMoE} 
Given an i.i.d sample of $n$ observations,  
the unknown parameter vector $\bsvPsi$ can be estimated by maximizing the observed-data log-likelihood,  which, under the TMoE model, is given by:
\begin{equation}
\log L(\bsvPsi) =  \sum_{i=1}^n  \log  \sum_{k=1}^K \pi_k(\bsr_i;\bsalpha) ~ t\left(y_i; \mu(\bsx_i;\bsbeta_k), \sigma_k^2, \nu_k\right).
\label{eq: log-lik TMoE}
\end{equation}To perform this maximization, we first use the EM algorithm and then describe an extension based on the ECM algorithm \citep{meng_and_rubin_ECM_93}    as in \citet{LiuAndRubin95} for a single $t$ distribution, and as in \cite{Mclachlan98robustTmixture} and \cite{Peel2000robusTtmixture} for mixture of $t$-distributions.

\subsection{The EM algorithm for the TMoE model} 
\label{ssec: EM TMoE}
To maximize the log-likelihood function (\ref{eq: log-lik TMoE}) for the TMoE model, the EM algorithm starts with an initial parameter vector $\bsvPsi^{(0)}$ and alternates between the E- and M- steps until convergence. The E-step computes the expected completed data log-likelihood (the $Q$-function) and the M-Step maximize it. 
 From the hierarchical representation of the TMoE (\ref{eq: hierarchical representation TMoE}), the complete data consist of the responses $(y_1,\ldots,y_n)$ and their corresponding covariates $(\bsx_1,\ldots,\bsx_n)$ and $(\bsr_1,\ldots,\bsr_n)$, as well as the latent variables $(w_1,\ldots,w_n)$ and the latent component labels $(z_1,\ldots,z_n)$. Thus, the complete-data log-likelihood of $\bsvPsi$  is given by:
{\small \begin{eqnarray}
\log L_c(\bsvPsi)  &=&  \sum_{i=1}^n \sum_{k=1}^K Z_{ik} \big[ \log\left(\Pro\left(Z_i=k|\bsr_i\right)\right) +   \log\left(f\left(w_i|Z_{ik}=1\right)\right)+ \log\left(f\left(y_i|w_i,Z_{ik}=1,\bsx_i\right)\right)\big] \nonumber \\
& = & \log L_{1c}(\bsalpha) + \sum_{k=1}^K \big[\log L_{2c}(\bstheta_k) + \log L_{3c}(\nu_k)\big],
\label{eq: complete log-likelihood TMoE}
\end{eqnarray}}where $\bstheta_k = (\bsbeta_k^T,\sigma^2_k)^T$,
{\small \begin{eqnarray}
\log L_{1c}(\bsalpha) &= & \sum_{i=1}^{n} \sum_{k=1}^K Z_{ik} \log \pi_k(\bsr_i;\bsalpha), \label{eq: L1c TMoE}\\
\log L_{1c}(\bstheta_k) &=& \sum_{i=1}^{n} Z_{ik} \Big[- \frac{1}{2}\log (2 \pi) -  \frac{1}{2}\log (\sigma^2_k) -  \frac{1}{2} w_i d^2_{ik}\Big], \label{eq: L2c TMoE}\\
\log L_{3c}(\nu_k) &=& \sum_{i=1}^{n} Z_{ik}  \Big[ - \log \Gamma \left(\frac{\nu_k}{2}\right) + \left(\frac{\nu_k}{2}\right) \log \left(\frac{\nu_k}{2}\right)    +  \left(\frac{\nu_k}{2}-1\right)  \log (w_i)
 - \left(\frac{\nu_k}{2}\right) w_i\Big]. \label{eq: L3c TMoE}
\end{eqnarray}}
\subsection{E-Step}
\label{ssec: E-Step TMoE}
The E-Step of the EM algorithm for the TMoE calculates the $Q$-function, that is the conditional expectation of the complete-data log-likelihood (\ref{eq: complete log-likelihood TMoE}),  given the observed data and a current parameter estimation $\bsvPsi^{(m)}$, $m$ being the current iteration.
It can be seen from (\ref{eq: L1c TMoE}), (\ref{eq: L2c TMoE}) and (\ref{eq: L3c TMoE}) that computing the $Q$-function requires the following conditional expectations:
\begin{eqnarray*}
\tau_{ik}^{(m)} &=& 
\E_{{\bsvPsi^{(m)}}}\left[Z_{ik}|y_i,\bsx_i,\bsr_i \right], \label{eq: E[Zik|yi] defintion TMoE}\\
w_{ik}^{(m)} &=& \E_{{\bsvPsi^{(m)}}}\left[W_{i}|y_i, Z_{ik}=1,\bsx_i,\bsr_i\right], \label{eq: E[Wi|yi,Zik] definition TMoE}\\
e_{1,ik}^{(m)} &=& \E_{{\bsvPsi^{(m)}}}\left[\log(W_{i})|y_i, Z_{ik}=1,\bsx_i,\bsr_i\right]\cdot \label{eq: E[log Wi|yi,Zik] definition TMoE}
\end{eqnarray*}
It follows that the $Q$-function is given by:
\begin{equation}
Q(\bsvPsi;\bsvPsi^{(m)})=Q_{1}(\bsalpha;\bsvPsi^{(m)})+\sum_{k=1}^K \left[Q_{2}(\bstheta_k,\bsvPsi^{(m)}) + Q_{3}(\nu_k,\bsvPsi^{(m)})\right],
\label{eq: Q-function decomposition TMoE}
\end{equation}
where
{\small \begin{eqnarray*}
Q_{1}(\bsalpha;\bsvPsi^{(m)}) &= & \sum_{i=1}^{n} \sum_{k=1}^K \tau^{(m)}_{ik} \log \pi_k(\bsr_i;\bsalpha), \label{eq: Q_alpha TMoE}\\ 
Q_{2}(\bstheta_k;\bsvPsi^{(m)}) &=& \sum_{i=1}^{n} \tau^{(m)}_{ik}
\Big[- \frac{1}{2}\log (2 \pi) -  \frac{1}{2}\log (\sigma^2_k) -  \frac{1}{2} ~ w^{(m)}_{ik} d^2_{ik}\Big].\label{eq: Q_Psik TMoE}\\
Q_{3}(\nu_k;\bsvPsi^{(m)}) &=& \sum_{i=1}^{n} \tau^{(m)}_{ik}  \left[- \log \Gamma \left(\frac{\nu_k}{2}\right) + \left(\frac{\nu_k}{2}\right) \log \left(\frac{\nu_k}{2}\right)    
 - \left(\frac{\nu_k}{2}\right) ~ w^{(m)}_{ik} 
  +  \left(\frac{\nu_k}{2}-1\right) e^{(m)}_{1,ik} \right]. \label{eq: Q_nuk TMoE}
\end{eqnarray*}}These conditional expectations are given as follows. 
First, the conditional expectation 
$\E_{{\bsvPsi^{(m)}}}\left[Z_{ik}|y_i,\bsx_i,\bsr_i \right]$, which corresponds to the posterior component memberships, is given by:
\begin{eqnarray}
\tau_{ik}^{(m)} &=& \frac{\pi_k(\bsr_i;\bsalpha^{(m)}) t(y_i;\mu(\bsx_i;\bsbeta_k^{(m)}),{\sigma^2_k}^{(m)}, \nu^{(m)}_k)}{f(y_i|\bsr_i,\bsx_i;\bsvPsi^{(m)})}\cdot
\label{eq: posterior prob TMoE}
\end{eqnarray}
Then, it can be easily shown (see for example \citet{Mclachlan98robustTmixture}, \citet{Peel2000robusTtmixture} and \citet{LiuAndRubin95} for details) that:
{\footnotesize\begin{eqnarray}
\!\!\!\!\!\!\!\!\!\! \E_{{\bsvPsi^{(m)}}}\left[W_{i}|y_i, Z_{ik}=1,\bsx_i,\bsr_i \right] &\!\!\!\! =\!\!\!\! & \frac{\nu^{(m)}_k+1}{\nu^{(m)}_k+{d^2_{ik}}^{(m)}} 
= w^{(m)}_{ik}, \label{eq: E[Wi|yi,Zik] expression TMoE}\\
\!\!\!\!\!\!\!\!\!\! \E_{{\bsvPsi^{(m)}}}\left[\log(W_{i})|y_i, Z_{ik}=1,\bsx_i,\bsr_i \right] &\!\!\!\! =\!\!\!\! & \log\left(w^{(m)}_{ik}\right) + \left\{\psi\left(\frac{\nu^{(m)}_k +1}{2}\right) -  \log\left(\frac{\nu^{(m)}_k + 1}{2}\right) \right\}  
= e^{(m)}_{1,ik}, \label{eq: E[log Wi|yi,Zik] expression TMoE}
\end{eqnarray}}where $\psi(x) = \left\{\partial \Gamma(x)/\partial x \right\}/\Gamma(x)$ is the Digamma function.

\subsection{M-Step}
\label{ssec: M-Step TMoE}In the M-step, as it can be seen from (\ref{eq: Q-function decomposition TMoE}), the $Q$-function can be maximized by independently maximizing $Q_{1}(\bsalpha;\bsvPsi^{(m)})$, and, for each $k$, $Q_{2}(\bsPsi_k;\bsvPsi^{(m)})$, $Q_{3}(\nu_k;\bsvPsi^{(m)})$, with respect to $\bsalpha$, $\bsPsi_k$ and $\nu_k$, respectively.
Thus, on the $(m+1)$th iteration of the EM algorithm, the model parameters are updated as follows.

\paragraph{M-Step 1}Calculate $\bsalpha^{(m+1)}$ by maximizing $Q_{1}(\bsalpha;\bsvPsi^{(m)})$ w.r.t $\bsalpha$: 
\begin{equation}
\bsalpha^{(m+1)} =  \arg \max_{\bsalpha} Q_{1}(\bsalpha;\bsvPsi^{(m)}).
\label{eq: arg max_alpha Q-function TMoE}
\end{equation}
Unlike the case of the standard $t$ mixture model (e.g., \citet{Mclachlan98robustTmixture, Peel2000robusTtmixture})
and  $t$ regression mixture model
\citep{Wei2012,Bai2012,Ingrassia2012}, for which the mixing proportions are not predictor-depending and their update is done in closed form, for the proposed TMoE does, there is no a a closed form solution to update the gating network parameters. This is performed by Iteratively Reweighted Least Squares (IRLS).

\paragraph{The Iteratively Reweighted Least Squares (IRLS) algorithm:}
\label{par: IRLS M-Step SNMoE}
The IRLS algorithm is used to maximize $Q_{1}(\bsalpha,\bsvPsi^{(m)})$  with respect to the parameter $\bsalpha$ in the M-Step at each iteration $m$ of the EM algorithm.   
The IRLS is a Newton-Raphson algorithm and  consists in starting with an initial vector $\bsalpha^{(0)}$, and, at the $(l+1)$th iteration of the IRLS, updating the estimation of $\bsalpha$ as follows:
\begin{equation}
\bsalpha^{(l+1)}=\bsalpha^{(l)}-\Big[\frac{\partial^2 Q_{1}(\bsalpha,\bsvPsi^{(m)})}{\partial \bsalpha \partial \bsalpha^T}\Big]^{-1}_{\bsalpha=\bsalpha^{(l)}} \frac{\partial Q_{1}(\bsalpha,\bsvPsi^{(m)})}{\partial \bsalpha}\Big|_{\bsalpha=\bsalpha^{(l)}}
\label{eq: IRLS update}
\end{equation}
where $\frac{\partial^2 Q_{1}(\bsalpha,\bsvPsi^{(m)})}{\partial \bsalpha \partial \bsalpha^T}$ and $\frac{\partial Q_{1}(\bsalpha,\bsvPsi^{(m)})}{\partial \bsalpha}$ are respectively the Hessian matrix and the gradient vector of $Q_{1}(\bsalpha,\bsPsi^{(m)})$. At each IRLS iteration the Hessian and the gradient are evaluated at $\bsalpha = \bsalpha^{(l)}$ and are  computed analytically similarly as in \citet{chamroukhi_et_al_NN2009}. 
The parameter update $\bsalpha^{(m+1)}$ in (\ref{eq: arg max_alpha Q-function TMoE}) is taken at convergence of the IRLS algorithm (\ref{eq: IRLS update}).  
Then,  for  $k=1\ldots,K$: 

\paragraph{M-Step 2} Calculate $\bstheta_k^{(m+1)}$ by maximizing $Q_{2}(\bstheta_k;\bsvPsi^{(m)})$ 
w.r.t $\bstheta_k=(\bsbeta^T_k,\sigma^2_k)^T$. This is achieved by first maximizing $Q_{2}(\bstheta_k;\bsvPsi^{(m)})$  w.r.t $\bsbeta_k$ and then w.r.t $\sigma^2_{k}$.
For the  $t$ mixture of linear experts (TMoLE) case where the expert means have the form  (\ref{eq: linear regression mean}), this maximization is performed analytically and provides the following updates: 
\begin{eqnarray}
\bsbeta_k^{(m+1)}  &=& \Big[\sum_{i=1}^{n}\tau^{(m)}_{ik} w_{ik}^{(m)}  \bsx_i\bsx^T_i \Big]^{-1} \sum_{i=1}^{n} 
 \tau^{(q)}_{ik}w_{ik}^{(m)}  y_i  \bsx_i,
\label{eq: beta_k update for TMoE}\\
{\sigma^2_{k}}^{(m+1)} &= &
\frac{1}{\sum_{i=1}^n\tau_{ik}^{(m)}}\sum_{i=1}^n\tau_{ik}^{(m)} w_{ik}^{(m)} \left(y_i - {\bsbeta^T_{k}}^{(m+1)}\bsx_i\right)^2.
\label{eq: sigma2k update TMoE}
\end{eqnarray}
 Here, we note that, following \citet{Kent1994multivariateT}  in the
case of ML estimation for single component $t$ distribution and \citet{Mclachlan98robustTmixture, Peel2000robusTtmixture} for mixture of multivariate $t$ distributions, the EM algorithm can be modified slightly by replacing the divisor $\sum_{i=1}^n\tau_{ik}^{(m)}$ in (\ref{eq: sigma2k update TMoE})  by $\sum_{i=1}^n\tau_{ik}^{(m)} w_{ik}^{(m)}$. This modified algorithm may converge faster than the conventional EM algorithm.

\paragraph{M-Step 3}Calculate $\nu_k^{(m+1)}$ by maximizing $Q_{3}(\nu_k;\bsvPsi^{(m)})$ w.r.t $\nu_{k}$.  
The degrees of freedom update $\nu^{(m+1)}_k$ is therefore obtained by iteratively solving the following equation for $\nu_k$:
\begin{eqnarray}
& & - \psi \left(\frac{\nu_k}{2}\right) + \log \left(\frac{\nu_k}{2}\right) + 1 +
\frac{1}{\sum_{i=1}^n\tau_{ik}^{(m)}}\sum_{i=1}^n\tau_{ik}^{(m)} \left( \log(w^{(m)}_{ik}) - w^{(m)}_{ik} \right) 
\nonumber \\
& &
 + \psi\left(\frac{\nu^{(m)}_k +1}{2}\right) -  \log\left(\frac{\nu^{(m)}_k + 1}{2}\right) = 0.
\label{eq: nuk update TMoE}
\end{eqnarray}This scalar non-linear equation can be solved  with a root finding algorithm, such
as Brent's method \citep{Brent1973}.
 
It is obvious to see that, as mentioned previously,  if the number of degrees of freedom $\nu_k$ approaches infinity for all $k$, then the parameter updates  for the TMoE  model are exactly those of the NMoE model (since $w_{ik}$ tends to $1$ in that case). The TMoE model constitutes therefore a robust generalization of the NMoE model, which is able to model data with density heaving longer tails than those of the NMoE model. 

After deriving the EM algorithm for the parameter estimation of the TMoE model, now we describe an ECM extension.

\subsection{The ECM algorithm for the TMoE model}
\label{ssec: ECM TMoE}
Following the ECM extension of the EM algorithm for a single $t$ distribution proposed by  \citet{LiuAndRubin95} and the one of the EM algorithm for the $t$-mixture model  \citep{Mclachlan98robustTmixture, Peel2000robusTtmixture}, the EM algorithm for the TMoE model can also be modified to give an ECM version by adding  an additional E-Step between the two M-steps 2 and 3. 
This additional E-step consists in taking the parameter vector $\bsvPsi$ with $\bstheta_k = \bstheta_k^{(m+1)}$ instead of $\bstheta_k^{(m)}$, that is
$$Q_3(\nu_k;\bsvPsi^{(m)}) = Q_3(\nu_k;\bsalpha^{(m)},\bstheta_k^{(m+1)},\nu_k^{(m)}).$$
Thus, the M-Step 3 in the above is replaced by a Conditional-Maximization (CM)-Step in which the degrees of freedom update (\ref{eq: nuk update TMoE}) is calculated with the conditional expectation (\ref{eq: E[Wi|yi,Zik] expression TMoE}) and (\ref{eq: E[log Wi|yi,Zik] expression TMoE}) computed with the updated parameters $\bsbeta_k^{(m+1)}$ and ${\sigma^2_{k}}^{(m+1)}$  respectively given by (\ref{eq: beta_k update for TMoE}) and (\ref{eq: sigma2k update TMoE}).

\bigskip
The TMoE handles therefore the problem of heavy tailed data possibly affected by outliers.  
It therefore provides a more robust modeling  framework for fitting MoE to data. 
In the next section, we show how to use the TMoE in fitting regression functions and clustering, and we discuss the question of model selection. 

\section{Prediction using the TMoE}
\label{sec: Prediction using the NNMoE}
The goal in regression  is to be able to make predictions for the response variable(s) given some new value of the predictor variable(s) on the basis of a model trained on a set of training data. 
In regression analysis using MoE, the aim is therefore to predict the response $y$ given new values of the predictors $(\bsx,\bsr)$, on the basis of a MoE model characterized by a parameter vector $\hat \bsvPsi$ inferred from a set of training data, here, by maximum likelihood via EM.  
These predictions can be expressed in terms of the predictive distribution of $y$, which is obtained by substituting the maximum likelihood parameter $\hat\bsvPsi$ into (\ref{eq: mixture of regressions})-(\ref{eq: multinomial logistic}) to give:
\begin{equation*}
f(y|\bsx,\bsr;\hat \bsvPsi) = \sum_{k=1}^K \pi_{k}(\bsr;\hat \bsalpha)  f_k(y|\bsx; \hat \bsvPsi_k).
\label{eq: predictive MoE}
\end{equation*}Using $f$, we might then predict $y$ for a given set of $\bsx$'s and $\bsr$'s as the expected value under $f$, 
that is by calculating the prediction $\hat y = \E_{{\it \hat\bsvPsi}}(Y|\bsr,\bsx)$. We thus need to compute the expectation of the MoE model.
It is easy to show (see for example Section 1.2.4 in \citet{sylvia_fruhwirth_book_2006}) that the mean and the variance of a MoE distribution of the form (\ref{eq: predictive MoE}) are respectively given by:
{\small
\begin{eqnarray}
\!\!\!\!\E_{{\it \hat\bsvPsi}}(Y|\bsr,\bsx) &\! =\! & \sum_{k=1}^K \pi_k(\bsr;\hat \bsalpha_{n}) \E_{{\it \hat\bsvPsi}}(Y|Z=k,\bsx),
\label{eq: mean of MoE}\\ 
\!\!\!\!\V_{{\it \hat\bsvPsi}}(Y|\bsr,\bsx) &\! =\! & \sum_{k=1}^K \pi_k(\bsr;\hat \bsalpha_{n})
\big[\left(\E_{{\it \hat\bsvPsi}}(Y|Z=k,\bsx)\right)^2  + \V_{{\it \hat\bsvPsi}}(Y|Z=k,\bsx) \big] - \big[\E_{{\it \hat\bsvPsi}}(Y|\bsr,\bsx)\big]^2,
\label{eq: variance of MoE}
\end{eqnarray}}where $\E_{{\it \hat\bsvPsi}}(Y|Z=k,\bsx)$ and $\V_{{\it \hat\bsvPsi}}(Y|Z=k,\bsx)$  are respectively the component-specific (expert) means and variances. 
 The mean and the variance for the MoE models described here are given as follows.

\paragraph{NMoE}For the NMoE model, the normal expert means and variances are respectively $\E_{{\it \hat\bsvPsi}}(Y|Z=k,\bsx) = \hat \bsbeta^T_{k} \bsx$
and 
$\V_{{\it \hat\bsvPsi}}(Y|Z=k,\bsx) = \hat \sigma^2_{k}$. 
 
\paragraph{TMoE} For the TMoE model, by using the expressions of the mean and the variance of the $t$ distribution, it follows that for the TMoE model, for $\hat \nu_{k}>1$, the expert means are 
$\E_{{\it \hat\bsvPsi}}(Y|Z=k,\bsx) = \hat \bsbeta^T_{k} \bsx$
and, for $\hat \nu_{k}>2$, the expert variances are $\V_{{\it \hat\bsvPsi}}(Y|Z=k,\bsx) = \frac{\hat \nu_{k}}{\hat \nu_{k} - 2} ~ \hat \sigma^2_{k}$.

\section{Model-based clustering using the TMoE} 
\label{sec: MBC using the NNMoE}

It is natural to utilize the MoE models for a model-based clustering perspective  to provide a partition of the regression data into $K$ clusters. 
Model-based clustering using the TMoE, as in MoE in general, consists in assuming that the observed data $\{\bsx_i,\bsr_i,y_i\}_{i=1}^n$ are  generated from a $K$ component mixture of $t$ experts with parameter vector $\bsvPsi$. The mixture components can be  interpreted as clusters and hence each cluster can be associated with a mixture component. 
The problem of clustering therefore becomes the one of estimating the MoE parameters $\bsvPsi$, which is performed here by using dedicated EM algorithms. 
Once the parameters are estimated, the provided posterior component memberships $\hat \tau_{ik}$ defined in (\ref{eq: posterior prob TMoE})  represent a fuzzy partition of the data. A hard partition of the data can then be obtained  by applying the optimal Bayes' allocation rule, that is:  
\begin{eqnarray}
\hat{z}_i = \arg \max_{k=1}^K \hat \tau_{ik}
\label{eq: MAP rule for clustering}
\end{eqnarray}where $\hat{z}_i$ represents the estimated cluster label for the $i$th observation.

\section{Model selection for the NNMoE}
\label{sec: Model selection for the NNMoE}
One of the issues in mixture model-based clustering is model selection.  
The problem of model selection for the TMoE model presented  here in its general form, is equivalent to the one of choosing the optimal number of experts $K$, the degree $p$ of the polynomial regression and the degree $q$ for the logistic regression. 
The optimal value  of $(K,p,q)$ can be computed by using some model
selection criteria such as the Akaike Information Criterion (AIC) \citep{AIC}, the Bayesian Information Criterion (BIC) \citep{BIC} or the Integrated Classification Likelihood criterion (ICL)  \citep{ICL}, etc. 
The AIC and BIC are penalized observed data log-likelihood criteria which can be defined as functions to be maximized and are respectively given by:
\begin{eqnarray*}
\mbox{AIC}(K,p,q) &=& \log L(\hat{\bsPsi}) - \eta_{\bsvPsi},\\
\mbox{BIC}(K,p,q) &=&  \log L(\hat{\bsvPsi}) - \frac{\eta_{\bsvPsi} \log(n)}{2}.
\end{eqnarray*}The ICL criterion consists in a penalized complete-data log-likelihood and can be expressed as:
\begin{equation*}
\ICL(K,p,q) = \log L_c(\hat{\bsvPsi}) - \frac{\eta_{\bsvPsi} \log(n)}{2}.
\end{equation*}In the above, $\log L(\hat{\bsvPsi})$ and  $\log L_c(\hat{\bsvPsi})$ are respectively the incomplete (observed) data log-likelihood and the complete data log-likelihood, obtained at convergence of the E(C)M algorithm for the corresponding MoE model. The number of free parameters of the model $\eta_{\bsvPsi} $ is given by $\eta_{\bsvPsi} = K(p+q+3)-q-1$ for the NMoE model and $\eta_{\bsvPsi} = K(p+q+4)-q-1$ for the TMoE model.
 
However, note that in MoE it is common to use a gating functions modeled as logistic transformation of linear functions of the covariates, that is the covariate vector  in (\ref{eq: multinomial logistic}) is given by $\bsr_i = (1, r_i)^T$ (corresponding to $q=2$), $r_i$ being an univariate covariate variable. This is what we adopted in this work.
Moreover, for the case of linear experts, that is when the experts are linear regressors with parameter vector $\bsbeta_k$ for which the corresponding covariate vector $\bsx_i$ in  (\ref{eq: linear regression mean})  is given by $\bsx_i = (1, x_i)^T$ (corresponding to $p=2$), $x_i$ being an univariate covariate variable possibly different from $r_i$, the model selection reduces to choosing the number of experts $K$.
Here in the presented experiments we mainly consider this linear case for the expert components. Notice that the overall modeling problem is still non-linear and is adapted to fit non-linear regression functions. 

\section{Experimental study}  
\label{sec: Experimental study}

This section is dedicated to the evaluation of the proposed approach on simulated data and real-world data . 
We evaluated the performance of proposed EM algorithm by comparing it the standard normal MoE (NMoE) model \citep{jacobsME,jordanHME}  and the Laplace MoE of \citep{Nguyen2014-MoLE}\footnote{All the algorithms have been implemented in Matlab and the codes are available upon request from the author.} on both simulated and real-world data sets. 
%

\subsection{Initialization and stopping rules}
\label{ssec: initialization and stopping}
The parameters $\bsalpha_k$ ($k=1, \ldots, K-1$) of the mixing proportions are initialized randomly, 
including an initialization at the null vector for one run (corresponding to equal mixing proportions).
Then, the common parameters $(\bsbeta_k,\sigma^2_k)$ ($k=1,\ldots, K$) are initialized from a random partition of the data into $K$ clusters. This corresponds to fitting a normal MoE where the initial values of the parameters are respectively  given by  
(\ref{eq: beta_k update for NMoE}) and (\ref{eq: sigma2k update NMoE}) with the posterior memberships $\tau_{ik}$ replaced by the hard assignments $Z_{ik}$ issued from the random partition.
For the TMoE  model, the robustness parameters $\nu_k$ ($k=1,\ldots, K$) is initialized randomly in the range [1, 200]. For the LMoE model 
\begin{equation}
f(y|\bsr,\bsx;\bsvPsi) = \sum_{k=1}^K \pi_k(\bsr;\bsalpha) ~ \text{Laplace}(y; \mu(\bsx;\bsbeta_k), \lambda_k), 
\label{eq: LMoE model}
\end{equation}
the scale parameter $\lambda_k$ is initialized in a similar way as $\sigma^2_k$. 
Then, the  algorithms are stopped when the relative variation of the observed-data log-likelihood 
$\frac{\log L(\bsvPsi^{(m+1)})- \log L(\bsvPsi^{(m)})}{|\log L(\bsvPsi^{(m)})|}$ reaches a prefixed threshold (for example $\epsilon=10^{-6}$).  
For each model, this process is repeated 10 times and the solution corresponding the highest log-likelihood is finally selected.

\subsection{Experiments on simulation data sets}
In this section we perform an experimental study on simulated data sets to apply and assess the proposed model. 
Two sets of experiments have been performed. 
The first experiment aims at observing the effect of the sample size on the estimation quality and the second one aims at observing the impact of the presence of outliers in the data on the estimation quality, that is the robustness of the models.

\subsubsection{Experiment 1}
For this first experiment on simulated data, each simulated sample consisted of $n$ observations with increasing values of the sample size $n: 50, 100, 200, 500, 1000$.  
The simulated data are generated from a two component mixture of linear experts, that is $K=2, p=q=1$.
The covariate variables $(\bsx_i, \bsr_i)$ are simulated such that $\bsx_i = \bsr_i = (1,x_i)^T$ where $x_i$ is
simulated uniformly over the interval $(-1, 1)$. 
We consider each of the three models  (NMoE, LMoE, TMoE) for data generation, that is, given the covariates, the response $y_i|\{\bsx_i,\bsr_i;\bsvPsi\}$ is simulated according to the generative process of the models (\ref{eq: normal MoE}), (\ref{eq: LMoE model}), and (\ref{eq: TMoE}).
For each generated sample, we fit each of the four models.   
Thus, the results are reported for all the models with data generated from each of the two models. We consider the mean square error (MSE) between each component of the true parameter vector and the estimated one, which is given by $\parallel \bsvPsi_j - \hat{\bsvPsi}_j\parallel^2$.
%
The squared errors
 are averaged on 100 trials. 
The used simulation parameters $\bsvPsi$ for each model are given in Table \ref{tab: simulation parameters situation 1}.
\begin{table}[H]
\centering
\small
\begin{tabular}{l | l l l l l }
\hline
& \multicolumn{5}{c}{parameters} \\
\hline
\hline
component 1 & $\bsalpha_1=(0, 10)^T$&$\bsbeta_1=(0,1)^T$ & $\sigma_1=0.1$ & $\nu_1 = 5$ & $\lambda_1 = 0.1$\\
component 2 & $\bsalpha_2=(0, 0)^T$&$\bsbeta_2=(0,-1)^T$  &$\sigma_2=0.1$ & $\nu_2 = 7$ & $\lambda_2 = 0.1$\\
 \hline
\end{tabular}
\caption{Parameter values used in simulation.}
\label{tab: simulation parameters situation 1}
\end{table}

\subsubsection{Obtained results}
 
 Table \ref{tab. MSE for the paramters: TMoE->TMoE} shows the obtained results in terms of the MSE for the TMoE.
 One can observe that the parameter estimation error is decreasing as $n$ increases, which illustrates the convergence property of the maximum likelihood estimator of the model. 
 For details on the convergence property of the MLE for MoE, see for example \citep{Jiang_and_tanner_IEEEinft_99}.  
 One can also observe that the error decreases significantly  for $n\geq 500$, especially for the regression coefficients and the scale parameters.
{\setlength{\tabcolsep}{3pt
\begin{table}[htbp]
\centering
{\small
\begin{tabular}{c c c  c c c c c c c c}
\hline
param. & $\alpha_{10}$ & $\alpha_{11}$ & $\beta_{10}$ & $\beta_{11}$ & $\beta_{20}$ & $\beta_{21}$ & $\sigma_{1}$& $\sigma_{2}$ & $\nu_{1}$ & $\nu_{2}$ \\ 
$n$		& & & & & & & & & &\\
 \hline
 \hline
$50$ 	& 1.3059 &  6.4611 & 0.0214130  & 0.0290114 & 0.0044140 & 0.0192600 &  0.0010655 & 0.0003317 & 37.956 &  11.722\\
$100$	& 1.2150 &  4.5056 & 0.0024706 &  0.0117546 & 0.0005275 & 0.0007891 &  0.0001450 & 0.0002301 & 6.1528 &  10.412\\
$200$	& 0.0341 &  3.8193 & 0.0001553 &  0.0007335 & 0.0002022 & 0.0005061 &  0.0000504 & 0.0000262 & 2.0975 &  6.3710\\
$500$	& 0.0356 &  2.2633 & 0.0000112 &  0.0000214 & 0.0001337 & 0.0002163 &  0.0000126 & 0.0000007 & 0.4859 &  5.4937\\
$1000$	& 0.0053 &  1.2510 & 0.0000018 &  0.0000258 & 0.0000005 & 0.0000427 &  0.0000126 & 0.0000004 & 0.0014 &  2.7844\\
 \hline
\end{tabular}
}
\caption{\label{tab. MSE for the paramters: TMoE->TMoE} MSE between each component of the estimated parameter vector of  the TMoE model and the actual one for a varying sample size $n$.}
\end{table}
}
In addition to the previously showed results, we plotted in Figures
\ref{fig. TwoClust-NMoE_NMoE}, \ref{fig. TwoClust-NMoE_LMoE} and \ref{fig. TwoClust-NMoE_TMoE} the estimated quantities provided by applying the proposed model and their true counterparts for $n=500$ for the same the data set which was generated according the normal MoE model. 
\begin{figure}[htbp]
   \centering 
   \begin{tabular}{cc}
   \includegraphics[width=5cm]{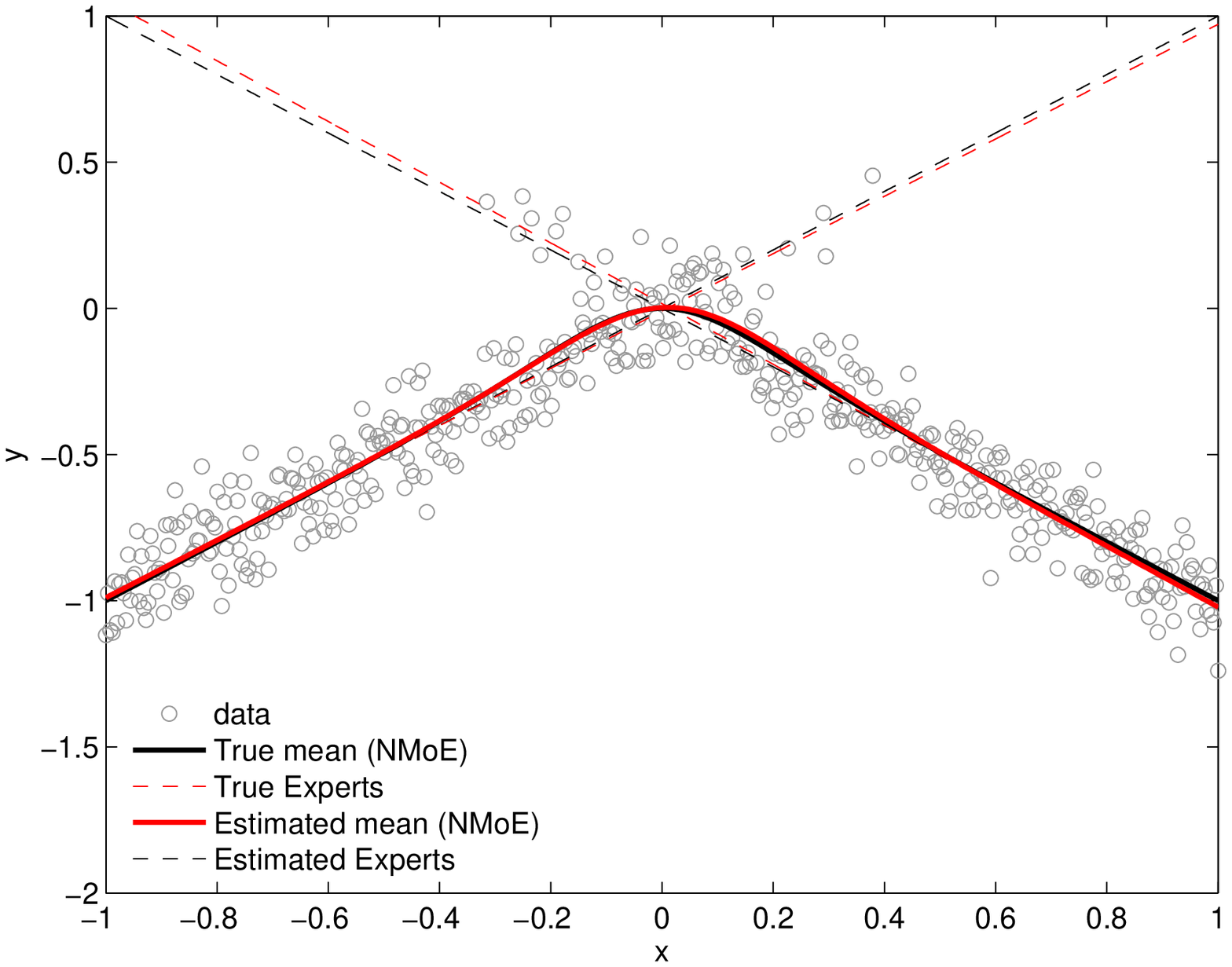}&  
   \includegraphics[width=5cm]{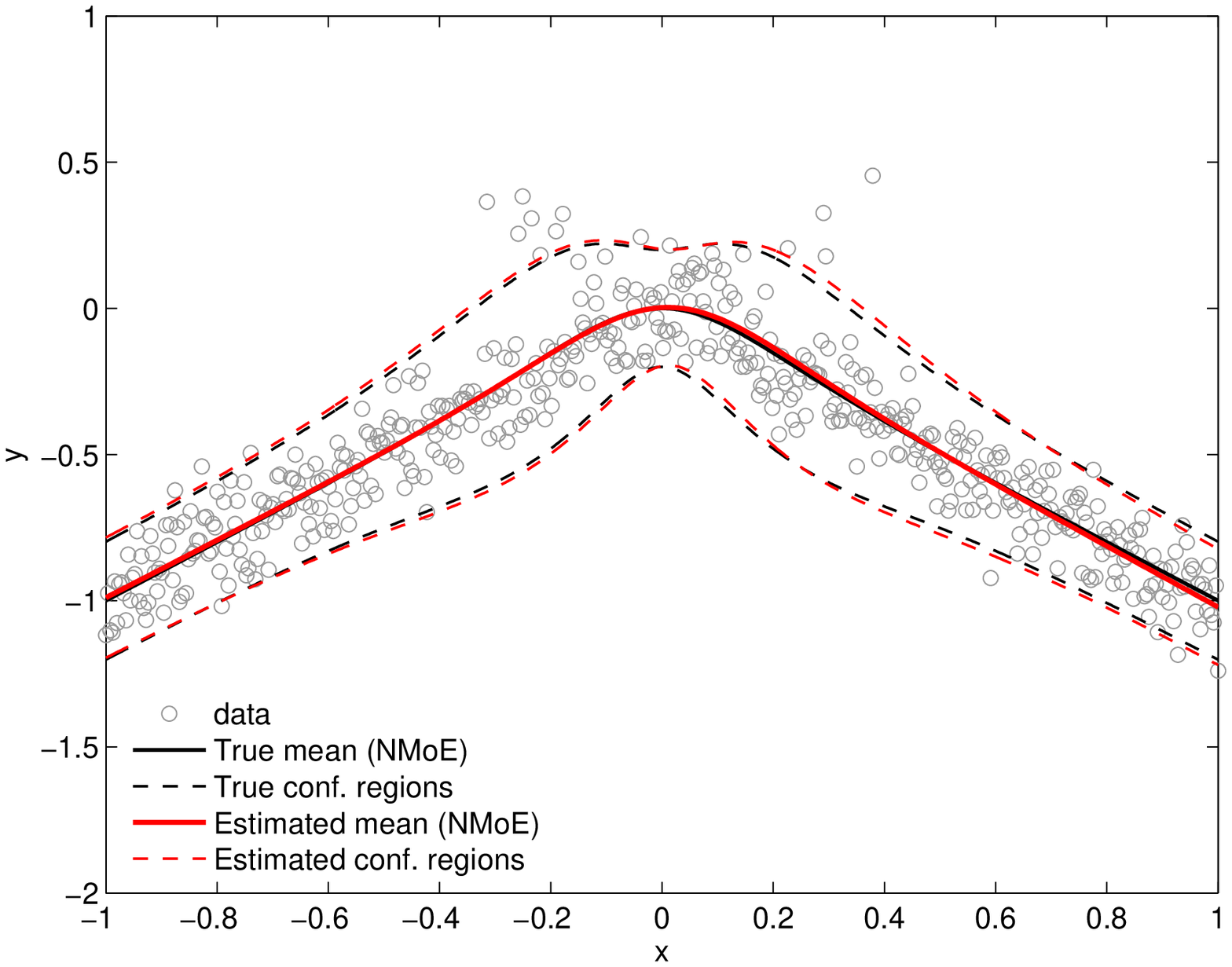}\\
\hspace{0.2cm}\includegraphics[width=4.8cm]{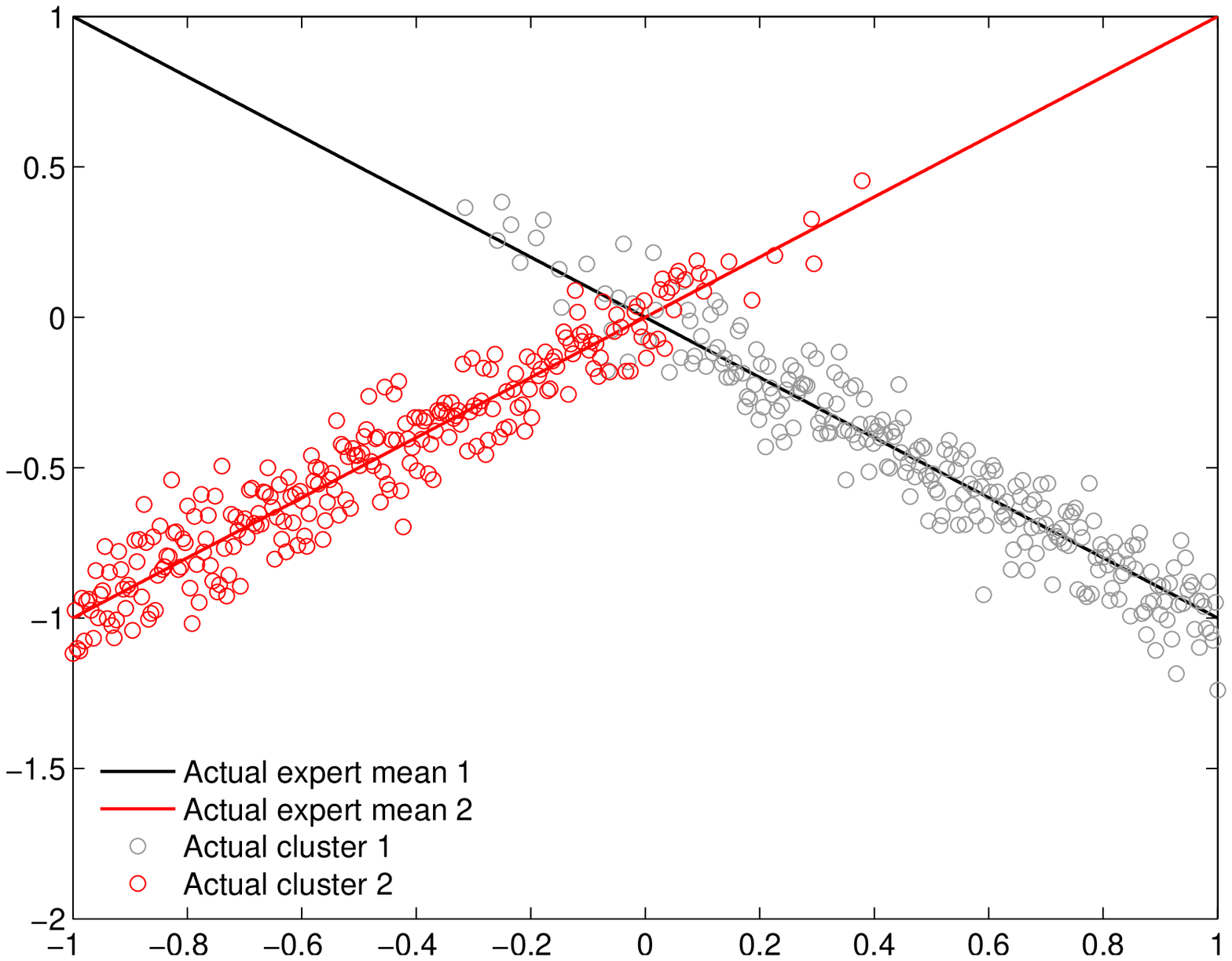}&
\hspace{0.2cm}\includegraphics[width=4.8cm]{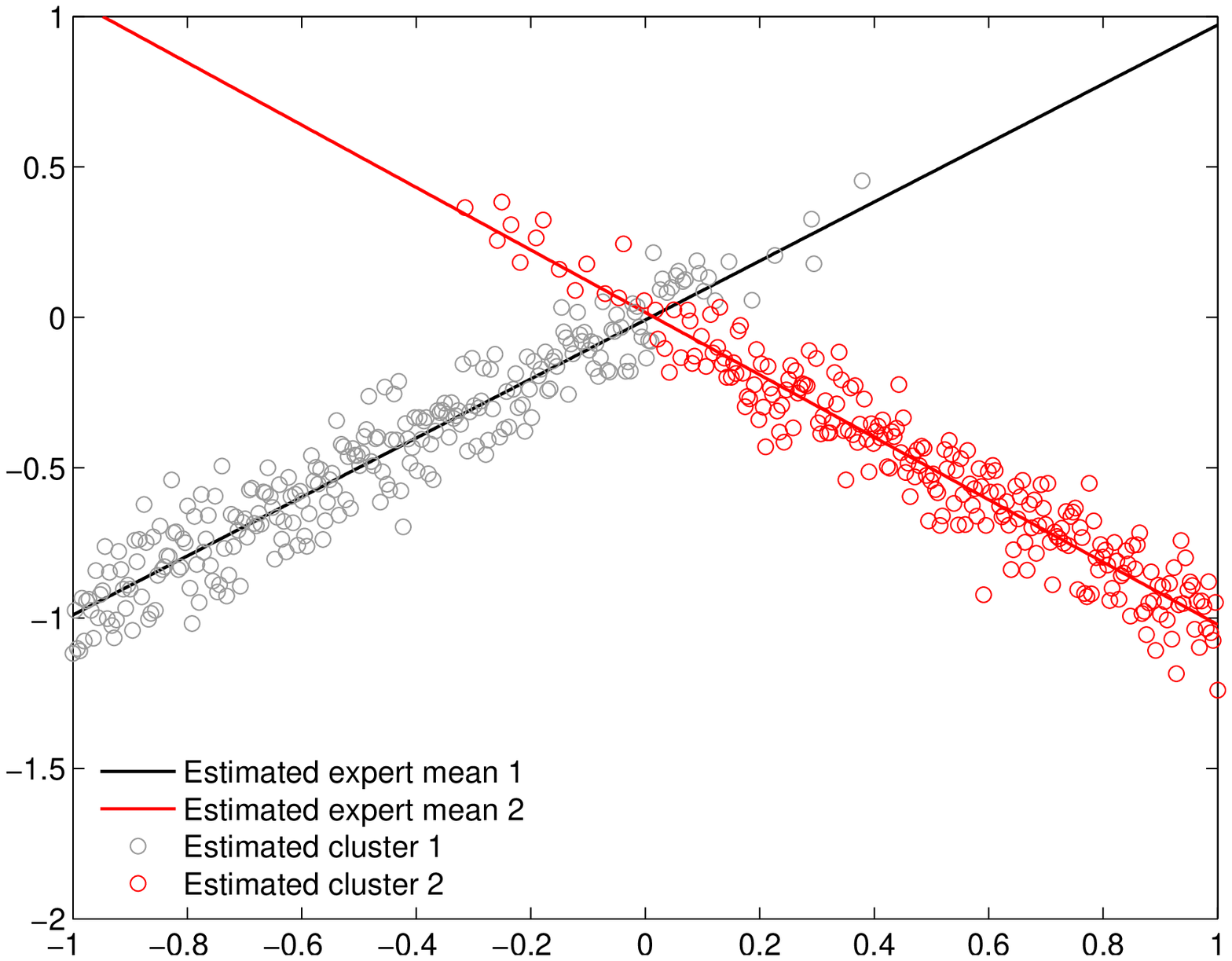}
   \end{tabular}
      \caption{\label{fig. TwoClust-NMoE_NMoE}Fitted NMoE model to a data set generated according to the NMoE model.}
\end{figure}
\begin{figure}[htbp]
   \centering  
   \begin{tabular}{cc}
   \includegraphics[width=5cm]{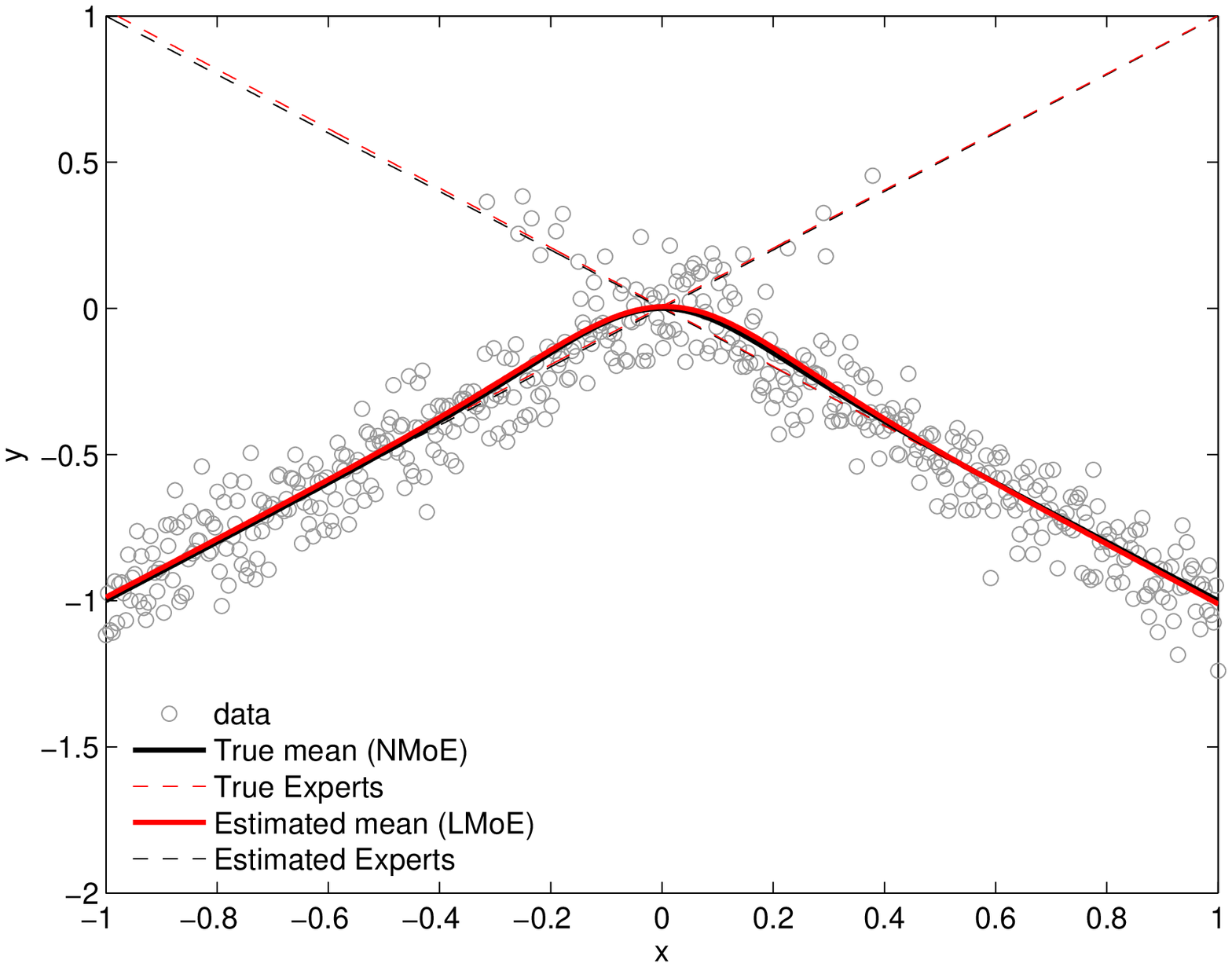}&  
   \includegraphics[width=5cm]{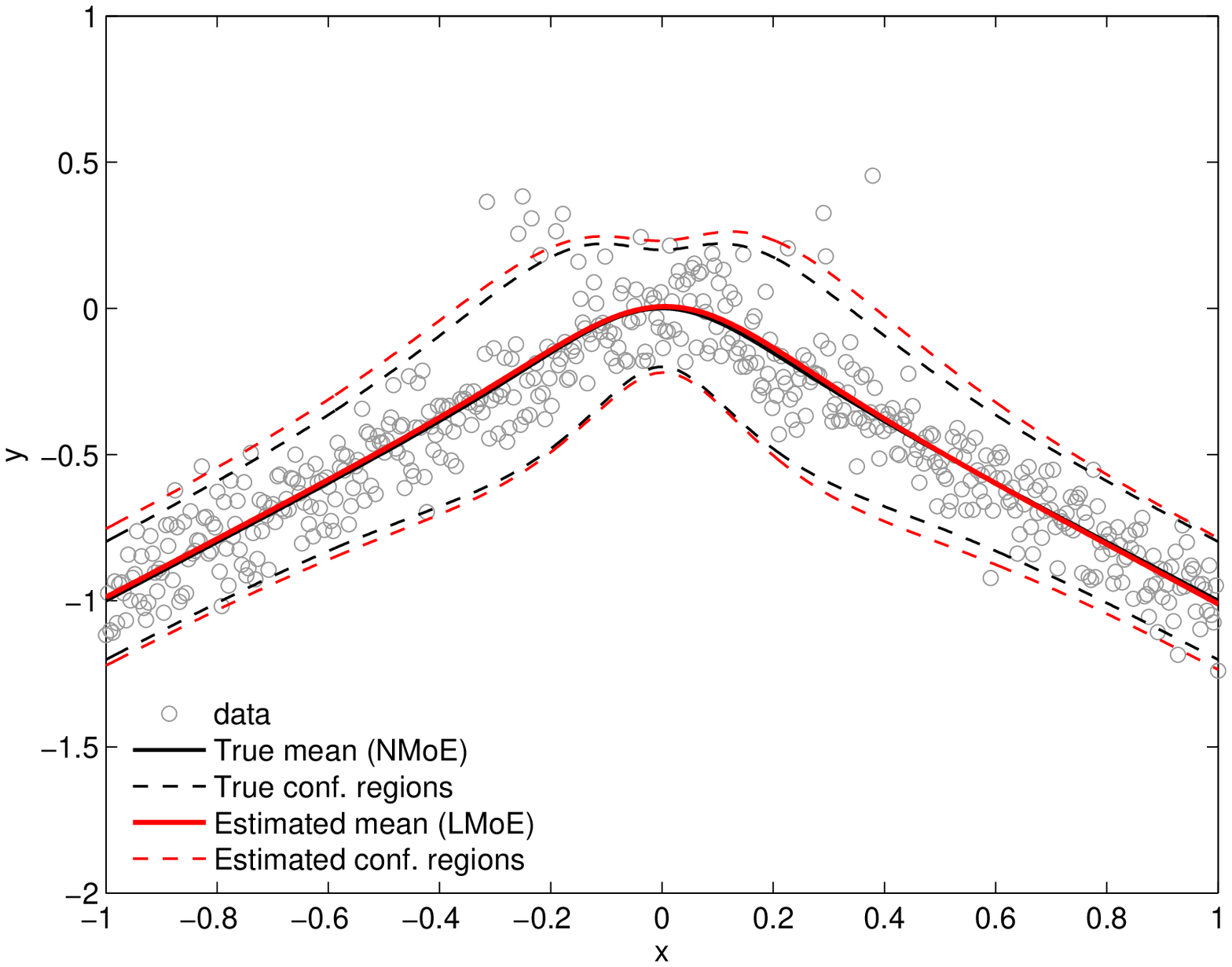}\\
\hspace{0.2cm}\includegraphics[width=4.8cm]{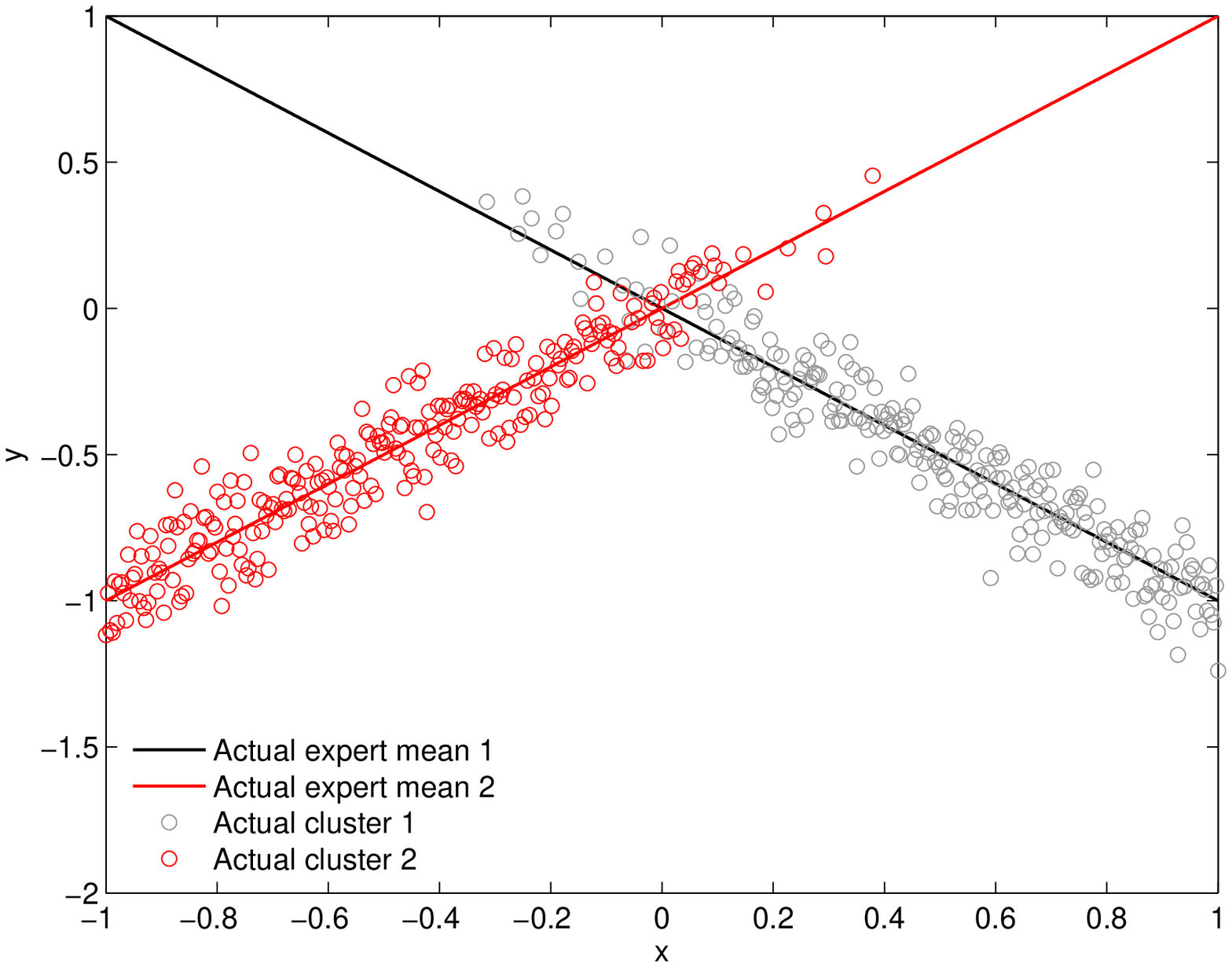}&
\hspace{0.2cm}\includegraphics[width=4.8cm]{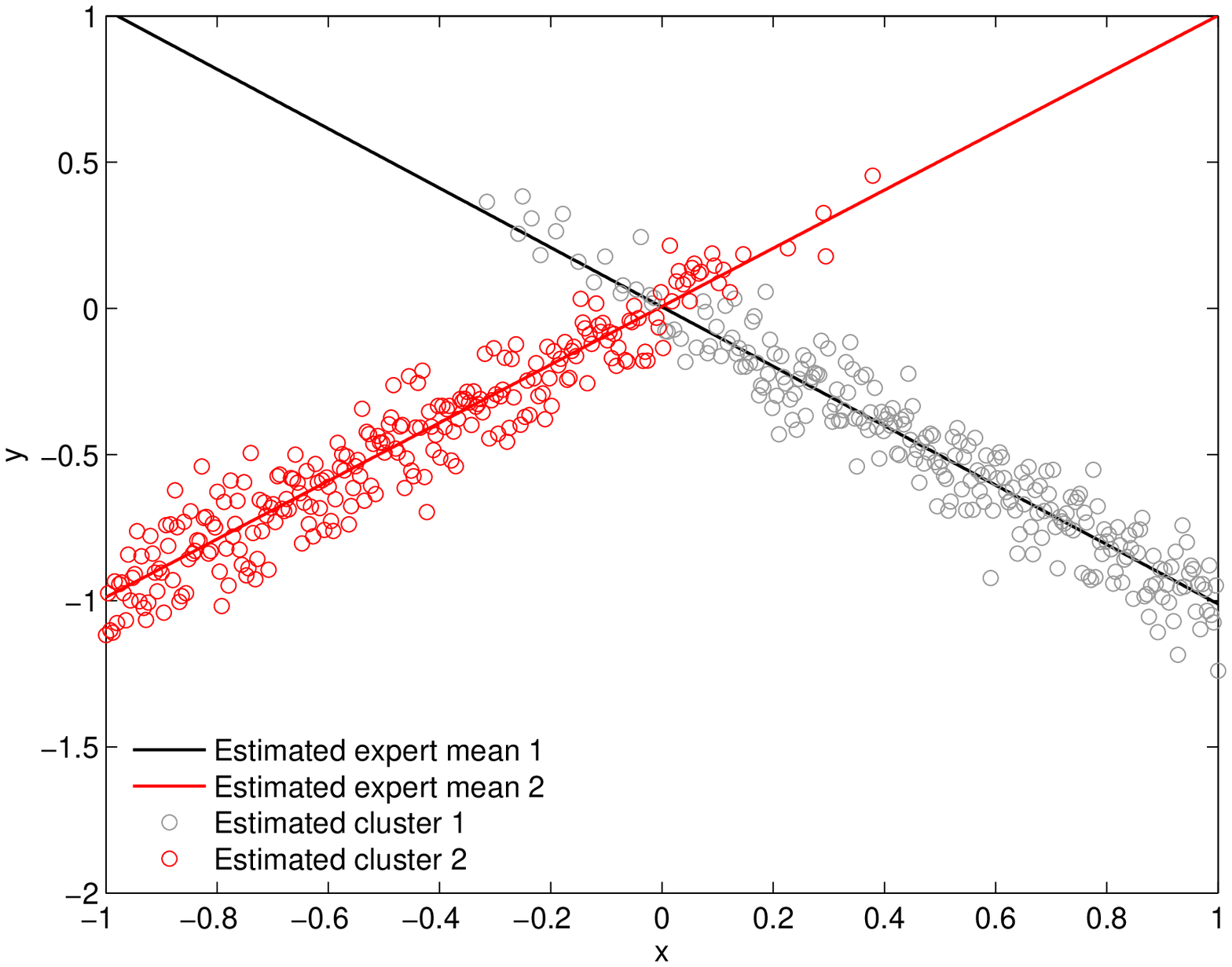}
   \end{tabular}
      \caption{\label{fig. TwoClust-NMoE_LMoE}Fitted  LMoE model to a data set generated according to the NMoE model.}
\end{figure}
\begin{figure}[htbp]
   \centering  
   \begin{tabular}{cc}
   \includegraphics[width=5cm]{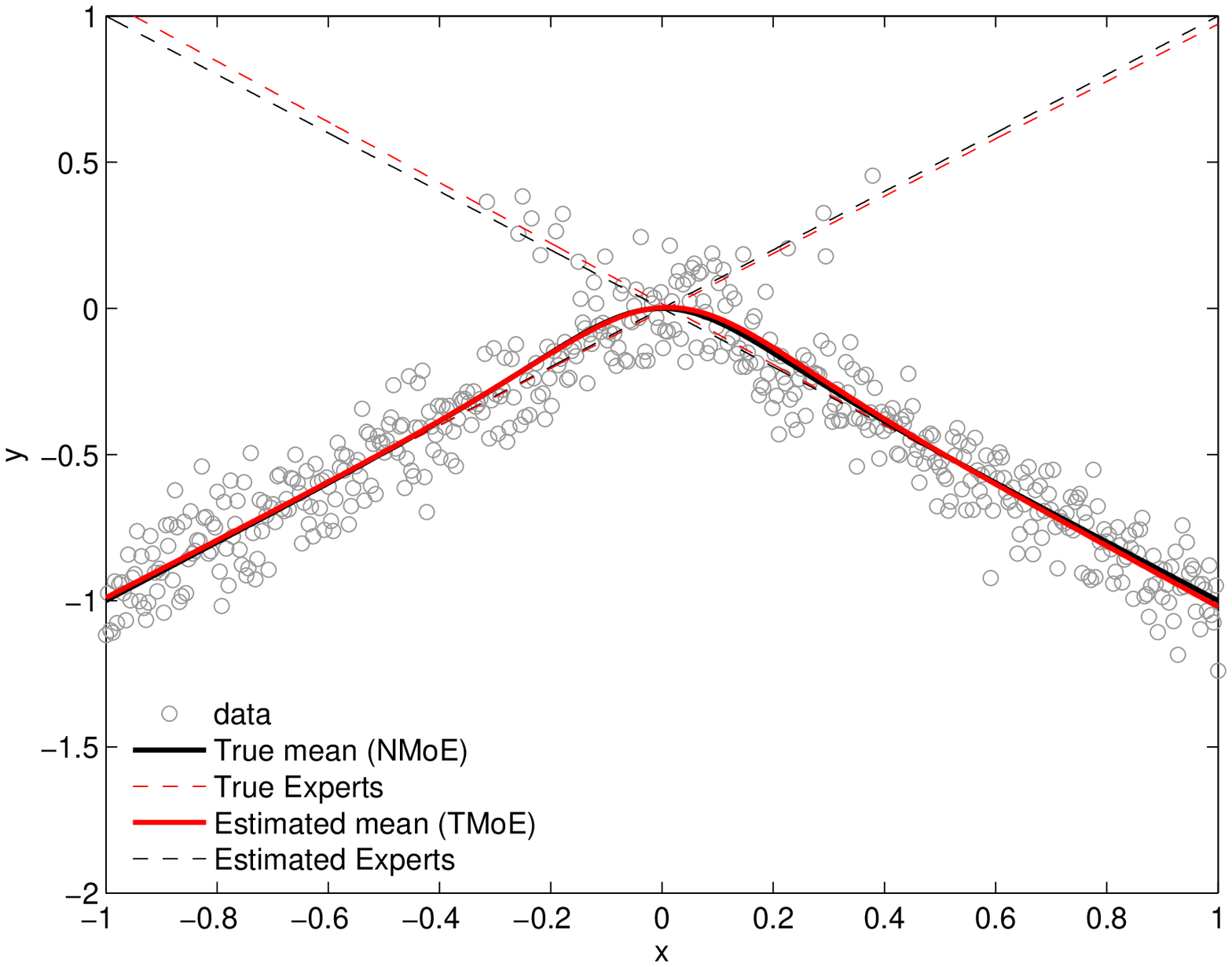}&  
   \includegraphics[width=5cm]{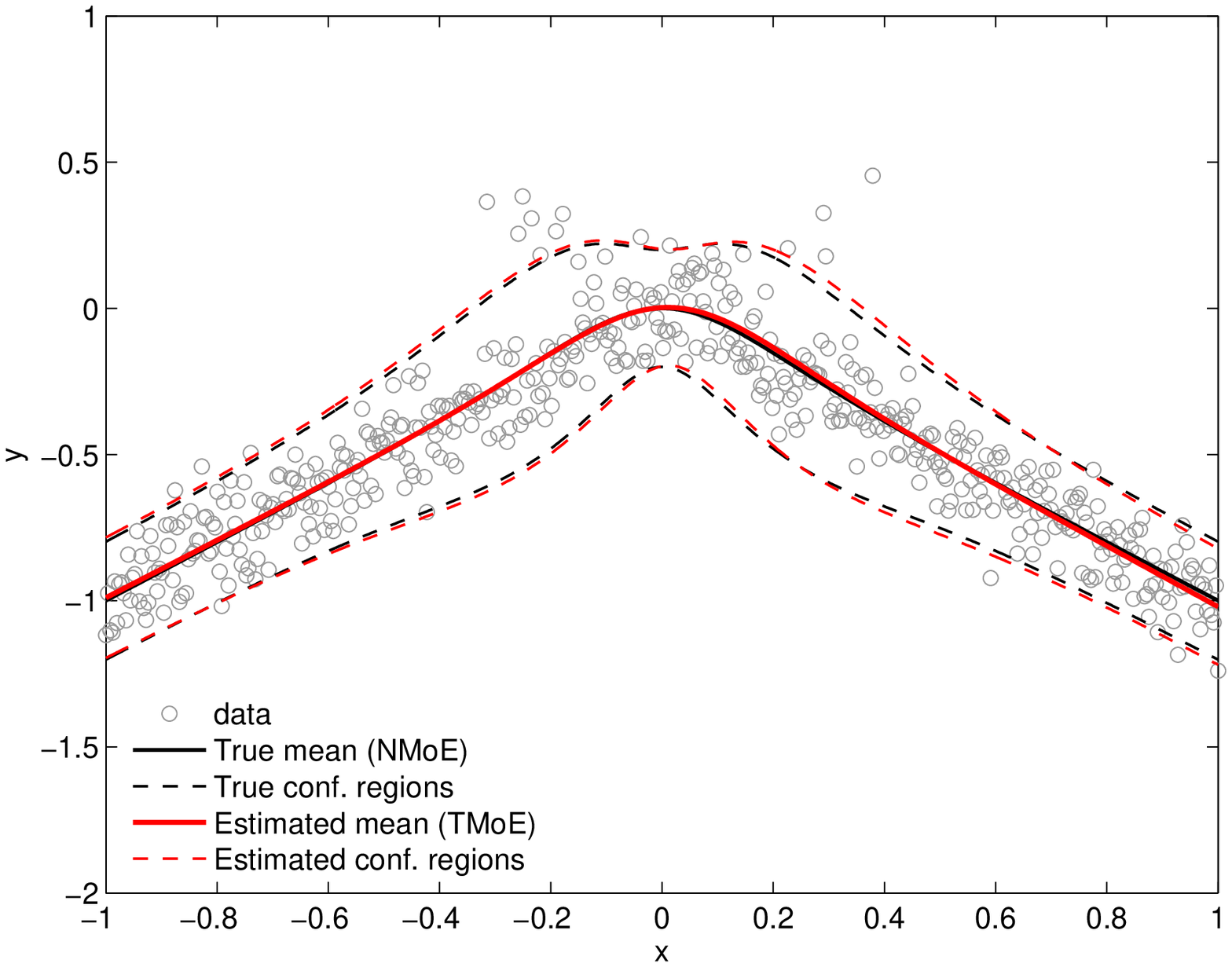}\\
\hspace{0.2cm}\includegraphics[width=4.8cm]{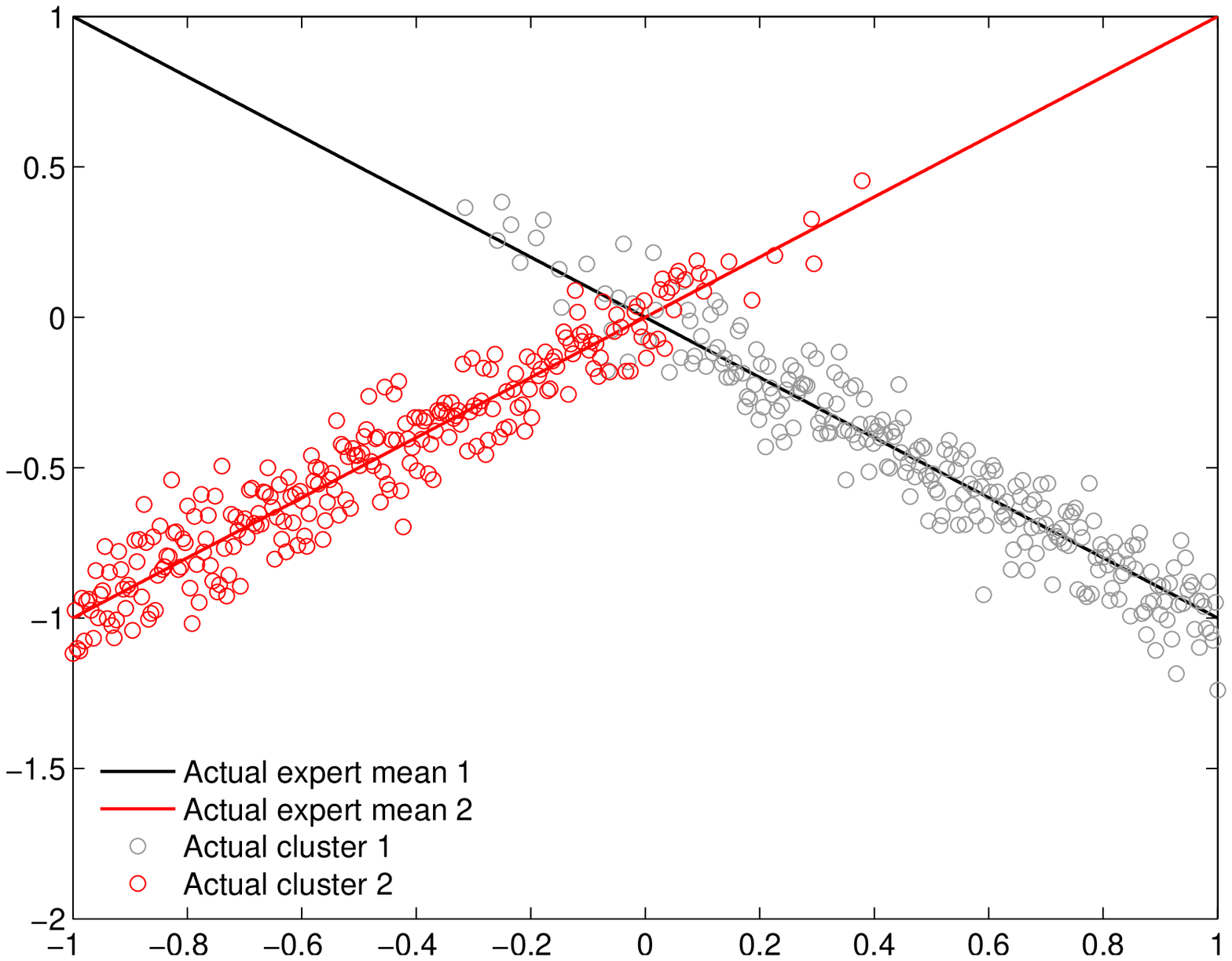}&
\hspace{0.2cm}\includegraphics[width=4.8cm]{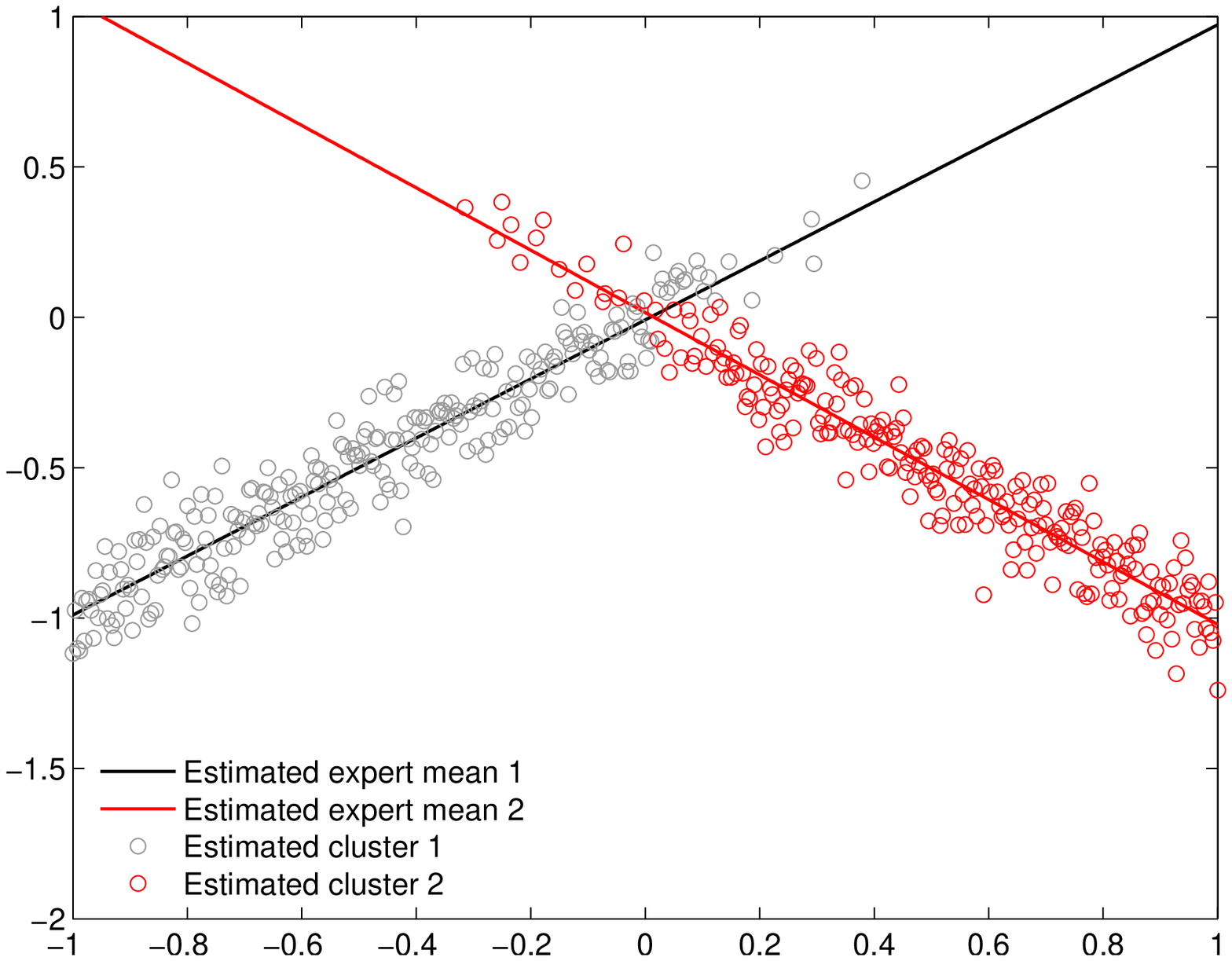}
   \end{tabular}
      \caption{\label{fig. TwoClust-NMoE_TMoE}Fitted  TMoE model to a data set generated according to the NMoE model.}
\end{figure}
The upper-left plot of each of these figures shows the estimated mean function, the estimated expert component mean functions, and the corresponding true ones.
The upper-right plot shows the estimated mean function and the estimated confidence region computed as plus and minus twice the estimated (pointwise) standard deviation of the model as presented in Section \ref{sec: Prediction using the NNMoE}, and their true counterparts. 
The bottom-left plot shows the true expert component mean functions and the true partition, and the bottom-right plot shows their estimated counterparts.
%

One can clearly see that the estimations provided by the proposed model are quasi identical to the true ones which  correspond to those of the NMoE model in this case. 
This provides an additional support to the fact that the proposed algorithm perform well and the  proposed TMoE model is a good generalization of the normal MoE (NMoE), as it clearly approaches the NMoE as shown in these simulated examples.
The proposed TMoE also provides quasi-identical results to the LMoE model. 

\subsubsection{Experiment 2}

In this experiment we examine the robustness of the proposed model to outliers  versus the standard NMoE one. 
For that, we considered each of the three models (NMoE, LMoE, TMoE) for data generation. For each generated sample, each of the two models in considered for the inference. The data were generated  exactly in the same way as in Experiment 1, except for some observations which were generated with a probability $c$ from a class of outliers. We considered the same class of outliers as in \citet{Nguyen2014-MoLE}, that is, the predictor $x$  is generated uniformly over the interval $(-1, 1)$ and the response $y$ is set the value $-2$. 
We apply the MoE models by setting the covariate vectors as before, that is,  $\bsx = \bsr  = (1, x)^T$. 
We considered varying probability of outliers $c = 0\%, 1\%, 2\%, 3\%, 4\%, 5\%$ and the sample size of the generated data is $n=500$. An example of simulated sample containing $5\%$ outliers is shown in Figure \ref{fig. TwoClust-Outliers-NMoE_NMoE}. 
As a criterion of evaluation of the impact of the outliers on the quality of the results, we considered the MSE between the true regression mean function and the estimated one. This MSE is calculated as
$\frac{1}{n}\sum_{i=1}^n\!\parallel \! \E_{{\it \bsvPsi}}(Y_i|\bsr_i,\bsx_i) - \E_{{\it \hat\bsvPsi}}(Y_i|\bsr_i,\bsx_i)\!\parallel^2$ where the expectations are computed as in Section \ref{sec: Prediction using the NNMoE}.
%
\subsubsection{Obtained results}

Table \ref{tab. MSE for the mean function - Noisy simulated data : All->all} shows,  for each of the two  models, the results in terms of mean squared error (MSE) between the true mean function and the estimated one, for an increasing number of outliers in the data.
First, one can see that, when there is no outliers ($c=0\%$), the error of the TMoE is less than those of the NMoE model, for the two situations, that is including the case where the data are not generated according to the TMoE model, which is somewhat surprising. This includes the case where the data are generated according to the NMoE model, for which the TMoE error is slightly less than the one of the NMoE model. 
Then, it can be seen that when there is outliers, the TMoE model clearly outperforms the NMoE model for all the situations. 
This confirms that the TMoE model is much more robust to outliers compared to the normal  one because the expert components in TMoE follow a robust distribution, that is the $t$ distribution. 
Furthermore, it can be seen that, when the number of outliers is increasing, the increase in the error of the NMoE model is more pronounced compared to the one of the TMoE  model. The error for   the TMoE may indeed slightly increase, remains stable or even slightly decreases in some situations when the data are generated according to the TMoE model. This supports the expected robustness of the TMoE
and the fact that the NMoE is  severely affected by outliers.  
To make comparison with the LMoE, whih is also clearly more robust that the NMoE, it can be seen that for some situations the LMoE provides better results compared to the TMoE, however, the overall results favorites the TMoE model, namely in the situation where the noise is relatively high (5\% of outliers). 
{\setlength{\tabcolsep}{6pt
\begin{table}[htbp]
\centering
{\small
\begin{tabular}{l l c c c c c c }
\hline
& \hspace{1cm}$c$& $0\%$ & $1\%$ & $2\%$ & $3\%$ & $4\%$ & $5\%$\\
\multicolumn{2}{c}{Model} & & & & & &\\
 \hline
 \hline 
\multirow{2}{*}{\rotatebox[origin=c]{0}{NMoE}}
& NMoE    & 	0.000178	& 0.001057 & 0.001241 & 0.003631 &	 0.013257 &	0.028966 \\
&LMoE		& 	\underline{0.000144} & 	\underline{0.000389} &	0.000686	&	 \underline{0.000153}	&	0.000296  	&	0.000121\\
& TMoE    & 	0.000168 & 0.000566 & \underline{0.000464} & 0.000221 &	 \underline{0.000263} &	\underline{0.000045} \\ 
 \hline 
 \hline  
\multirow{2}{*}{\rotatebox[origin=c]{0}{LMoE}}
& NMoE & 0.000287	&	0.003830 &	0.003740	&	0.010631	&	0.021247	&	0.026198\\
& LMoE & \underline{0.000126}	&	0.000378 &	\underline{0.000125}	&	0.000270	&	\underline{0.000165}	&	0.000605\\
& TMoE & 0.000183 &	\underline{0.000273} & 	0.000236 	&	\underline{0.000182}	&	0.000168	&  \underline{0.000070}\\
 \hline 
 \hline  
\multirow{2}{*}{\rotatebox[origin=c]{0}{TMoE}}
& NMoE    & 0.000257 & 0.0004660 & 0.002779 & 0.015692 & 0.005823 & 0.005419 \\
& LMoE 	& 0.000288	&	0.0004568 &	0.000205	&	\underline{0.000133}	&	\underline{0.000146}	&	0.000307 \\
& TMoE    & \underline{0.000252} & \underline{0.0002520} & \underline{0.000144} & 0.000157 & 0.000488 & \underline{0.000245}\\
\hline
\end{tabular}
}
\caption{\label{tab. MSE for the mean function - Noisy simulated data : All->all}MSE between the estimated mean function and the true one for each of the four  models for a varying probability $c$ of outliers for each simulation. The first column indicates the model used for generating the data and the second one indicates the model used for inference.}
\end{table} 
}
%
%
To highlight the robustness to noise of the TMoE model, in addition to the previously shown numerical results, 
figures 
\ref{fig. TwoClust-Outliers-NMoE_NMoE}, 
\ref{fig. TwoClust-Outliers-NMoE_LMoE}, 
and
\ref{fig. TwoClust-Outliers-NMoE_TMoE}
show an example of results obtained on the same data set by, respectively, the NMoE, the LMoE, and the TMoE. The data are generated by the NMoE model and contain $c=5\%$ of outliers. 

In this example, we clearly see that  the NMoE model is severely affected by the outliers. It provides a rough fit especially for the second component whose estimation is affected by the outliers.
However, one can  see that  the TMoE  model  provides a precise fit; the estimated mean functions and expert components are very close to the true ones. The TMoE is  robust to outliers, in terms of estimating the true model as well as in terms of estimating the true partition of the data (as shown in the middle plots).  The solution is also very close to the one provided by the LMoE model.
 %
Notice that for the TMoE  the confidence region is not shown because for this situation the estimated degrees of freedom are less than $2$ ($1.5985$ and   $1.5253$) for the TMoE; Hence the variance for the TMoE in that case is not defined (see Section \ref{sec: Prediction using the NNMoE}). 
The TMoE  model provides indeed components with small degrees of freedom corresponding to highly heavy tails, which allow to handle outliers in this noisy case.
\begin{figure}[H]
   \centering 
\begin{tabular}{cc}
   \includegraphics[width=5cm]{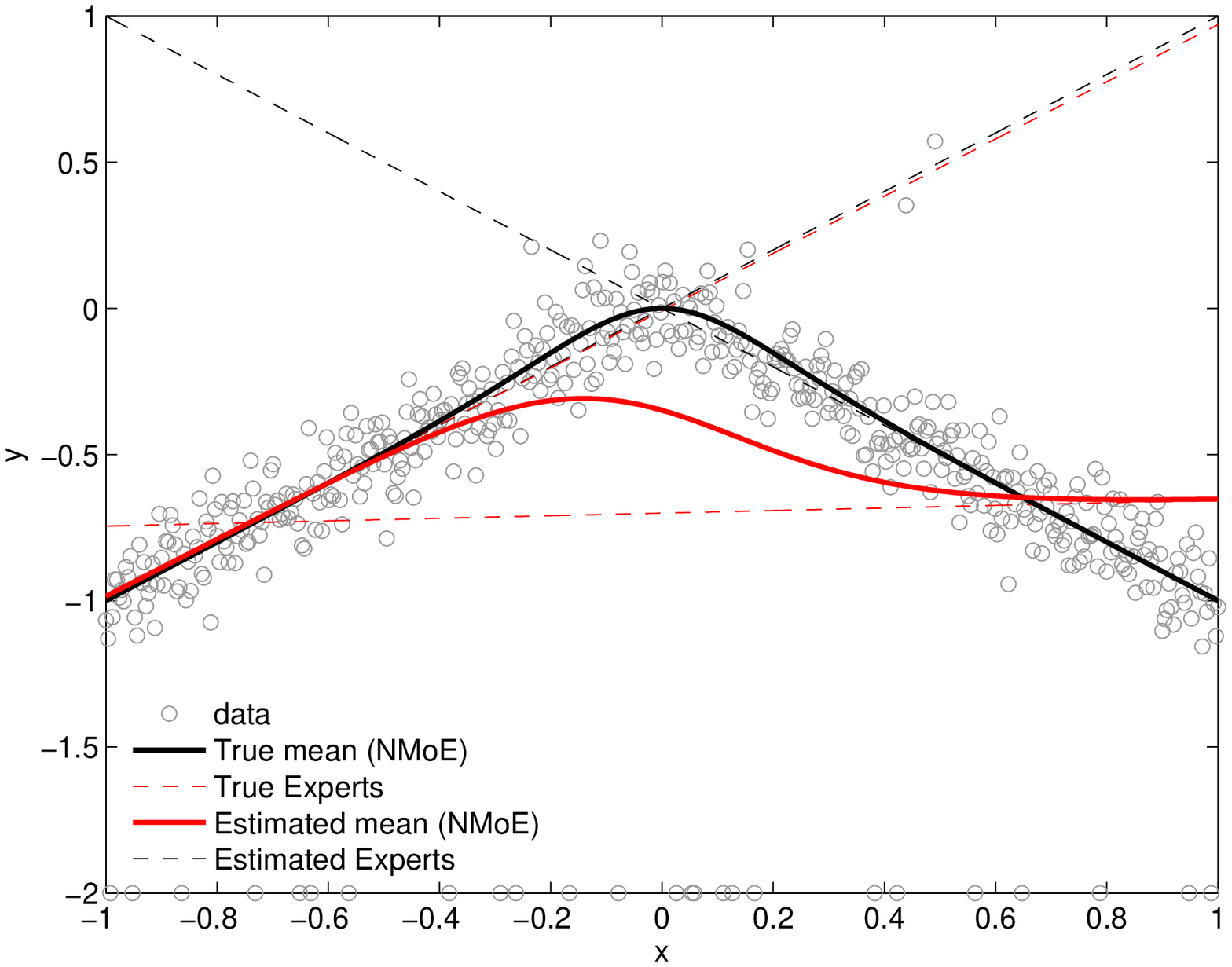}&  
   \includegraphics[width=5cm]{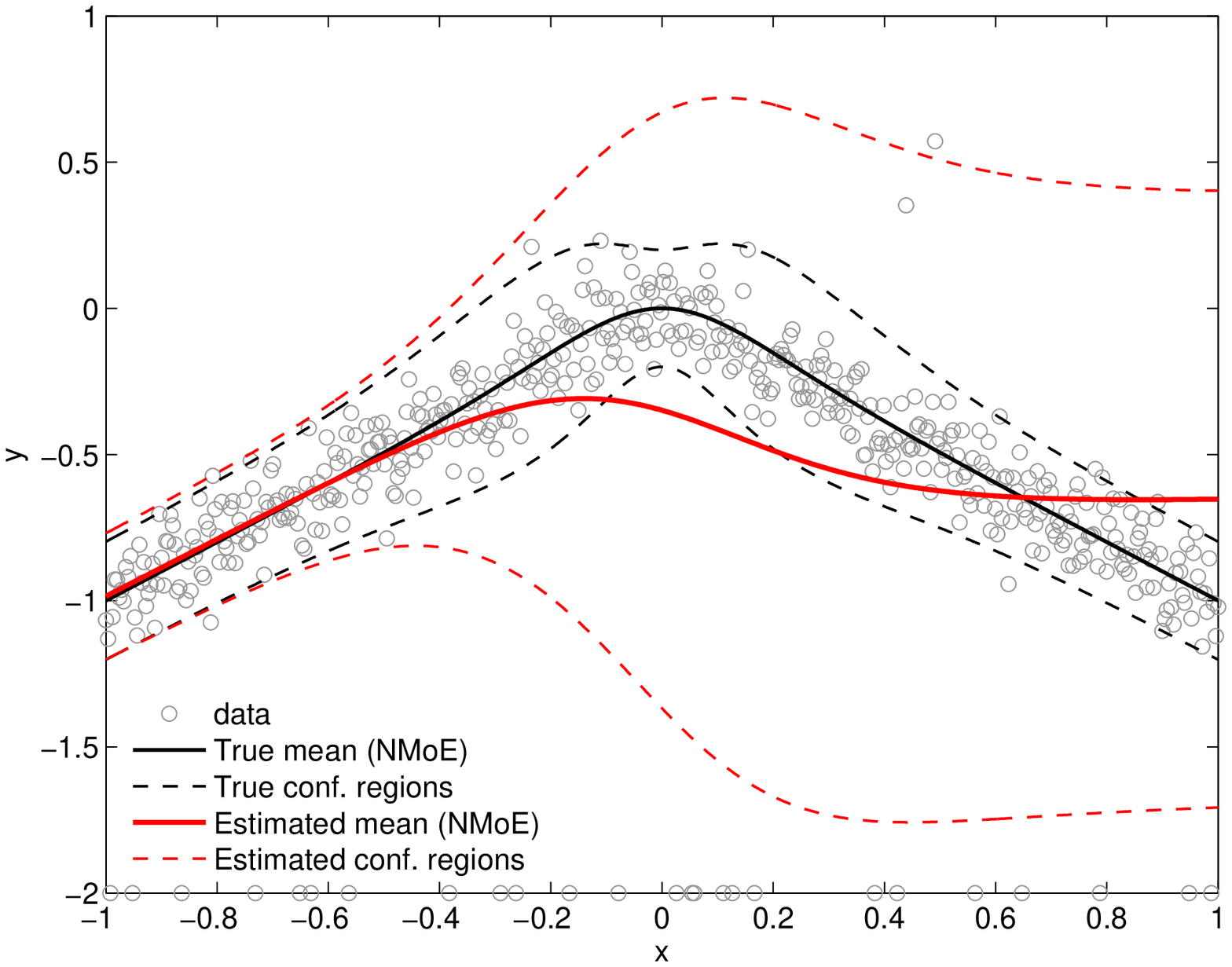}\\
   \hspace{0.2cm}\includegraphics[width=4.8cm]{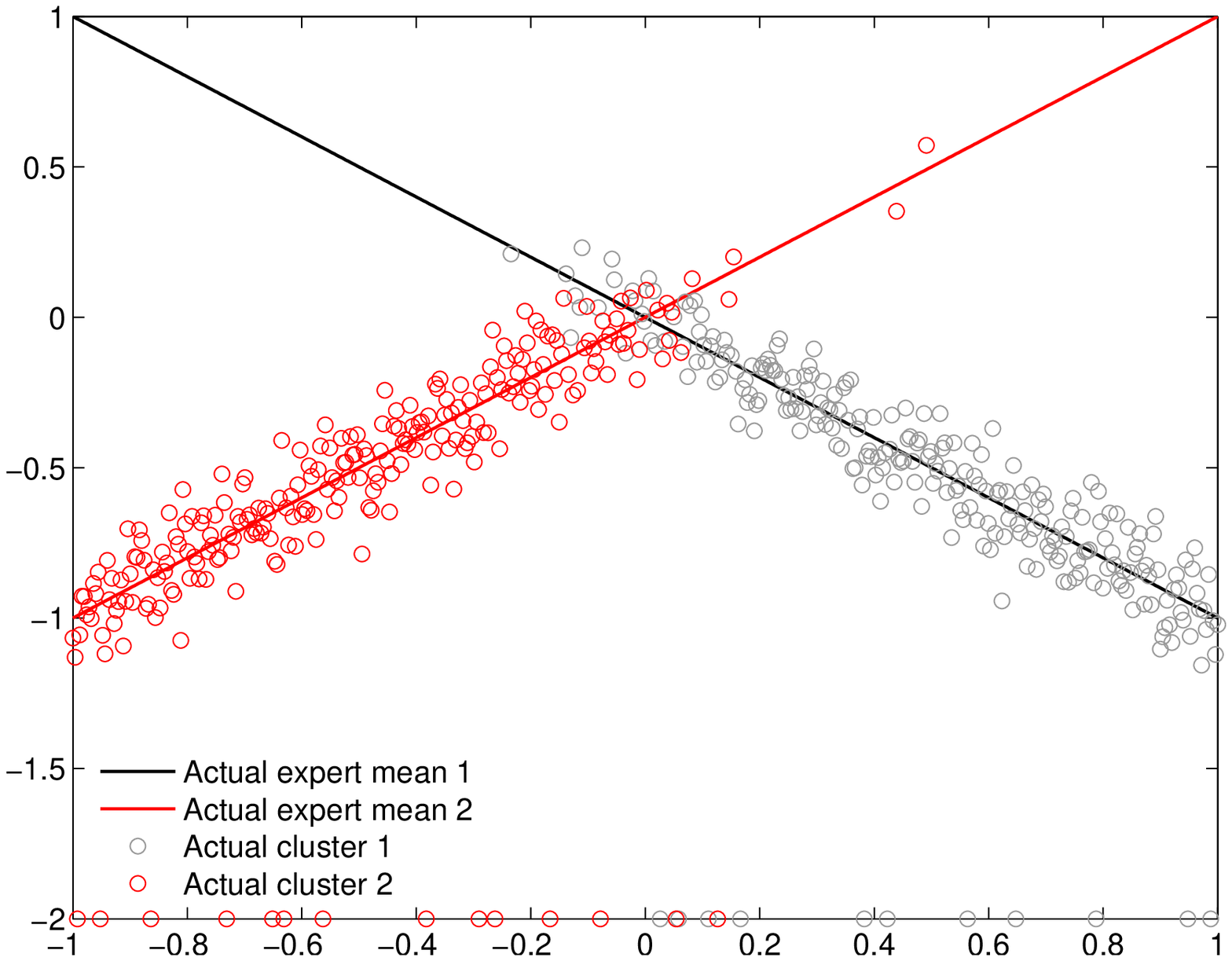}&
   \hspace{0.2cm}\includegraphics[width=4.8cm]{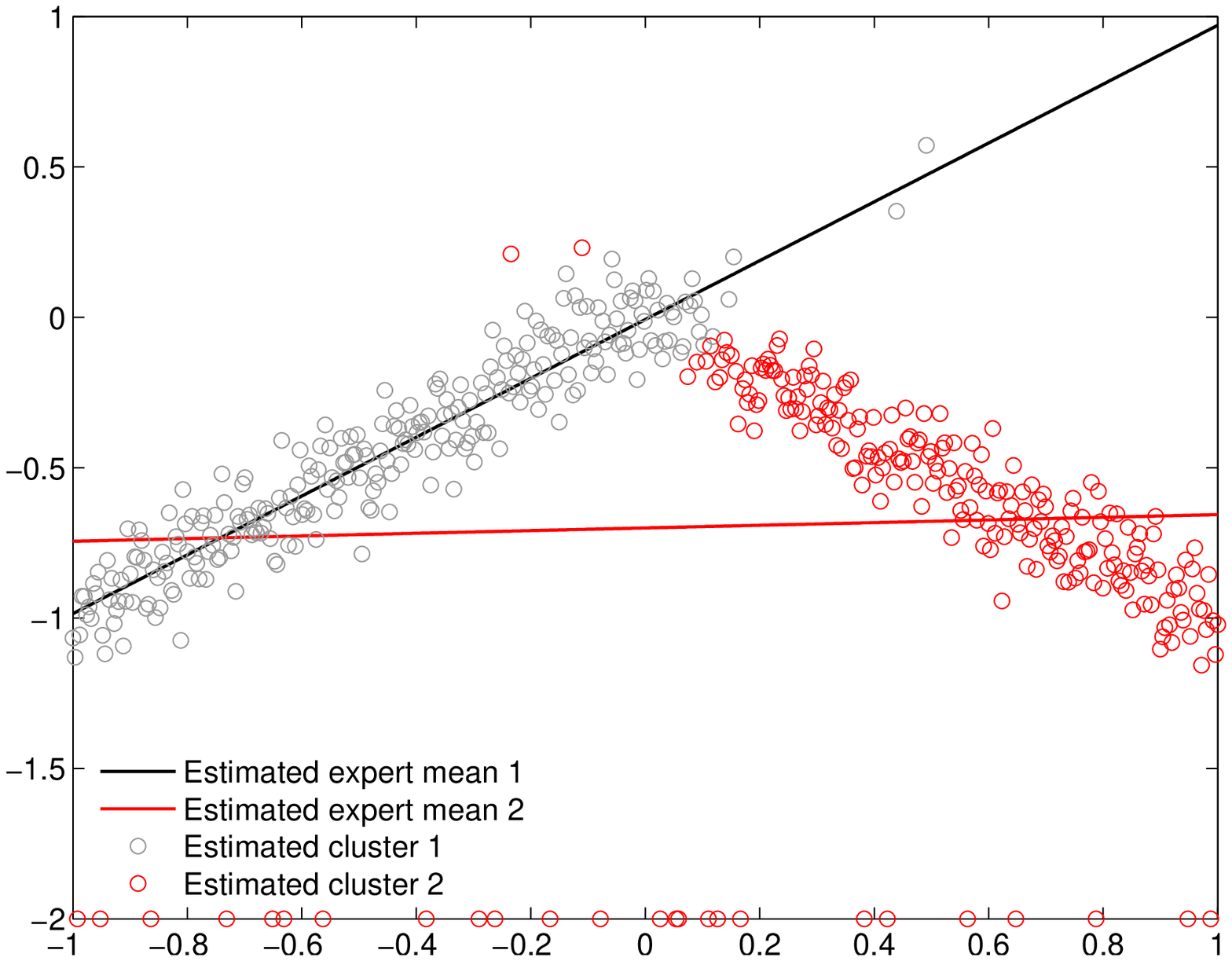} \\
   \includegraphics[width=5cm]{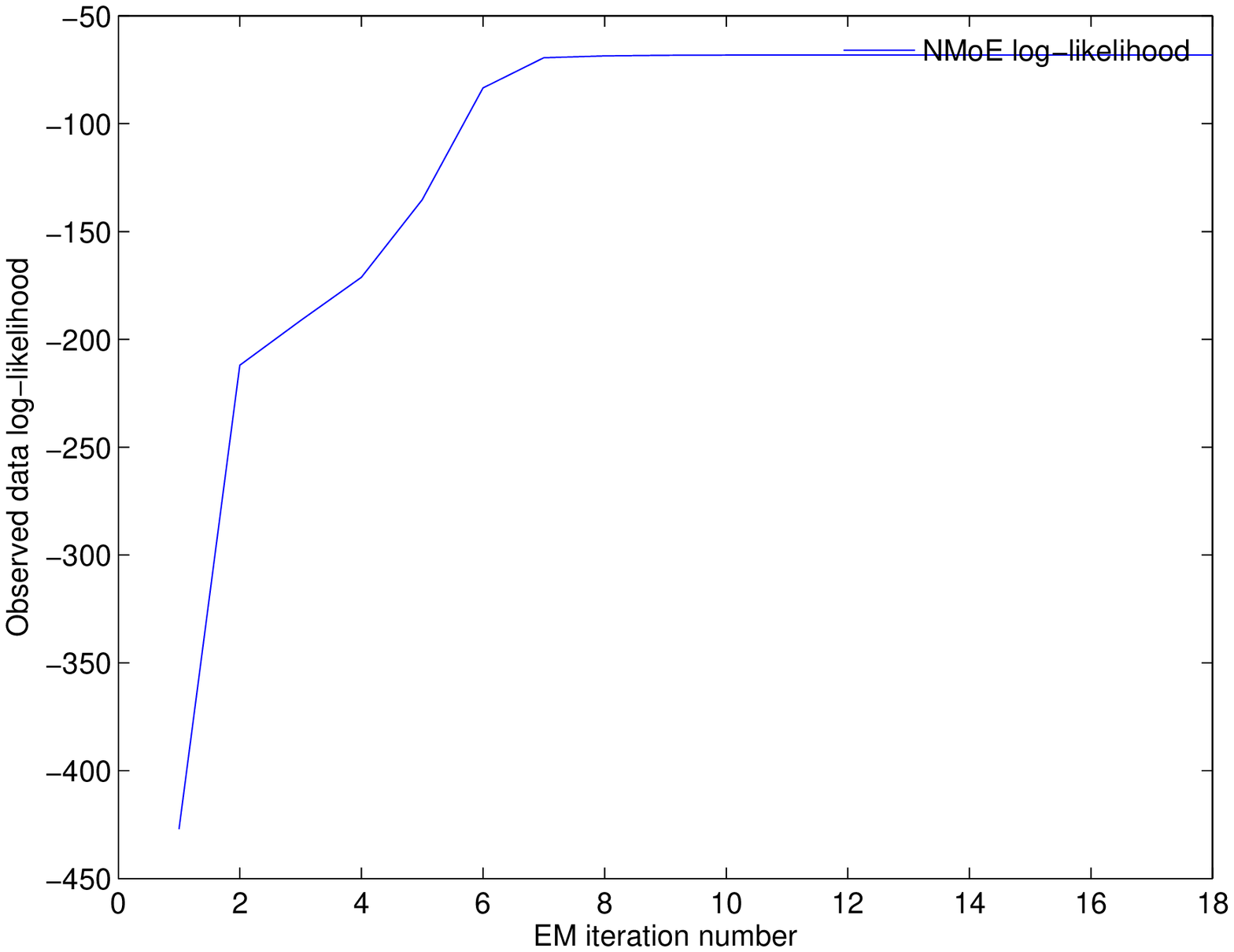} &
   \includegraphics[width=5cm]{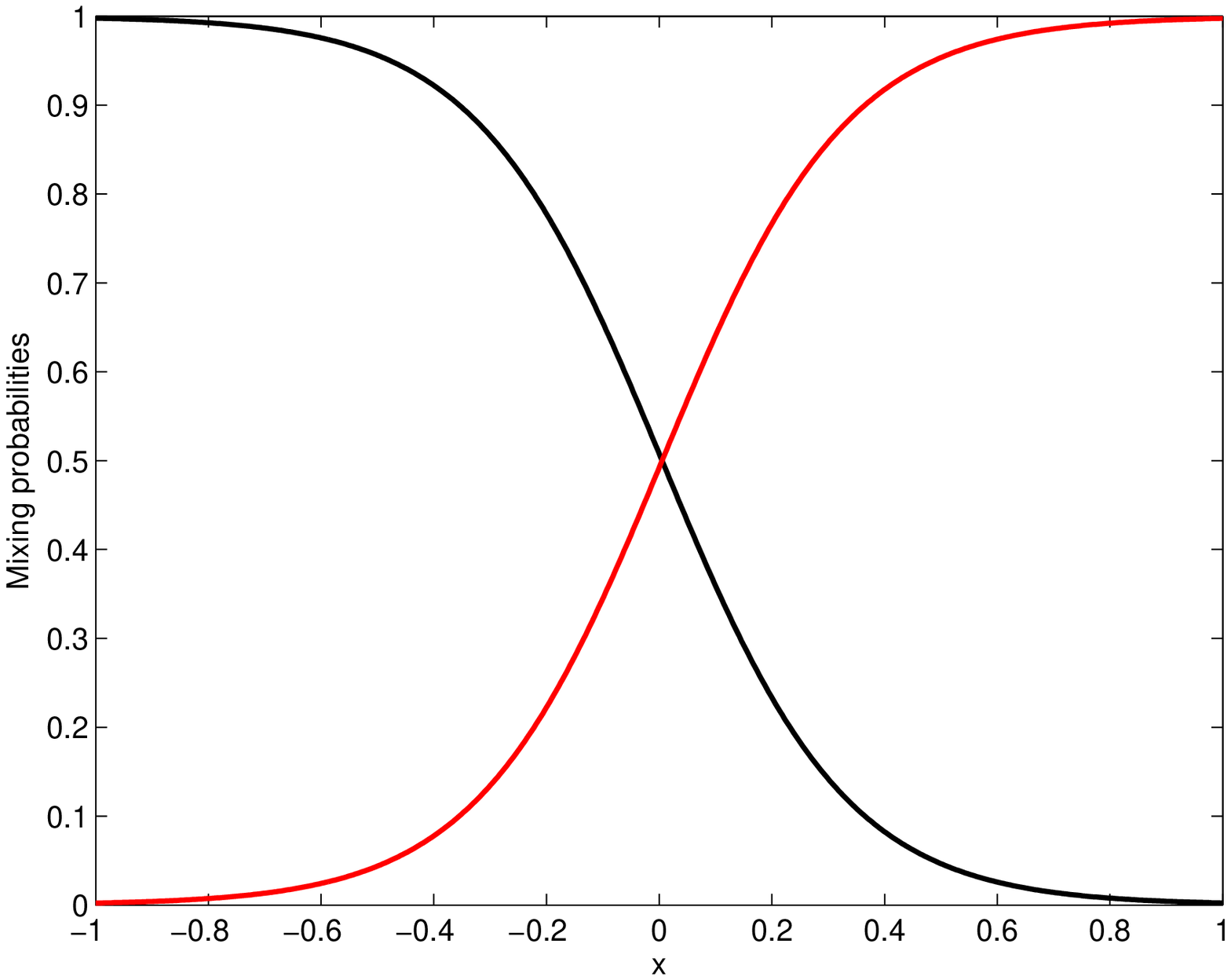}\\
   \end{tabular}
      \caption{\label{fig. TwoClust-Outliers-NMoE_NMoE}Fitted NMoE model to a data set of $n=500$ observations generated according to the NMoE model  and including $5\%$ of outliers.}
\end{figure}
%
%
\begin{figure}[H]
   \centering  
   \begin{tabular}{cc}
   \includegraphics[width=5cm]{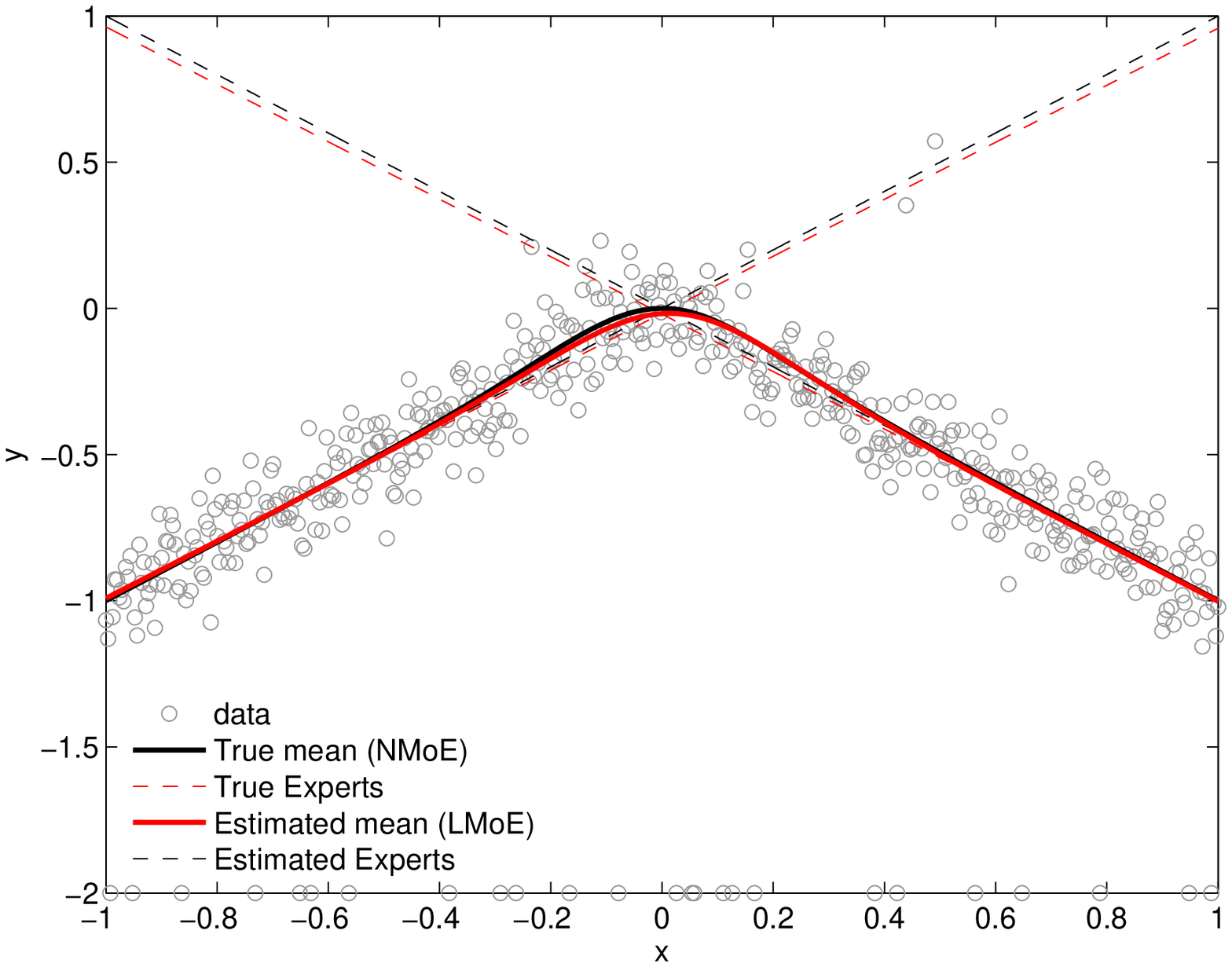}&  
   \includegraphics[width=5cm]{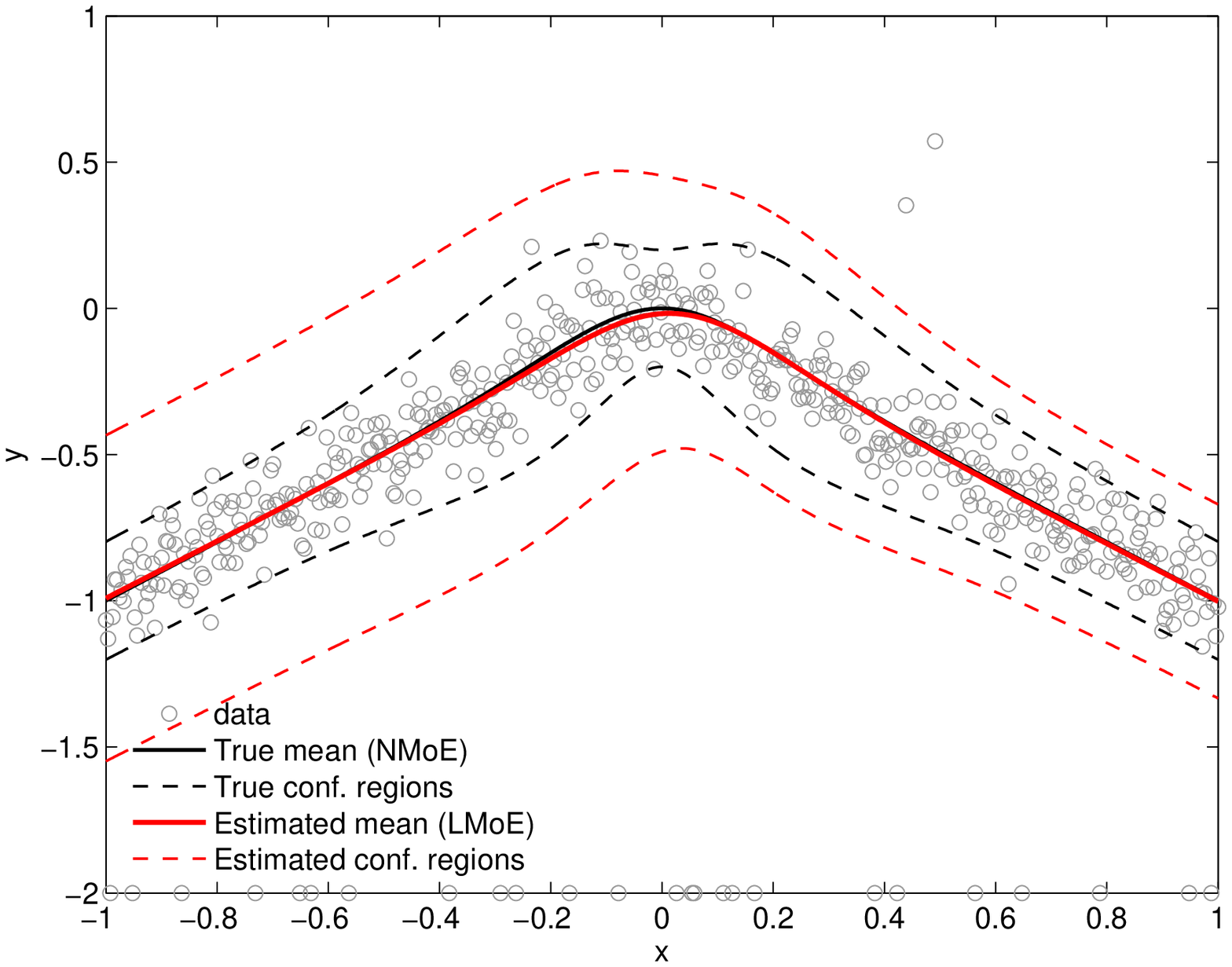}\\
   \includegraphics[width=5cm]{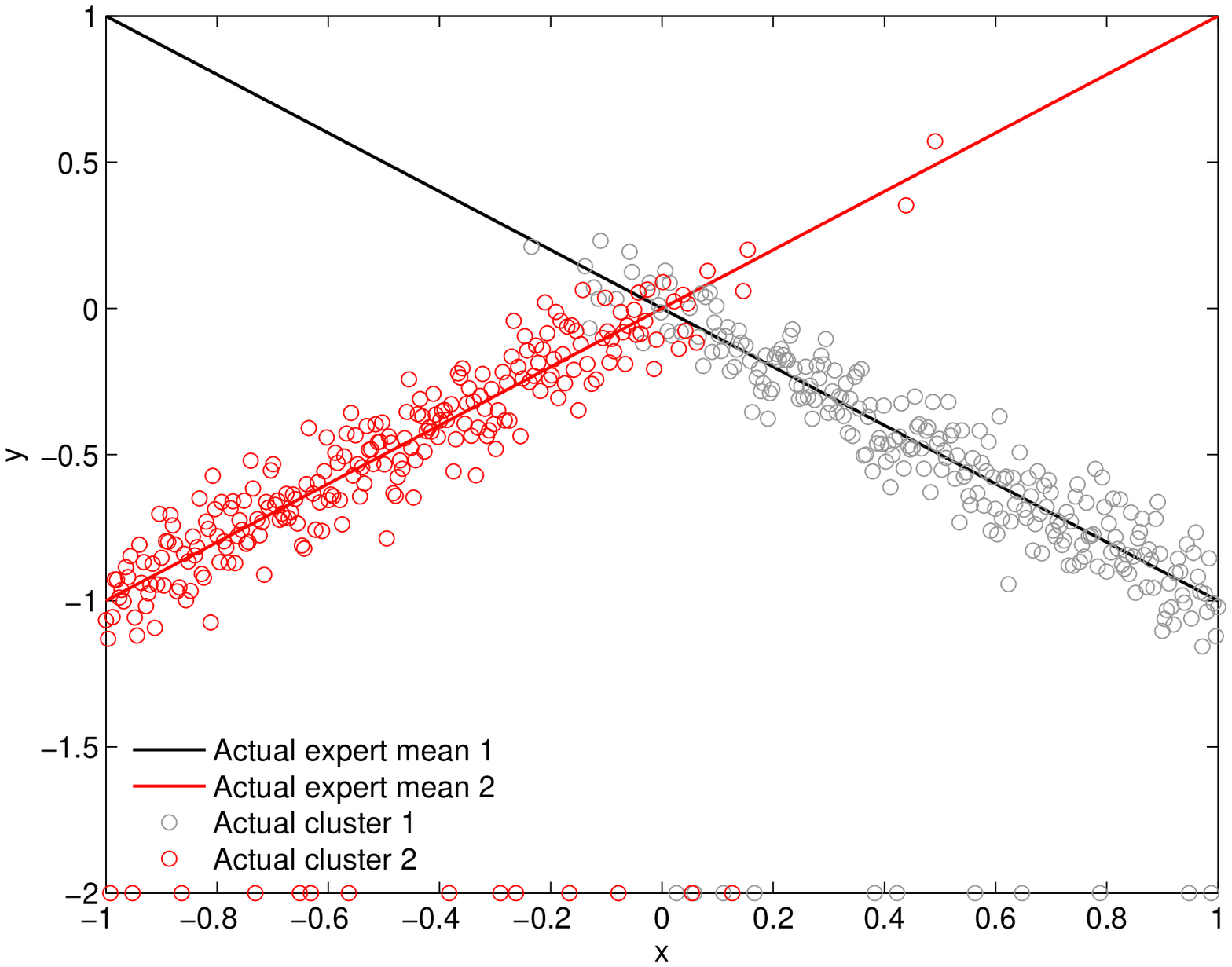}&
   \includegraphics[width=5cm]{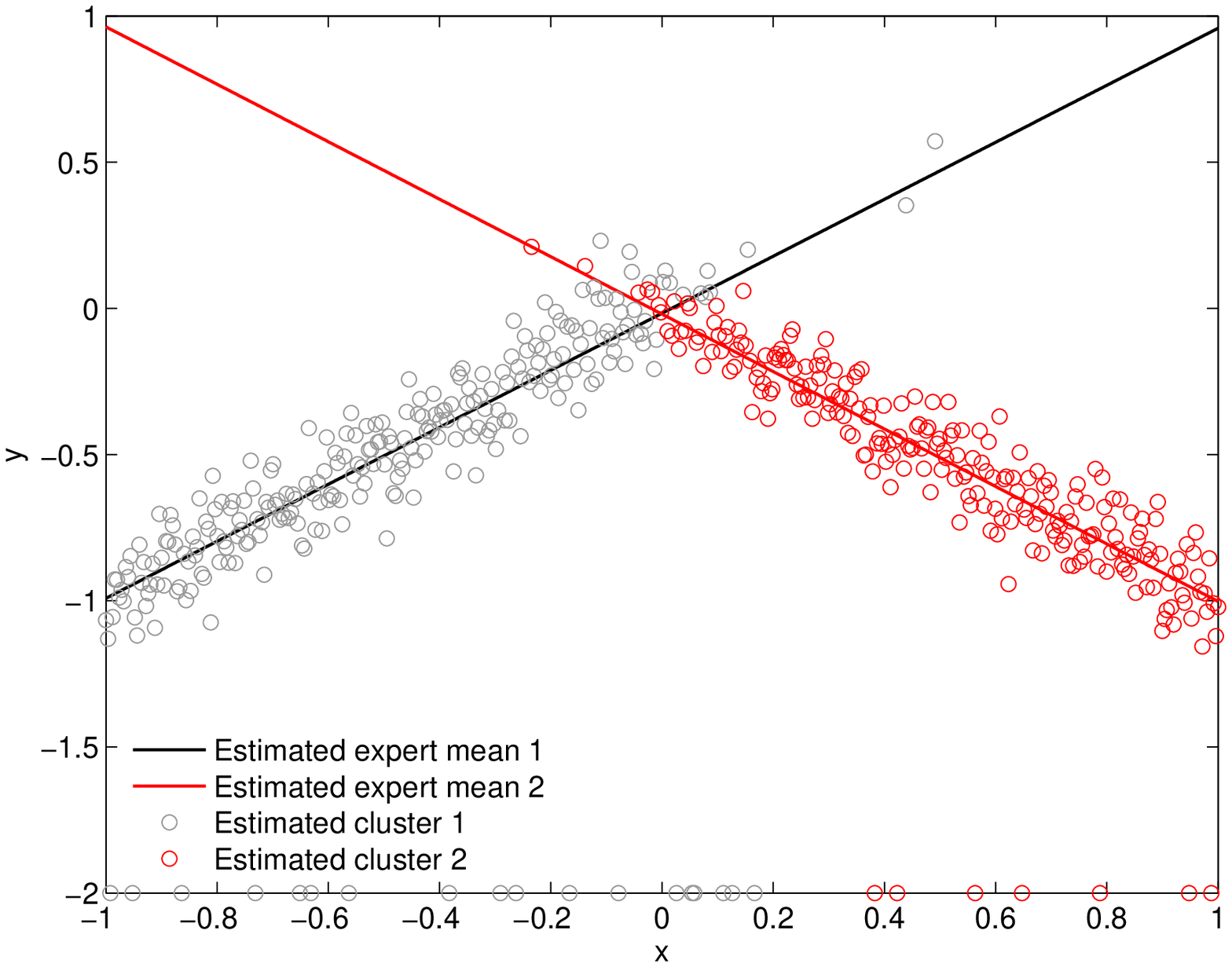}
   \\
   \includegraphics[width=5cm]{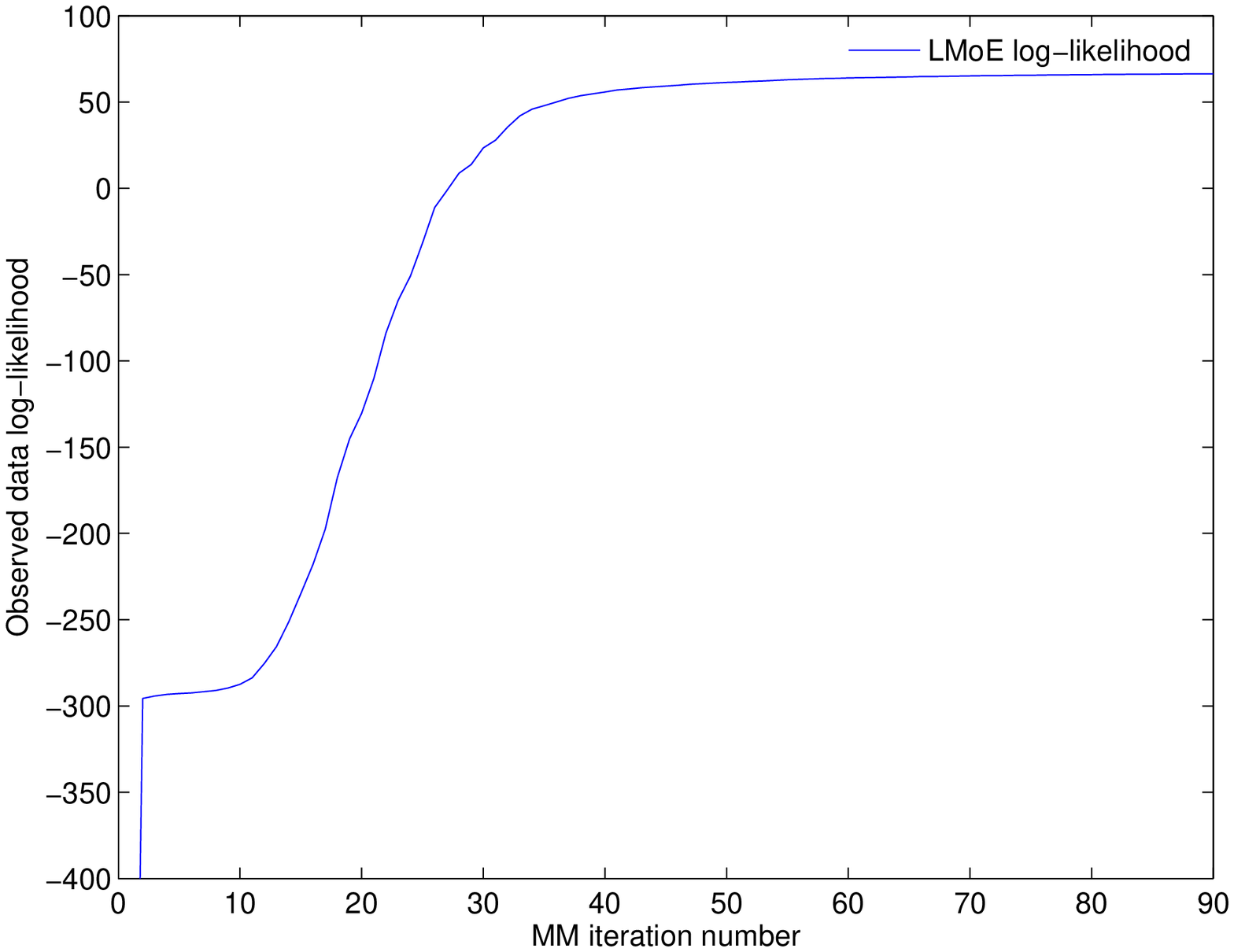} &
   \includegraphics[width=5cm]{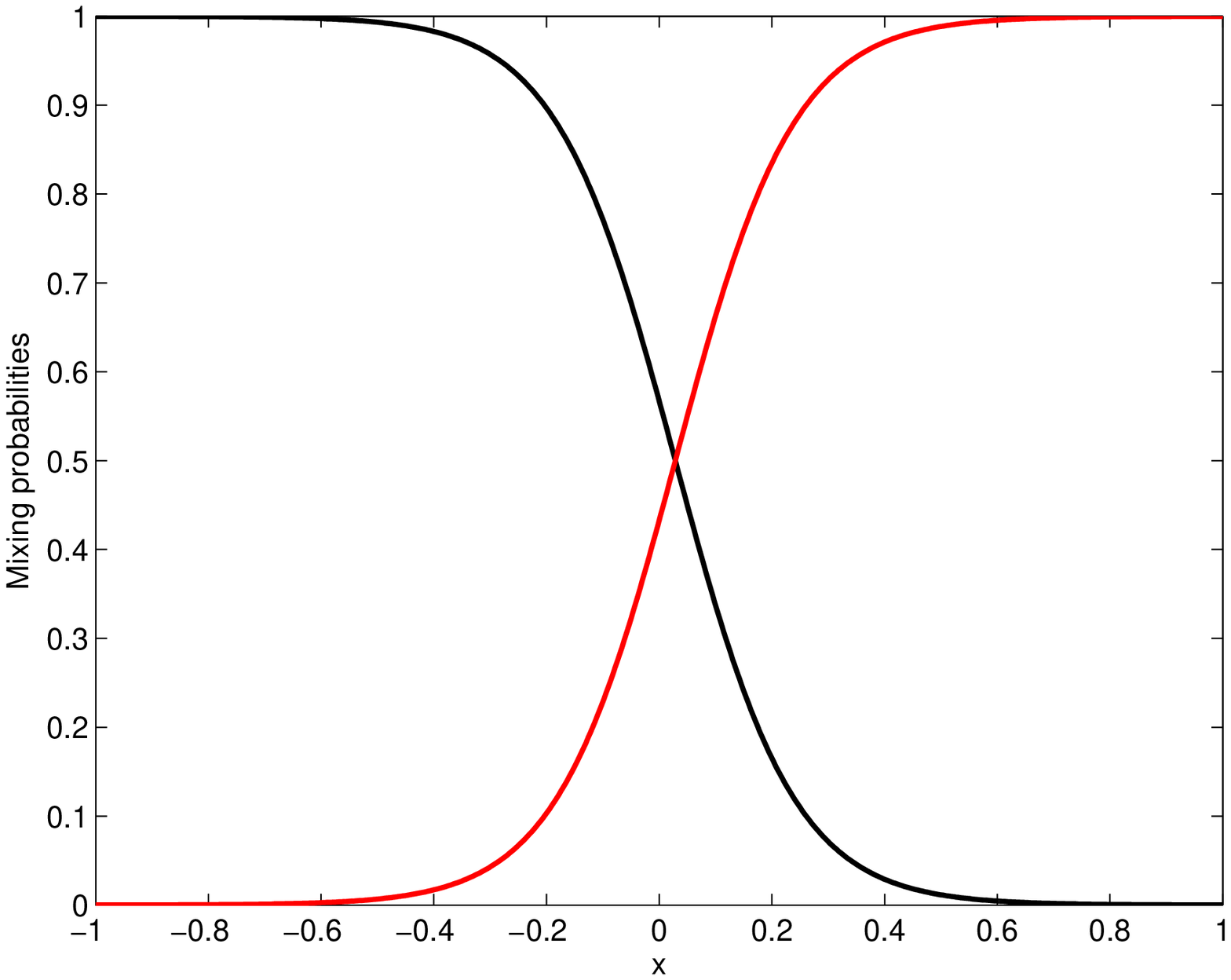}
   \end{tabular}
      \caption{\label{fig. TwoClust-Outliers-NMoE_LMoE}Fitted LMoE model to a data set of $n=500$ observations generated according to the NMoE model  and including $5\%$ of outliers (the same data set shown in Figure \ref{fig. TwoClust-Outliers-NMoE_NMoE}).}
\end{figure}
\begin{figure}[H]
   \centering  
   \begin{tabular}{cc}
   \includegraphics[width=5cm]{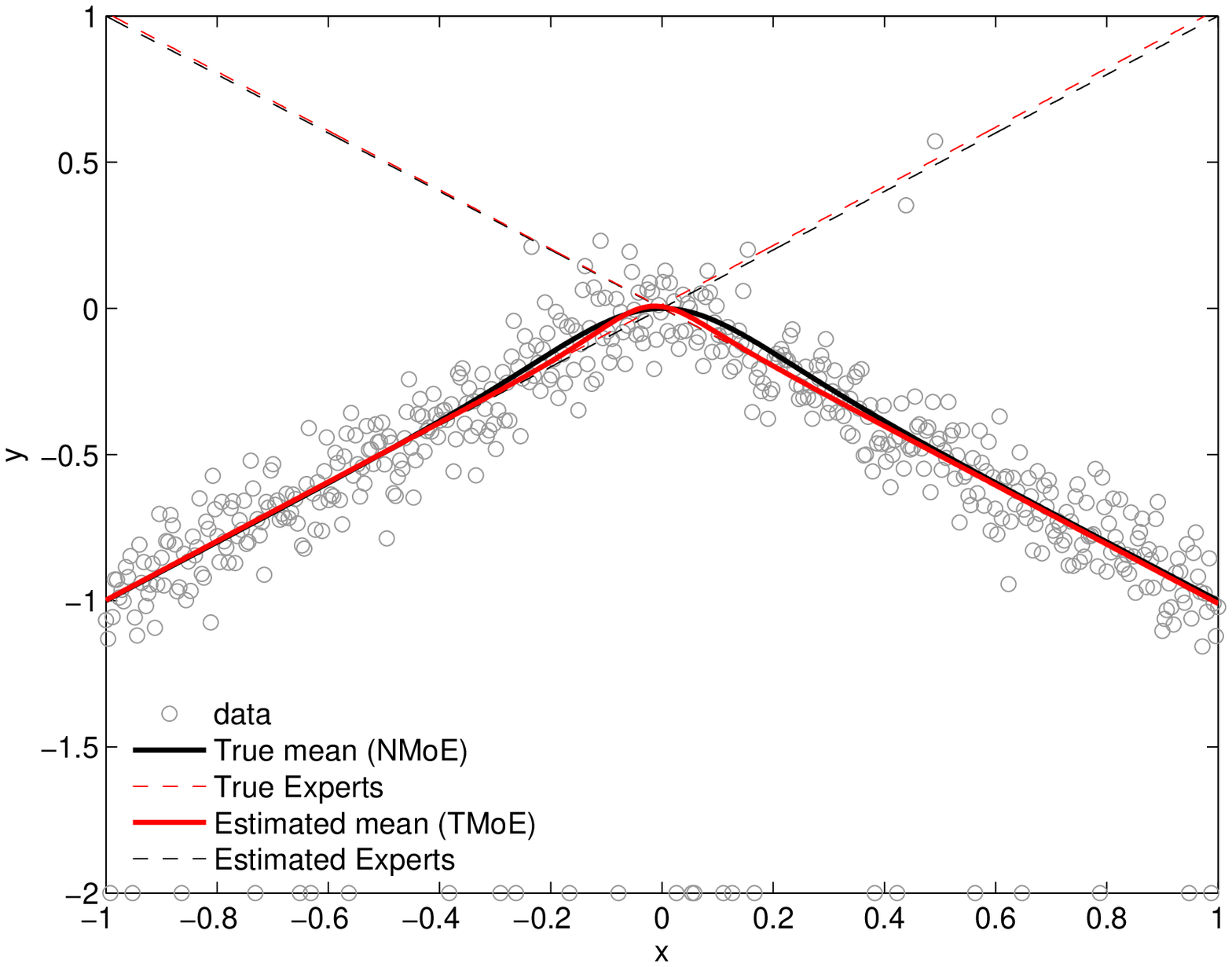}&  
   \includegraphics[width=5cm]{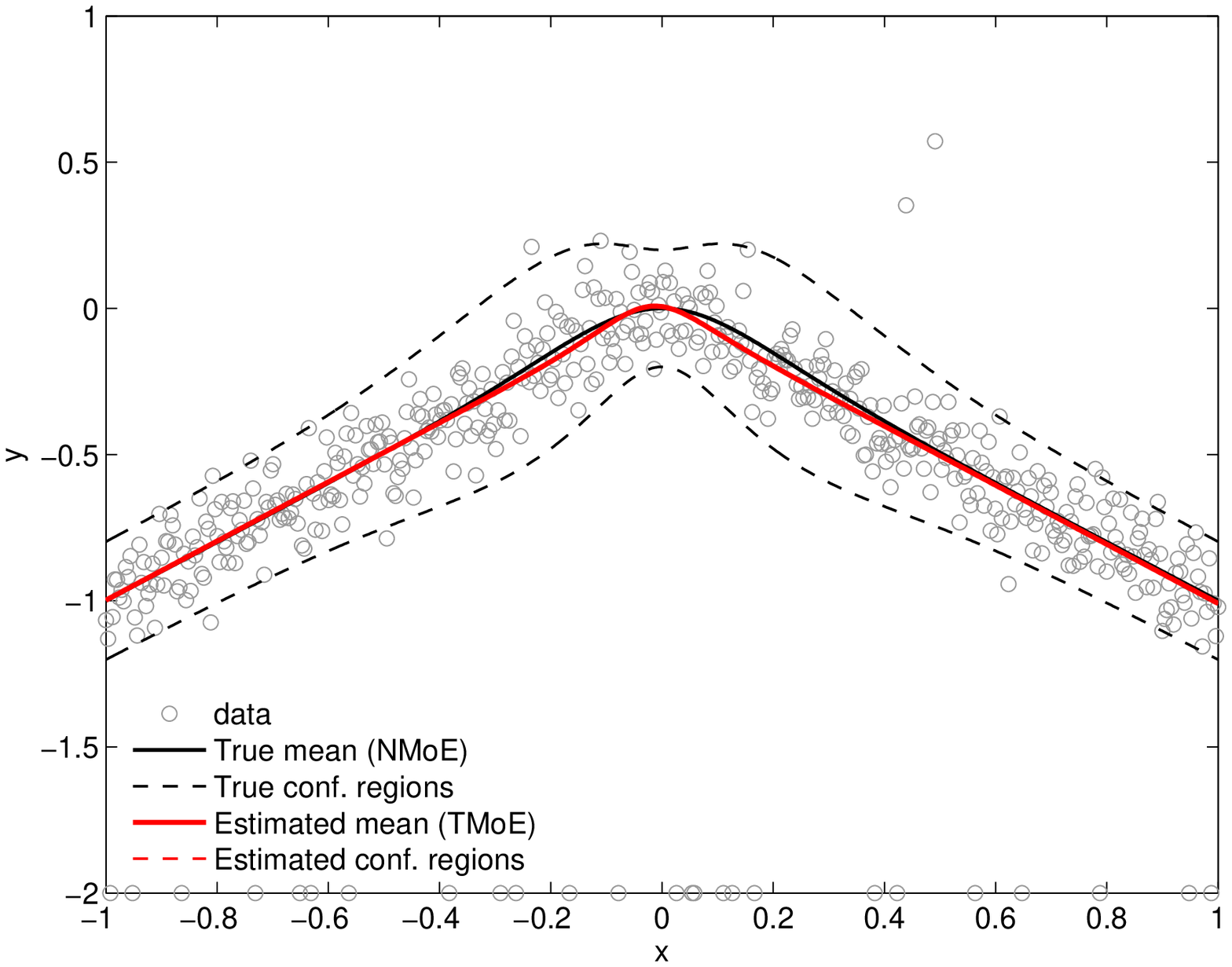}\\
   \hspace{0.2cm}\includegraphics[width=4.8cm]{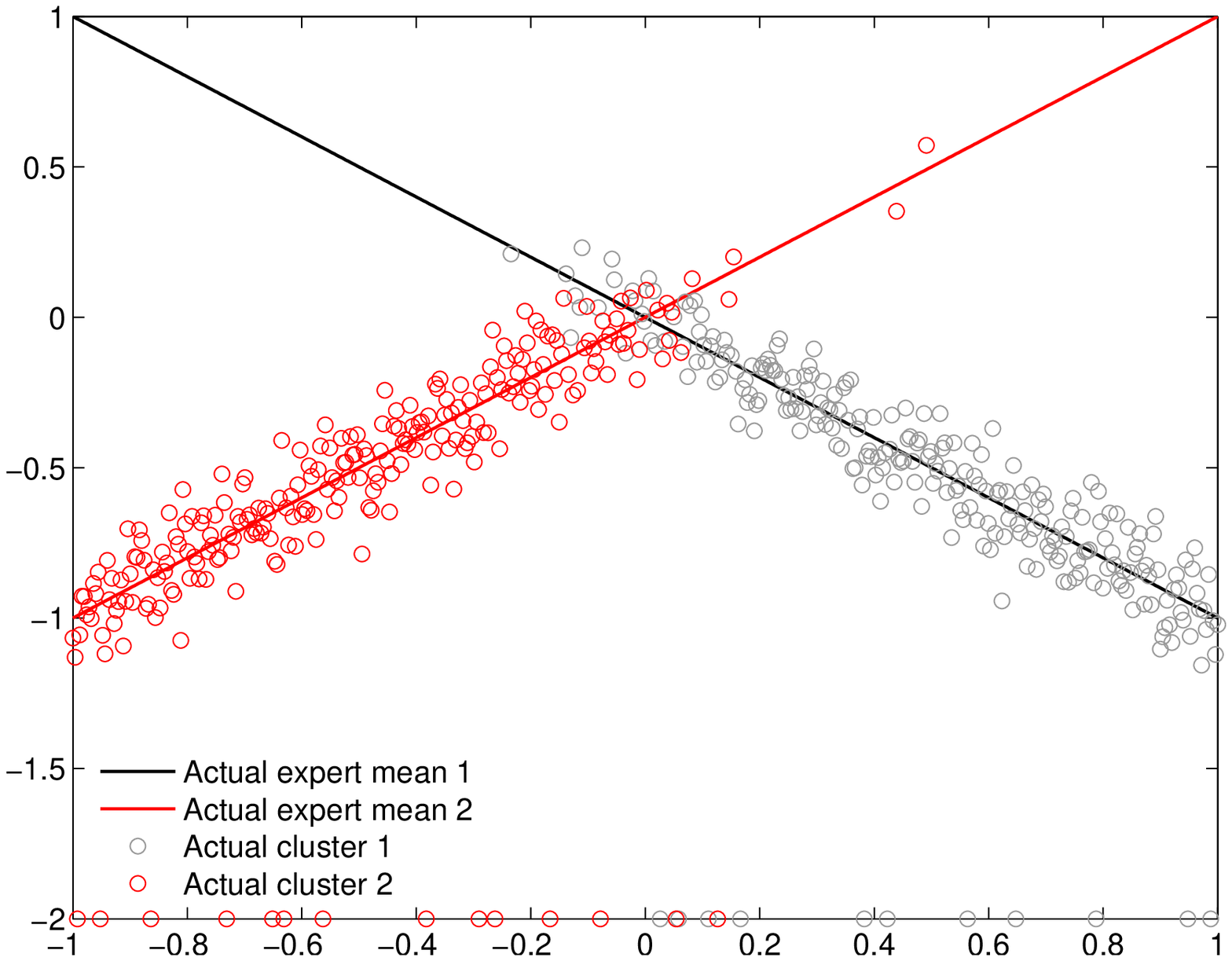}&
   \hspace{0.2cm}\includegraphics[width=4.8cm]{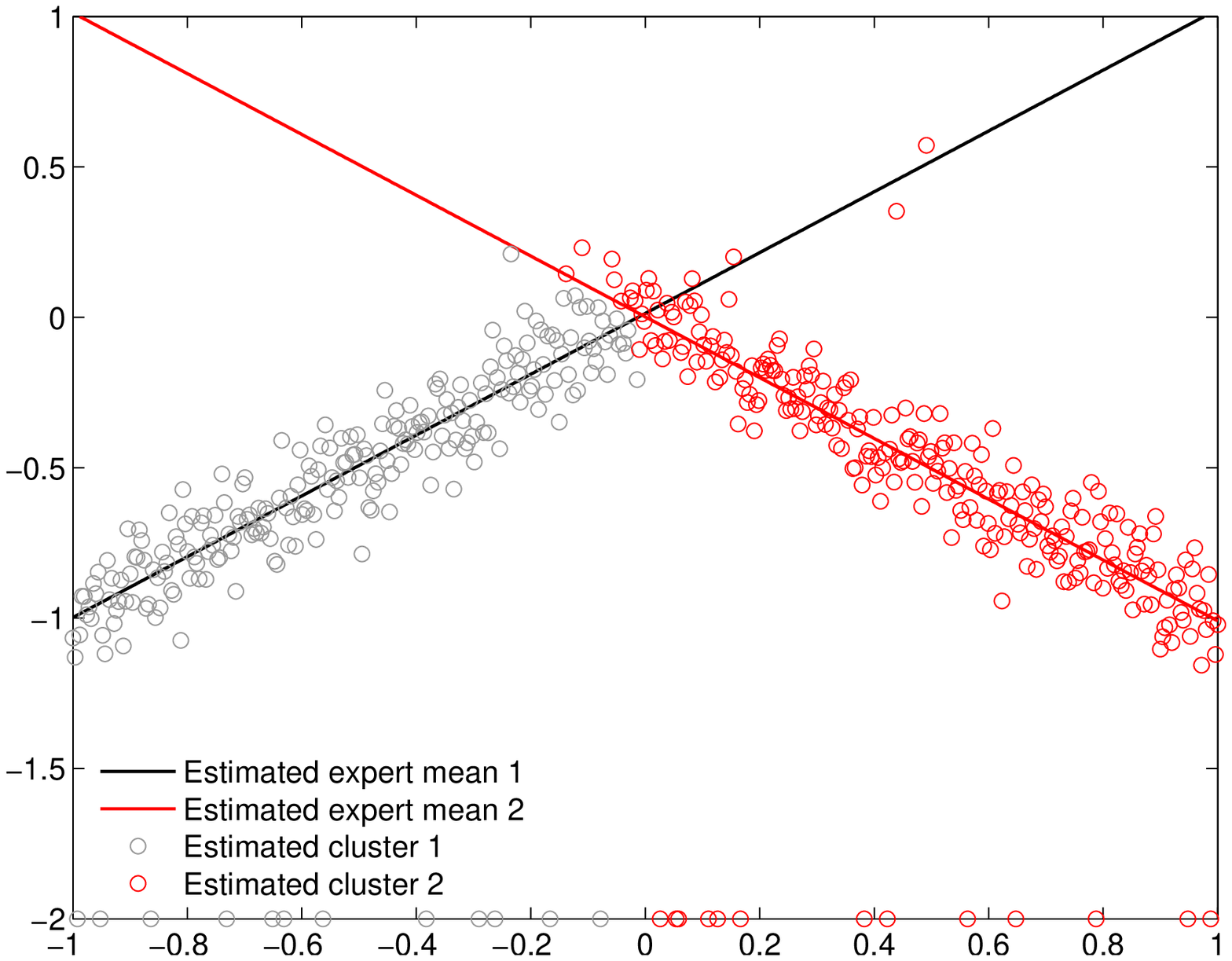}
   \\
   \includegraphics[width=5cm]{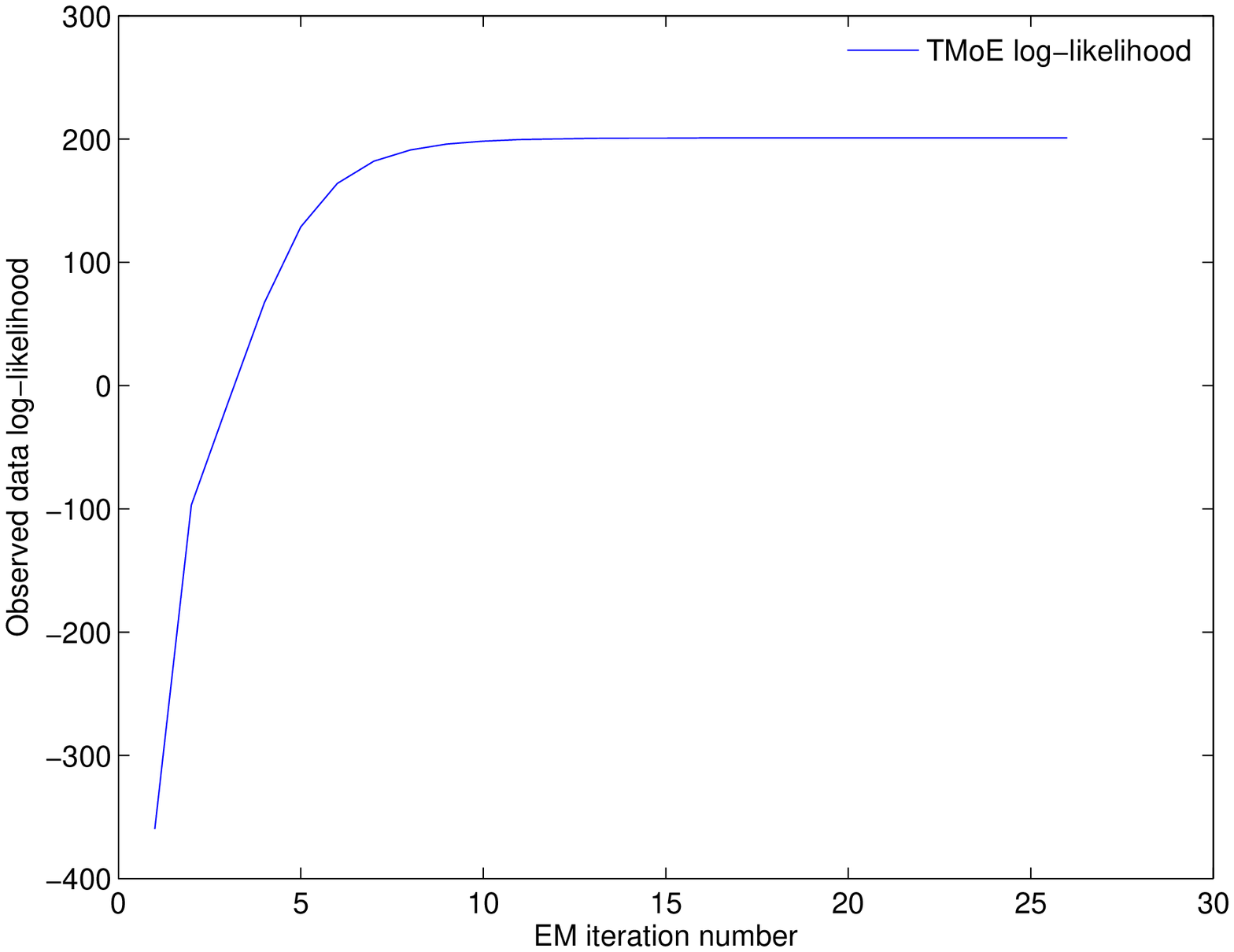} &
   \includegraphics[width=5cm]{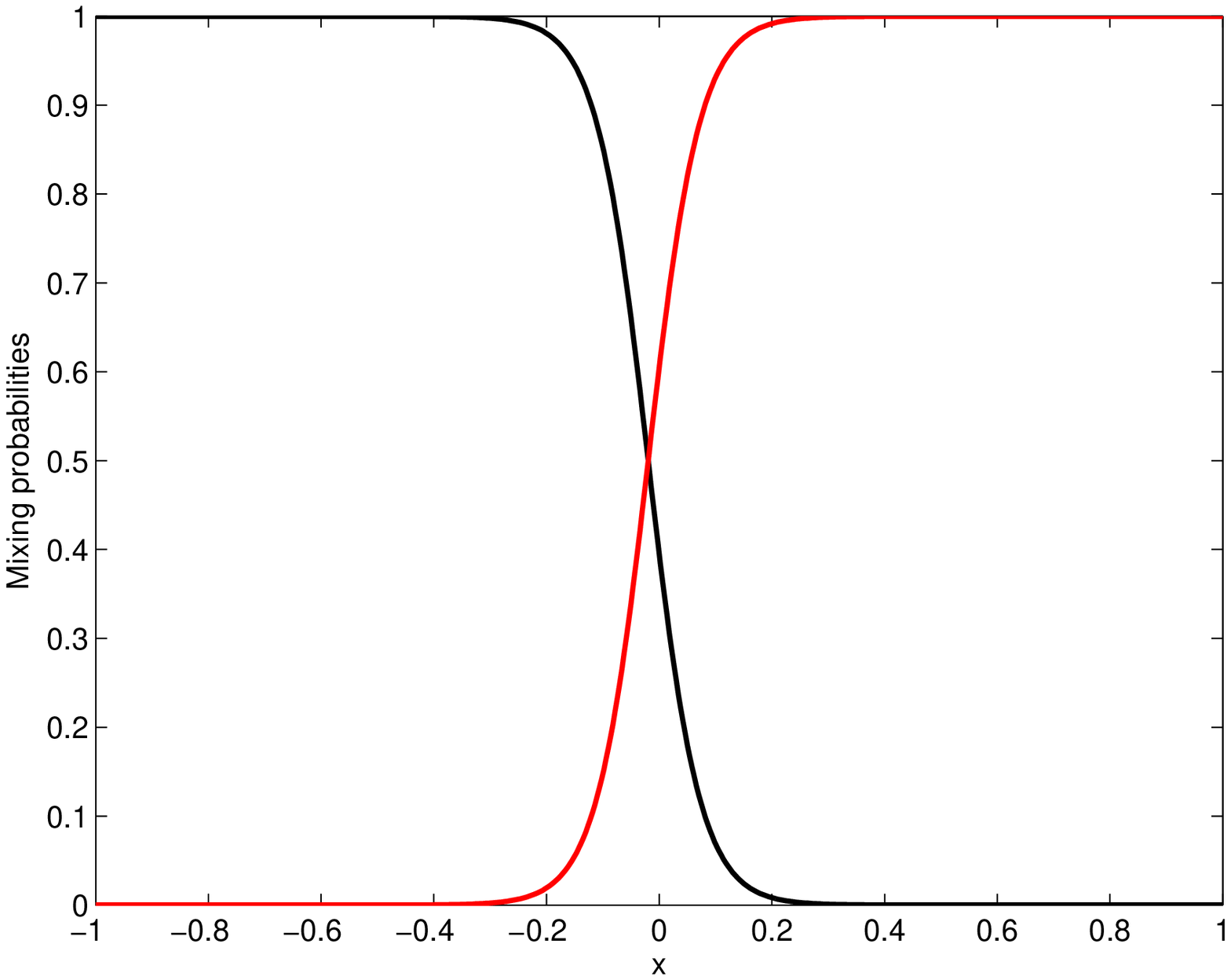}
   \end{tabular}
      \caption{\label{fig. TwoClust-Outliers-NMoE_TMoE}Fitted TMoE model to a data set of $n=500$ observations generated according to the NMoE model  and including $5\%$ of outliers (the same data set shown in Figure \ref{fig. TwoClust-Outliers-NMoE_NMoE}).}
\end{figure}
 %

\subsection{Application to two real-world data sets}

In this section, we consider an application to two real-world data sets: the tone perception data set and the temperature anomalies data set shown in Figure \ref{fig. Tone and temperature anomalies data}.
\begin{figure}[H]
   \centering 
   \begin{tabular}{cc}
   \includegraphics[width=6cm]{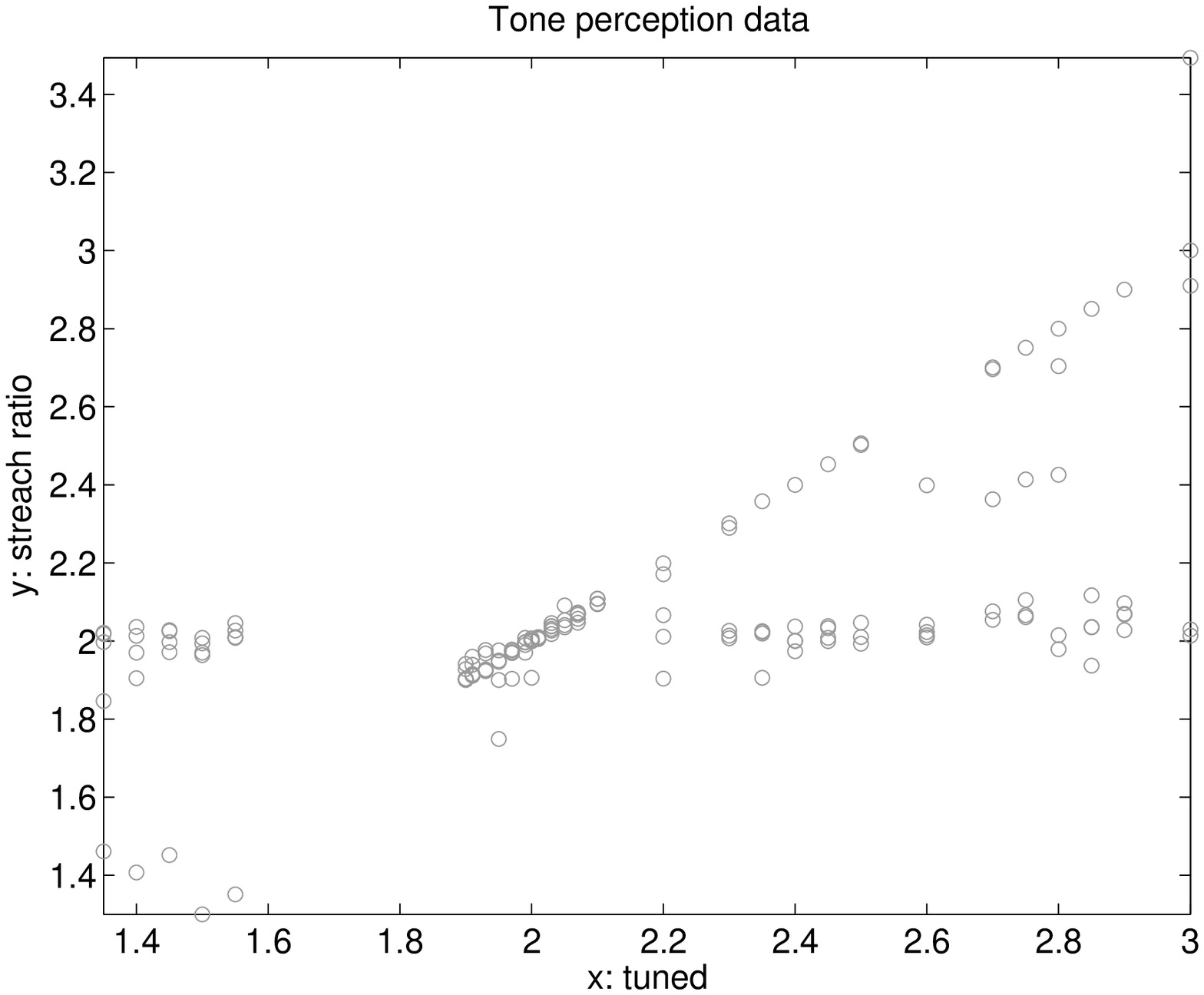} & 
   \includegraphics[width=6cm]{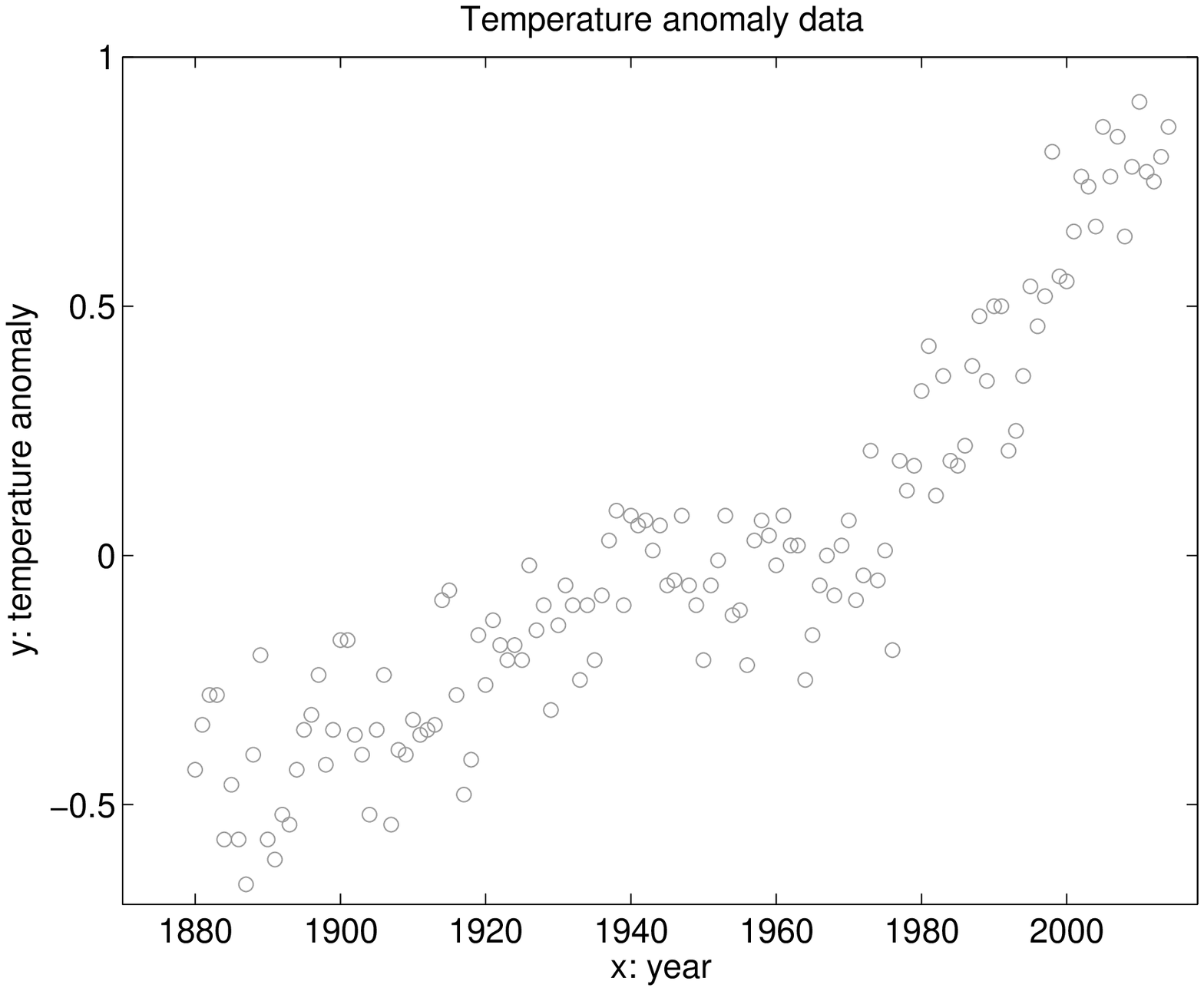} 
   \end{tabular}
      \caption{\label{fig. Tone and temperature anomalies data}Scatter plots of the tone perception data and the temperature anomalies data.}
\end{figure}

\subsubsection{Tone perception data set} The first analyzed data set is the real tone perception data set\footnote{Source: \url{http://artax.karlin.mff.cuni.cz/r-help/library/fpc/html/tonedata.html}} which goes back to \citet{Cohen1984}. It was recently studied by \citet{Bai2012} and \citet{Song2014} by using robust regression mixture models based on, respectively, the $t$ distribution and the Laplace distribution.
In the tone perception experiment, a pure fundamental tone was played to a trained musician. Electronically generated overtones were added, determined by a stretching ratio (``stretch ratio" = 2) which corresponds to the harmonic pattern usually heard in traditional definite pitched instruments. 
The musician was asked to tune an adjustable tone to the octave above the fundamental tone and a ``tuned'' measurement gives the ratio of the adjusted tone to the fundamental. 
The obtained data consists of $n=150$ pairs of ``tuned'' variables, considered here as  predictors  ($x$), and their corresponding ``strech ratio'' variables considered as responses ($y$). 
To apply the  MoE models, we set the response $y_i (i=1,\ldots,150)$ as the ``strech ratio'' variables and the covariates $\bsx_i = \bsr_i = (1,x_i)^T$ where $x_i$ is the ``tuned'' variable of the $i$th observation. 
We also follow the study in \citet{Bai2012} and \citet{Song2014} by using two mixture components. The model selection results, given later in Table \ref{tab. Model selection Tone data}, confirm two-components are present in the data when using the TMoE model and the Bayesian Information Criterion \citep{BIC}.
 
Figure  \ref{fig. Original Tone data and all models} shows the scatter plots of the tone perception data and the linear expert components of the fitted NMoE model, the LMoE model, and the proposed  TMoE model. 
One can observe that we obtain a reasonable fit with  the three models. 
But the one of the NMoE differs slightly from the one of the LMoE and the one of the TMoE (which are quasi-identical), and which, upon a visual inspection, can be seen more adapted by better fitting the two regression lines to the data. 
The two regression lines may correspond to correct tuning and tuning to the first overtone, respectively, as analyzed in \citet{Bai2012} (also see \citet{Song2014} for the analysis).
%
%
\begin{figure}[H]
   \centering 
   \begin{tabular}{ccc}
   \includegraphics[width=5cm]{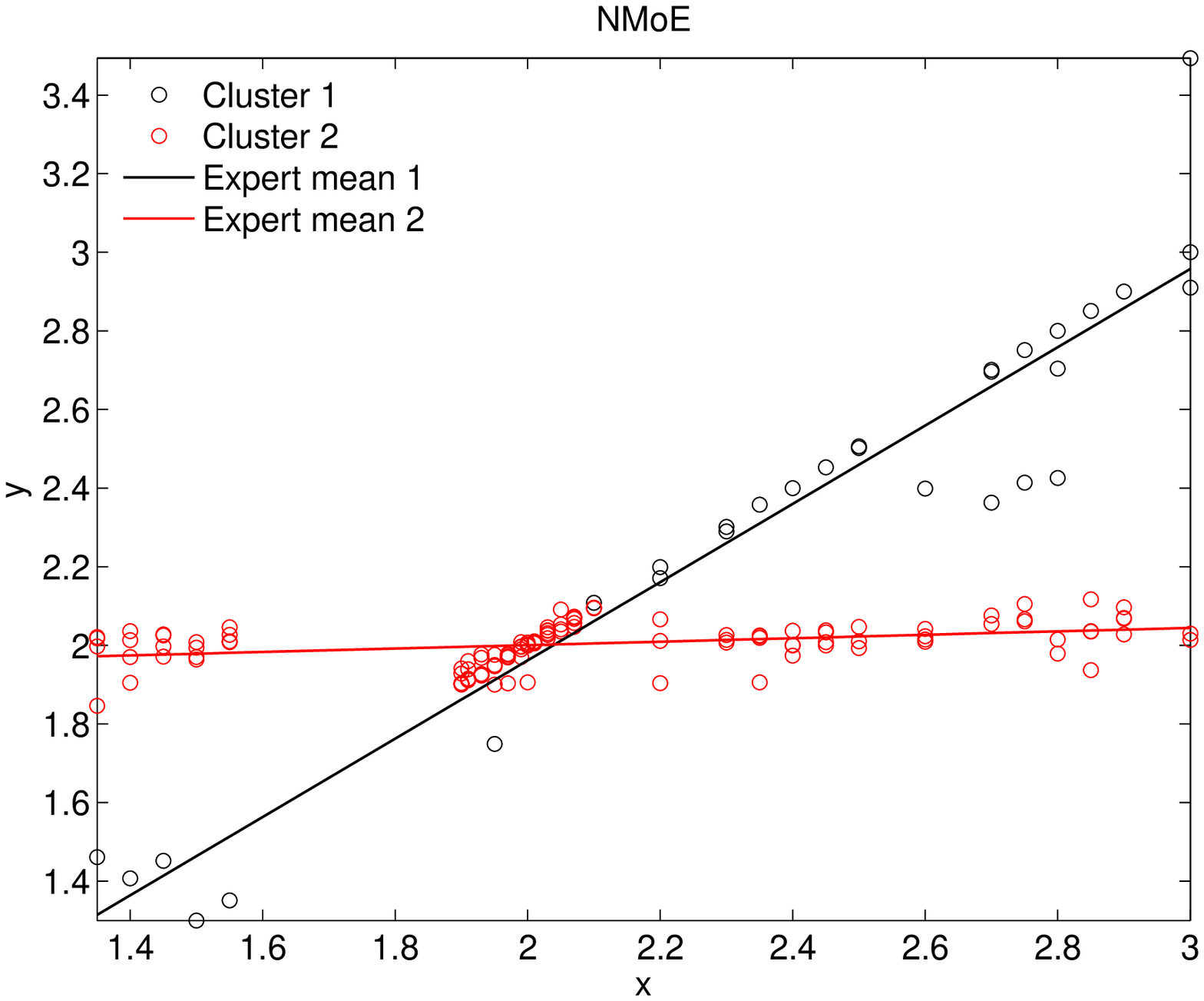} & 
\includegraphics[width=5cm]{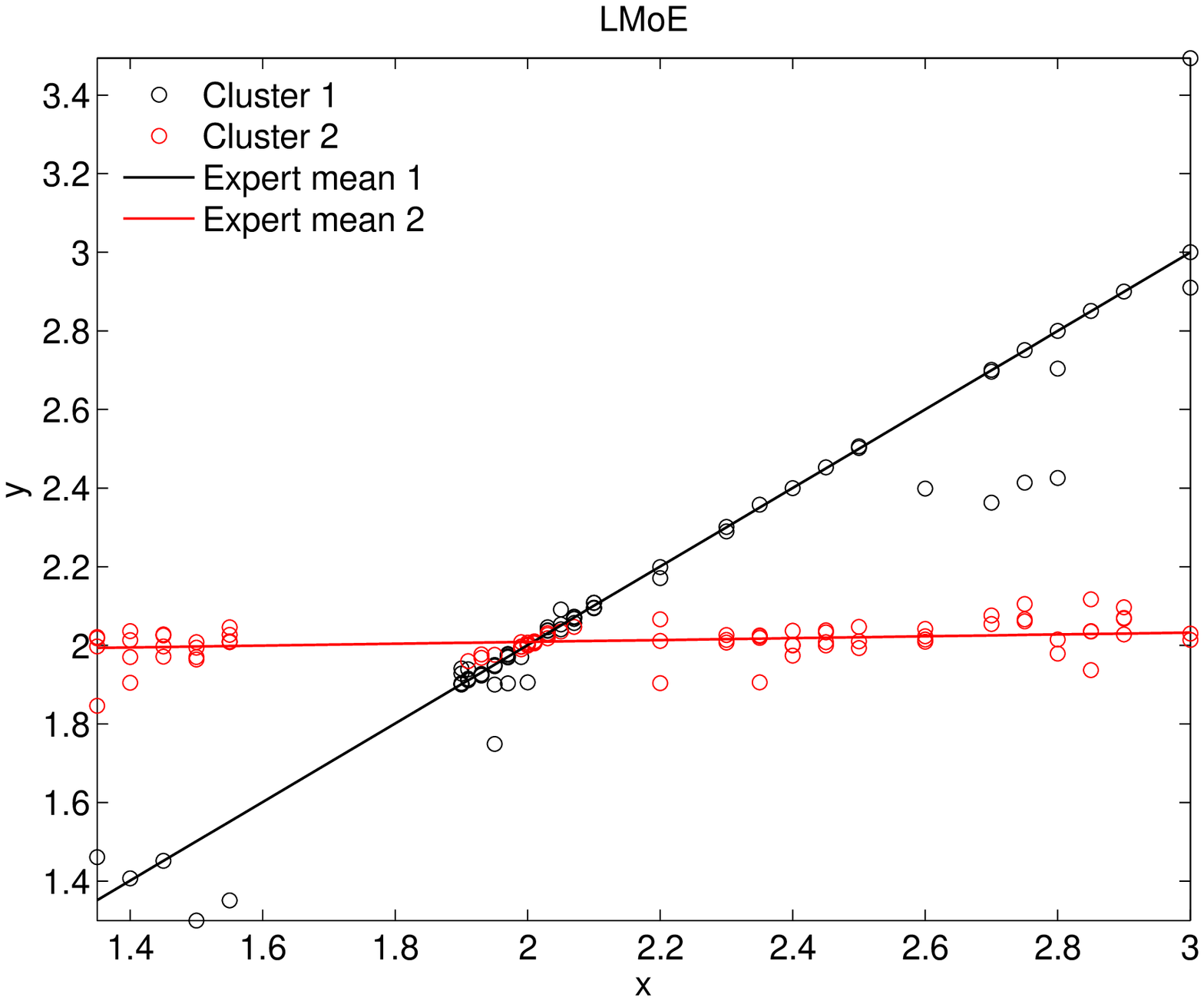} & 
\includegraphics[width=5cm]{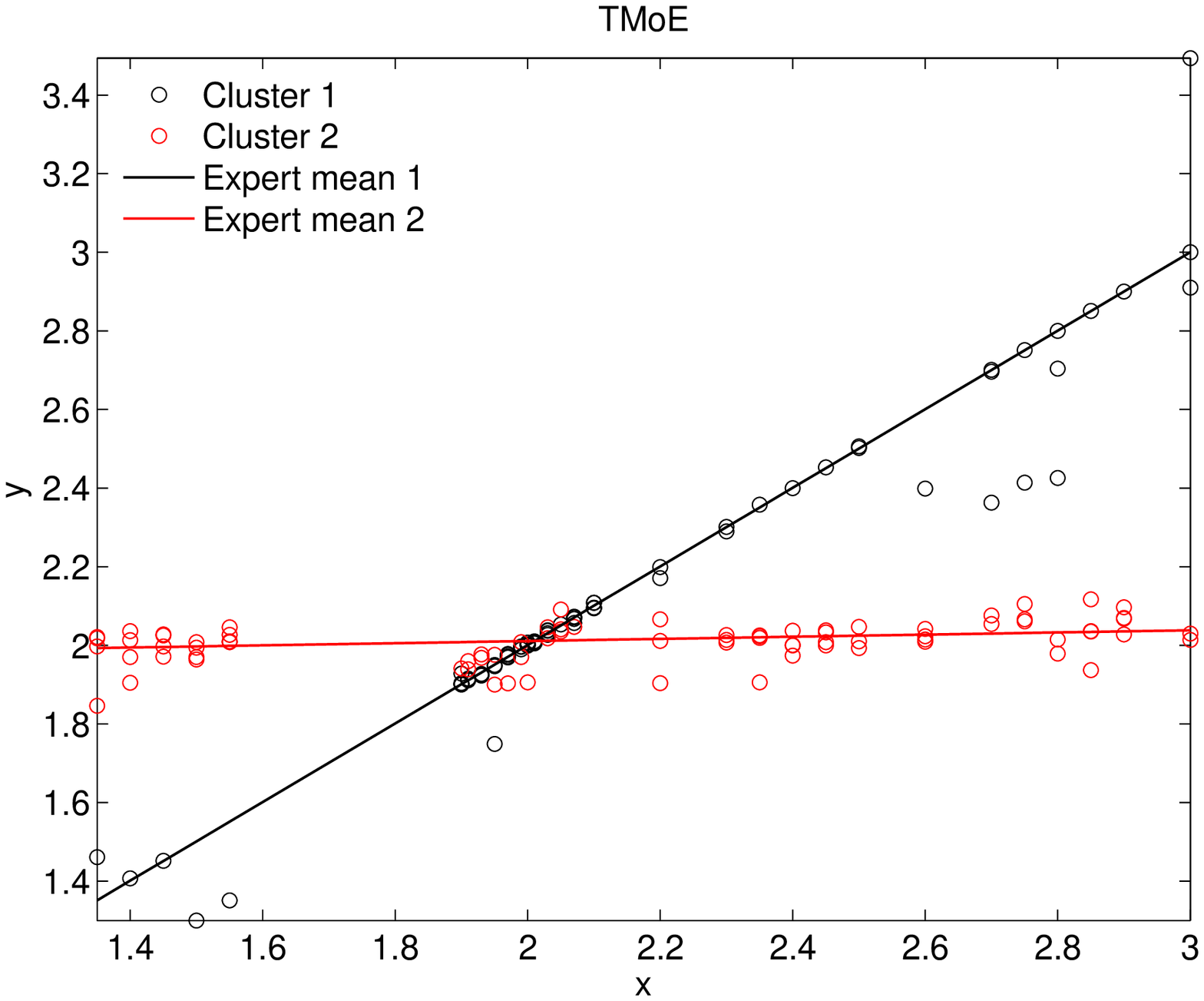} 
   \end{tabular}
      \caption{\label{fig. Original Tone data and all models}The fitted MoLE to the original tone data set with left: NMoE  solution, middle: LMoE solution, and right: TMoE model solution. The predictor $x$ is the actual tone ratio and the response $y$ is the perceived tone ratio.}
\end{figure}
%
Figure \ref{fig. Tone data and all loglik models} shows the log-likelihood profiles for each of the two models. It can namely be seen that training the $t$ MoE for this experiment may take more iterations than the normal  MoE model. The TMoE has indeed more parameters to estimate than the NMoE one, that is, the robustness parameters $\nu_k$. However, in terms of computing time, the models converge in only few seconds on a personal laptop (with 2,9 GHz processor and 8 GB memory). 
\begin{figure}[H]
   \centering 
   \begin{tabular}{ccc}
   \includegraphics[width=5cm]{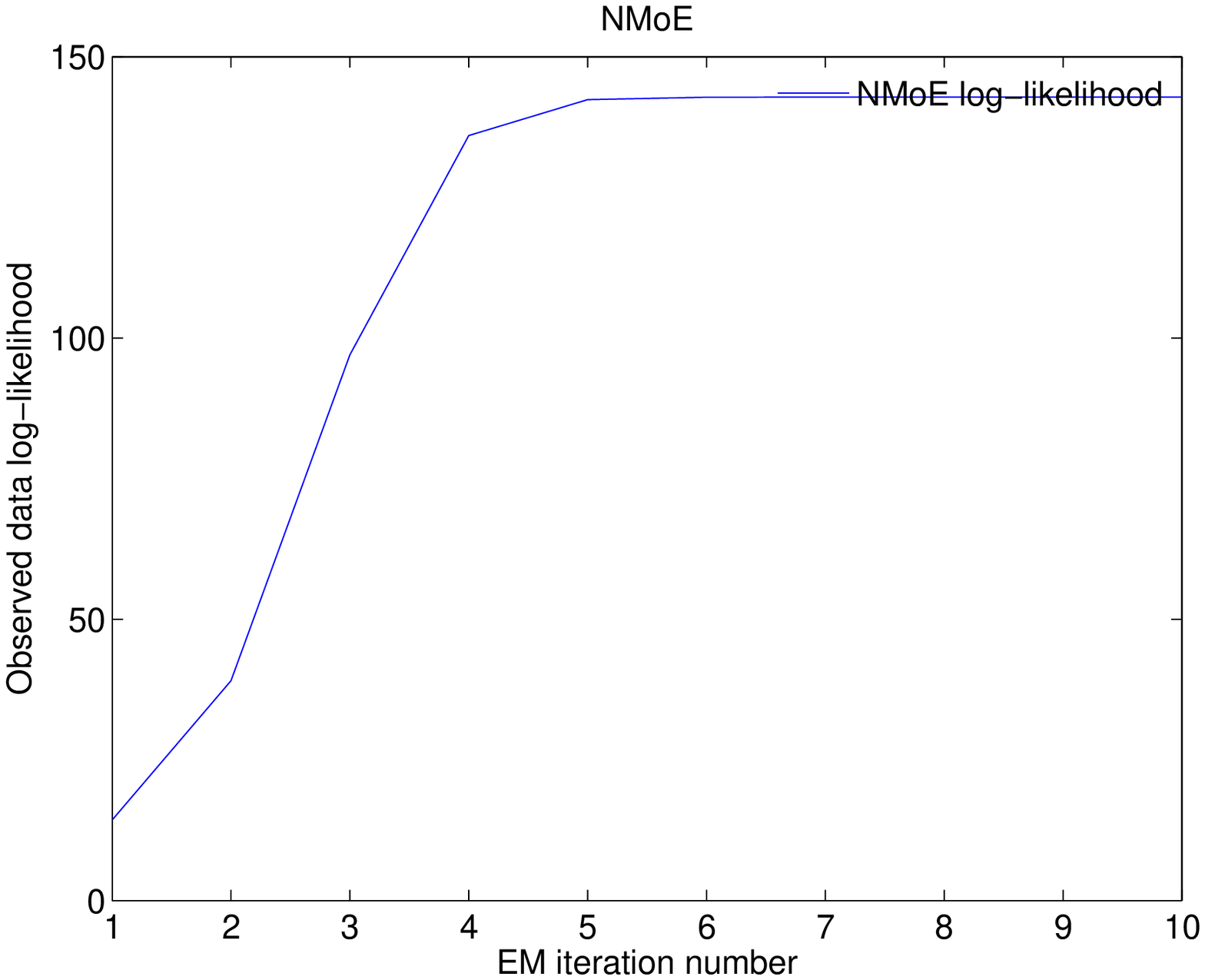} & 
   \includegraphics[width=5cm]{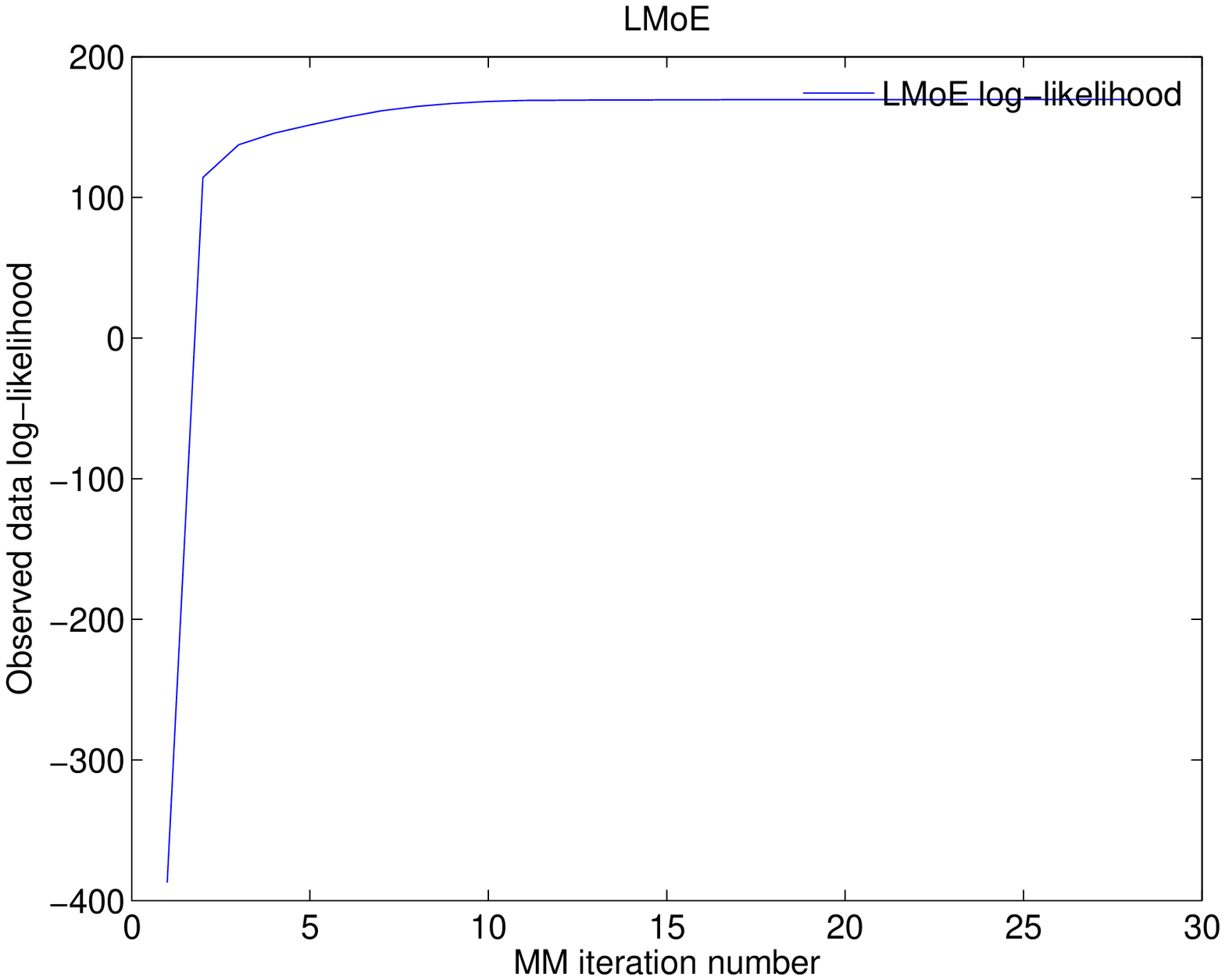} &
   \includegraphics[width=5cm]{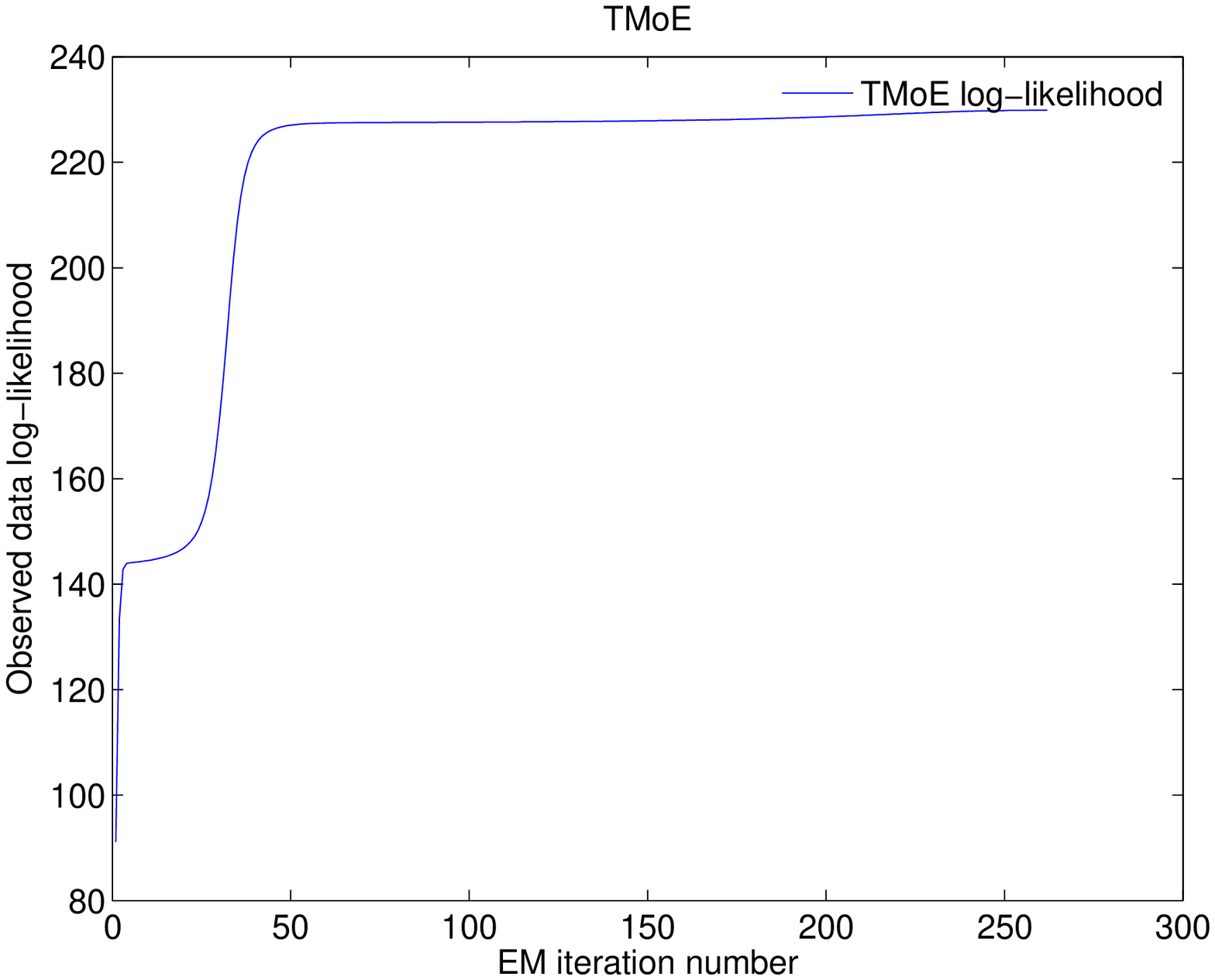}  
   \end{tabular}
      \caption{\label{fig. Tone data  and all loglik models}The log-likelihood during the iterations when fitting the MoLE models to the original tone data set. Left: NMoE model, Middle: LMoE model, Right: TMoE model.}
\end{figure}
The values of estimated parameters for the tone perception data set are given in Table \ref{tab. Estimated parameters for the tone perception data set}.
One can see that the regression coefficients are very similar for all the models, except for the first component of the NMoE model. This can be observed on the fit in Figure \ref{fig. Original Tone data and all models} where the first expert component for the NMoE model slightly differs from the corresponding one of both the LMoE model and the proposed TMoE model. In addition, it can be seen from the values of the common parameters that the LMoE and   the TMoE provide very close results. 
{\setlength{\tabcolsep}{4pt
\begin{table}[H]
\centering
{\small \begin{tabular}{l c c  c c c c c c c c c c}
\hline
param. & $\alpha_{10}$ & $\alpha_{11}$ & $\beta_{10}$ & $\beta_{11}$ & $\beta_{20}$ & $\beta_{21}$ & $\sigma_{1}$& $\sigma_{2}$ & $\lambda_1$ & $\lambda_2$ &  $\nu_{1}$ & $\nu_{2}$ \\ 
model		& & & & & & & & & & \\
 \hline
 \hline
NMoE  	&  -2.690 &	0.796 & 	-0.029 &	0.995 & 1.913 &	 0.043 &	0.137 & 0.047 & - & - & - & -	\\
LMoE	&  -0.460 &	0.087 & 	0.0036 &	 0.998 & 1.961 &	  0.023 & - & - &	0.049 & 0.030  &	 -       & - \\
TMoE    &  -0.058 &	-0.070& 	 0.002  &	0.999   & 1.956  &  0.027   &	0.002      & 0.029 & - & -   &	0.555 & 2.017	\\
\hline
\end{tabular}}
\caption{\label{tab. Estimated parameters for the tone perception data set}Values of the estimated MoE parameters for the original Tone perception data set.}
\end{table}
} 

We also performed a model selection procedure on this data set to choose the best number of MoE components for a number of components between 1 and 5. We used BIC, AIC, and ICL. Table  \ref{tab. Model selection Tone data} gives the obtained values of the  model selection criteria. 
One can see that for the NMoE model overestimate the number of components.  AIC performs poorly for all the models.
BIC provides the correct number of components for the three proposed TMoE model but seems to overestimate the number of components for the LMoE model (provides evidence for 3 components). ICL hesitates between 2 (the correct number) and 4 components for the TMoE model. One can conclude that the BIC is the criterion to be suggested for the analysis. Thus, from this experiment, it would be more adapted to use BIC with the proposed TMoE model. 
{\setlength{\tabcolsep}{2pt
\begin{table}[H]
\centering
{\small 
\begin{tabular}{l |ccc | ccc | ccc}
\hline
	& \multicolumn{3}{c|}{NMoE}	& \multicolumn{3}{c}{LMoE} & \multicolumn{3}{c}{TMoE} \\
\cline{2-10}
K			 &   BIC 	 & 	AIC 	&    ICL		 &   BIC 	 & 	AIC 	&    ICL	   &   BIC 	 & 	AIC 	&    ICL	\\
\hline
\hline
	1		& 1.8662  &  	6.3821    &	1.8662	& 36.8061  & 41.3220 &  \underline{-7.5160}& 71.3931 &  77.4143 &71.3931 \\
	2		&122.8050&  134.8476&  107.3840 &  149.6360 &	161.6786	& -20.0425& \underline{204.8241}&  219.8773&  186.8415\\
	3		&118.1939&  137.7630&   76.5249 &\underline{209.1995} & 228.7687 & -32.5691&199.4030  &223.4880 & 183.0389 \\
	4		&121.7031&  148.7989&   94.4606 & 204.3286 & \underline{231.4244} & -45.0957& 201.8046 & \underline{234.9216}&  \underline{187.7673}\\
	5		&\underline{141.6961}&\underline{176.3184} & \underline{123.6550} & 141.3988 & 176.0211  & -57.6223&187.8652 & 230.0141 & 164.9629\\
\hline 
\end{tabular}
\caption{\label{tab. Model selection Tone data}Choosing the number of expert components $K$ for the original tone perception data by using the information criteria BIC, AIC, and ICL. Underlined value indicates the highest value for each criterion.}
}
\end{table}}
  
\paragraph{Robustness to outliers}
Now we examine the sensitivity of the MoE models to outliers based on this real data set.  For this, we adopt the same scenario used in \citet{Bai2012} and \citet{Song2014} (the last and more difficult scenario) by adding 10 identical pairs $(0,4)$ to the original data set as outliers in the $y$-direction, considered as high leverage outliers. We apply the MoE models in the same way as before.

The left plot in Figure \ref{fig. Tone data with outliers and all models} show that the normal MoE is sensitive to outliers.
However,  compared to the normal regression mixture result in \citet{Bai2012}, and the Laplace regression mixture and the $t$ regression mixture results in \citet{Song2014}, the fitted NMoE is affected less severely by the outliers. 
This may be attributed to the fact that the mixing proportions here are depending on the predictors, which is not the case in these regression mixture models, namely the ones of \citet{Bai2012}, and  \citet{Song2014}.  One can also see that, even the regression mean functions are  affected severely by the outliers, the provided partitions are still reasonable and similar to those provided in the previous non-noisy case.
Then, the middle plot of in Figure \ref{fig. Tone data with outliers and all models} shows that the LMoE model is more robust to outliers compared to the NMoE model, however, the regression line is not very well adjusted to the data. 
However, the right plot in Figure \ref{fig. Tone data with outliers and all models}  clearly shows that the TMoE  provides a robust good fit, which is preferred to the LMoE solution. For the TMoE, the obtained fit is quasi-identical to the first one on the original data without outliers, shown in the right plot of Figure \ref{fig. Original Tone data and all models}.
Moreover, we notice that, as showed in \citet{Song2014}, for this situation with outliers, 
the $t$ mixture of regressions fails; The  fit is affected severely by the outliers.
However, for the proposed TMoE model, the ten high leverage outliers have no significant impact on the fitted experts.  
This is because here the mixing proportions depend on the inputs, which is not the case for the regression mixture model  described in \citet{Song2014}. 
%
%
%
\begin{figure}[H]
   \centering 
   \begin{tabular}{ccc}
   \includegraphics[width=5cm]{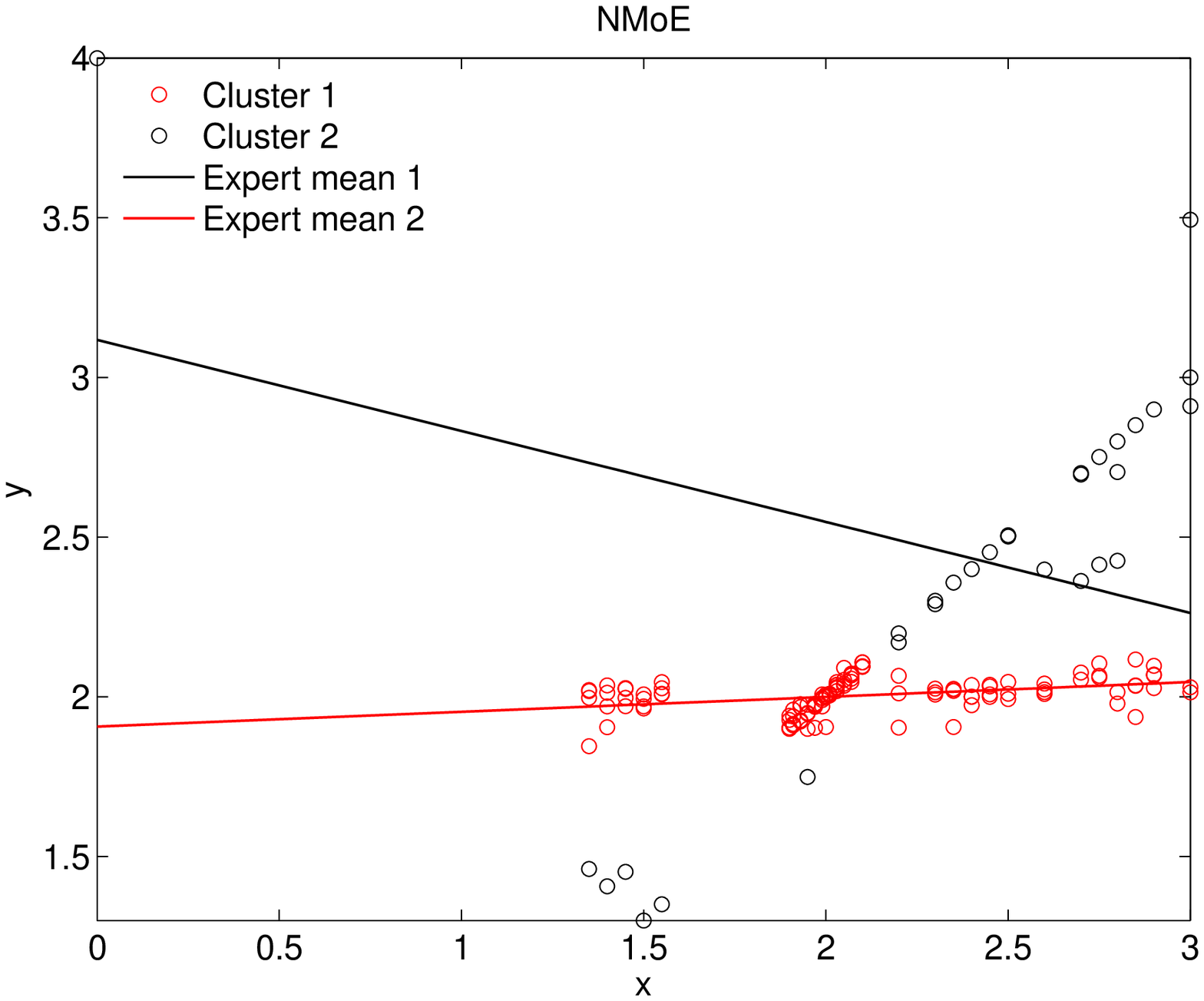} & 
   \includegraphics[width=5cm]{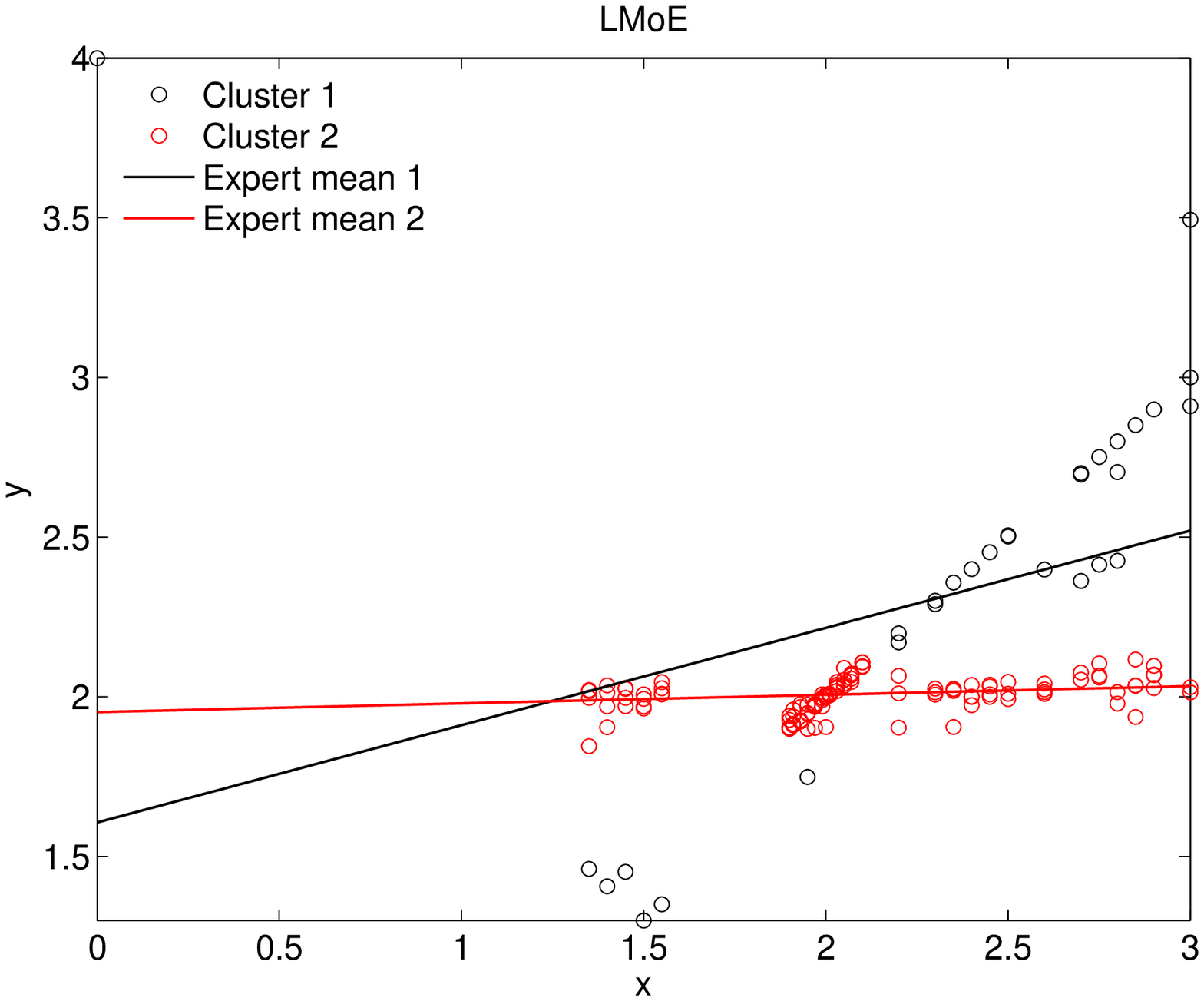} 
   \includegraphics[width=5cm]{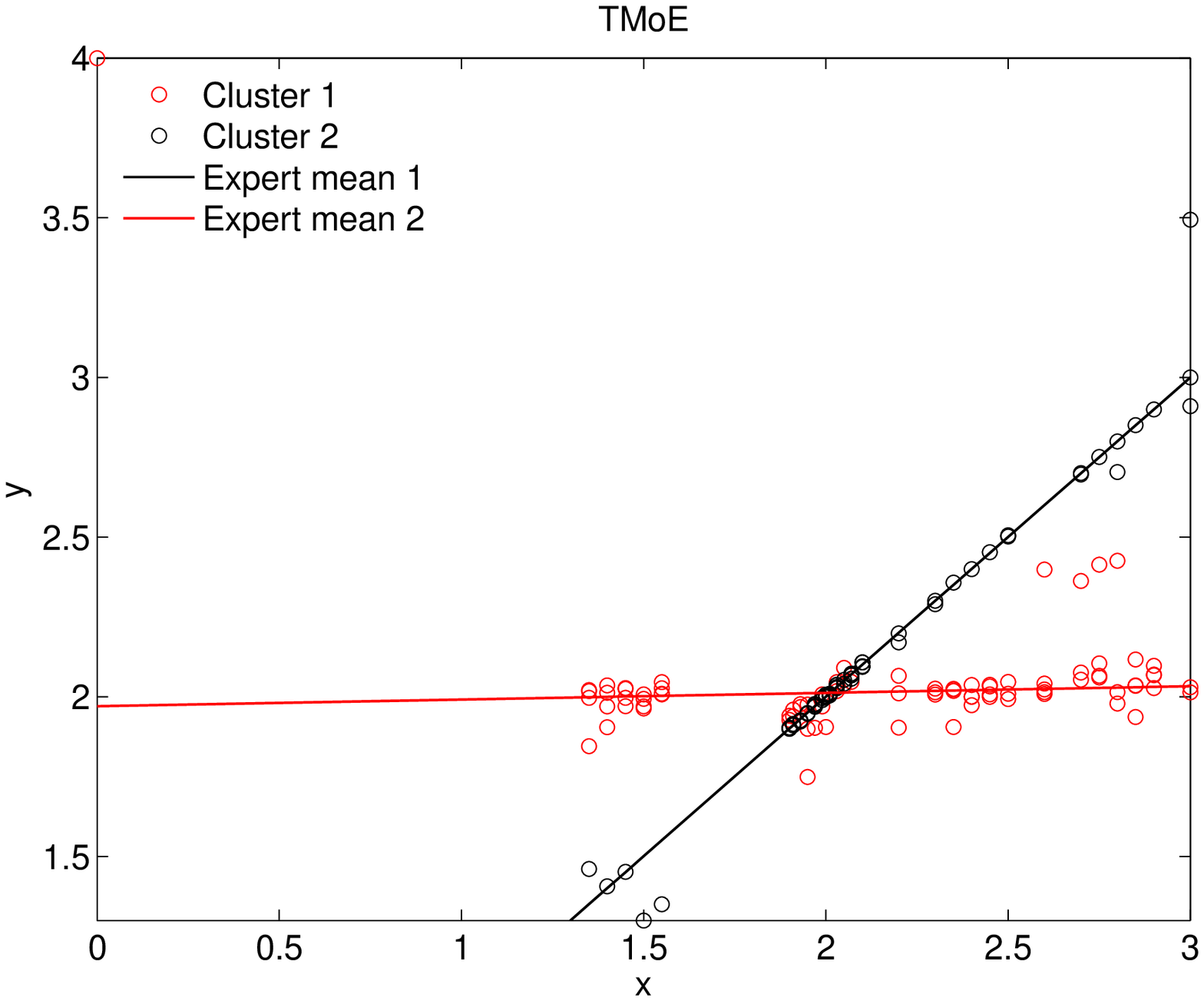} 
   \end{tabular}
      \caption{\label{fig. Tone data with outliers and all models}Fitting MoLE to the tone data set with ten added outliers $(0,4)$. Left: NMoE model fit, Middle: LMoE model fit, Right: TMoE model fit. The predictor $x$ is the actual tone ratio and the response $y$ is the perceived tone ratio.}
\end{figure}
Figure \ref{fig. Tone data with outliers and loglik} shows the log-likelihood profiles for each of the three models, which, while showing a similar behavior than the one in the case without outliers, show that the maximum likelihood value for the NMoE model is significantly less than the one in the case without outliers, compared to the best solution which is provided by the TMoE model. 
\begin{figure}[H]
   \centering 
   \begin{tabular}{ccc}
   \includegraphics[width=5cm]{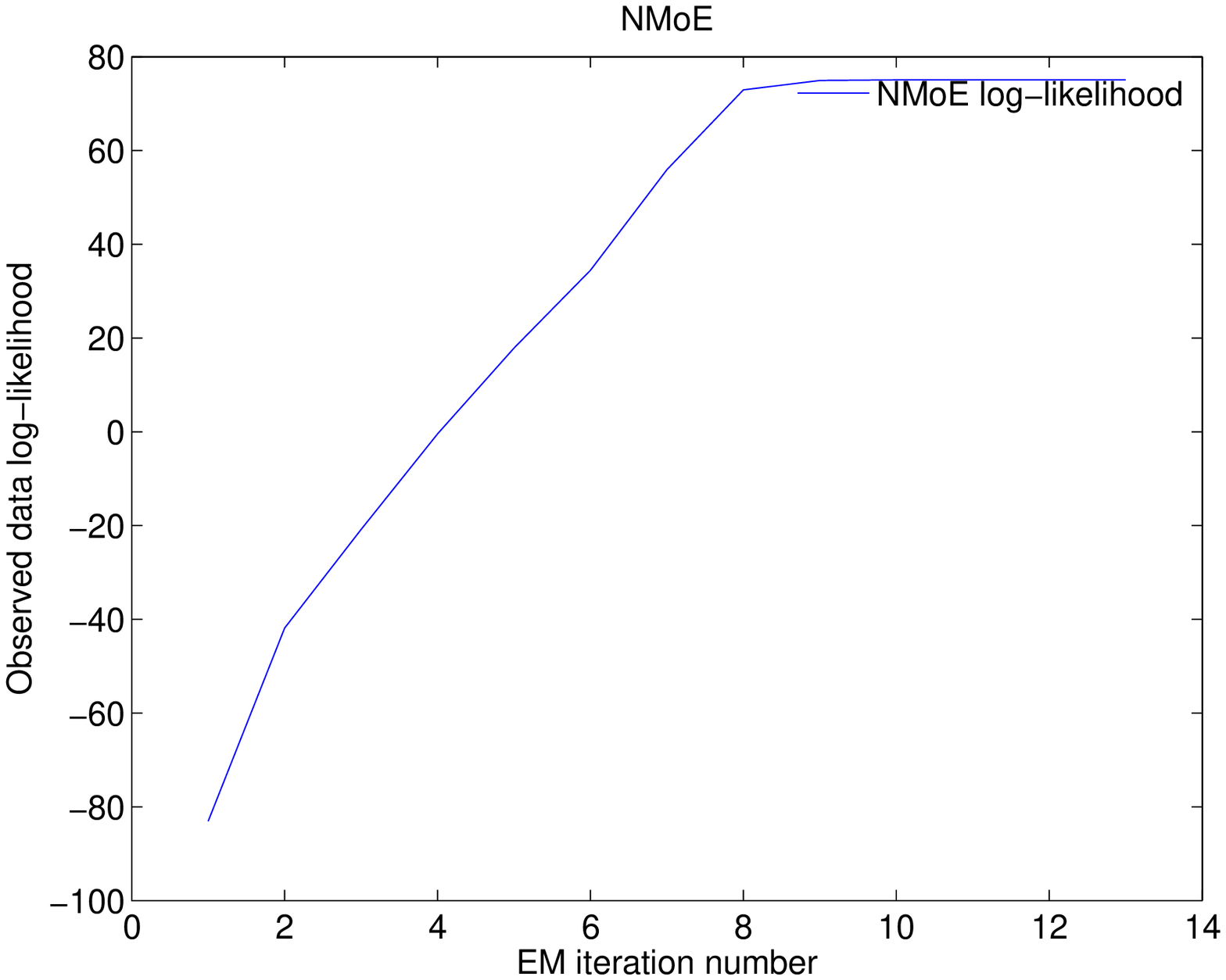} & 
   \includegraphics[width=5cm]{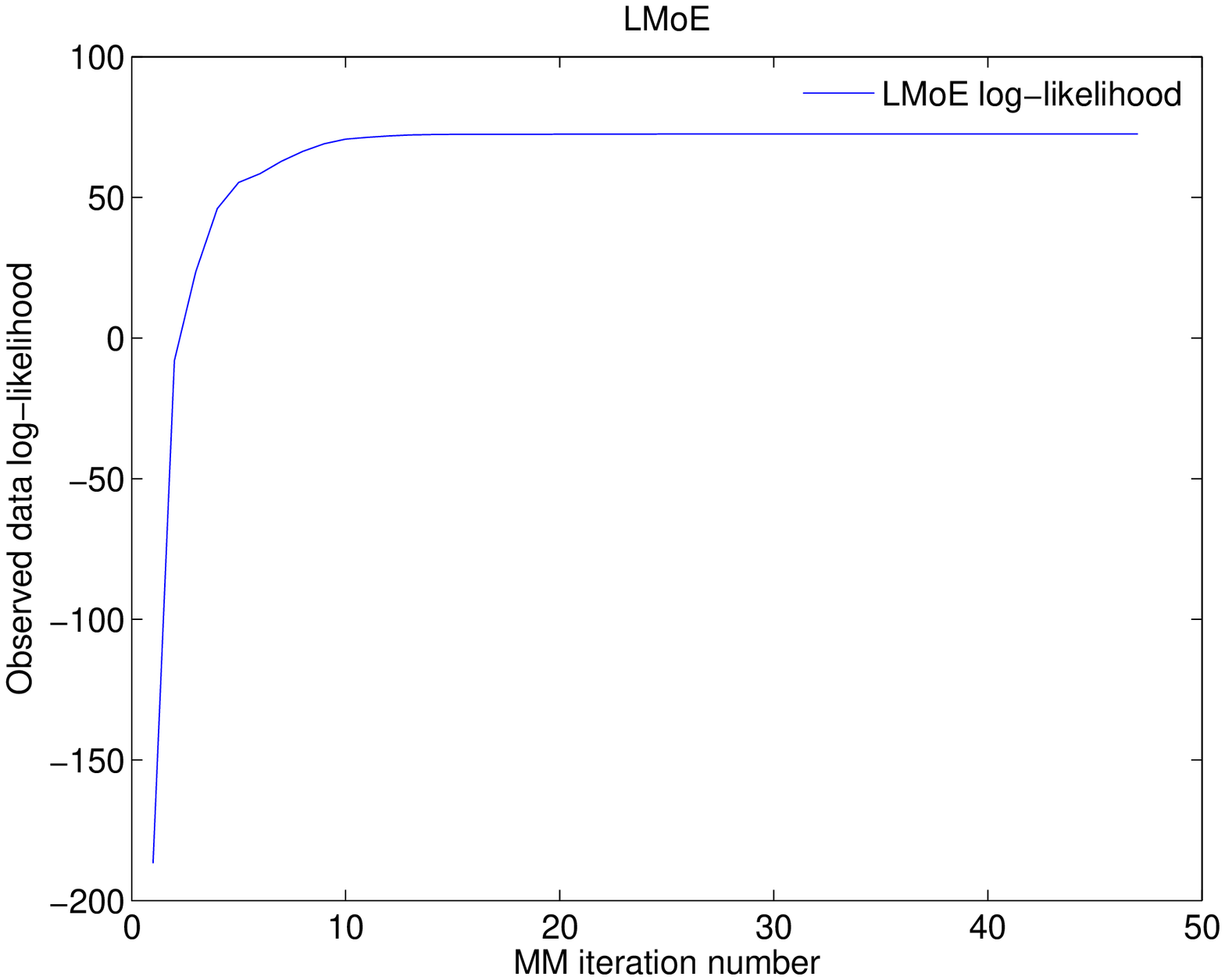} 
   \includegraphics[width=5cm]{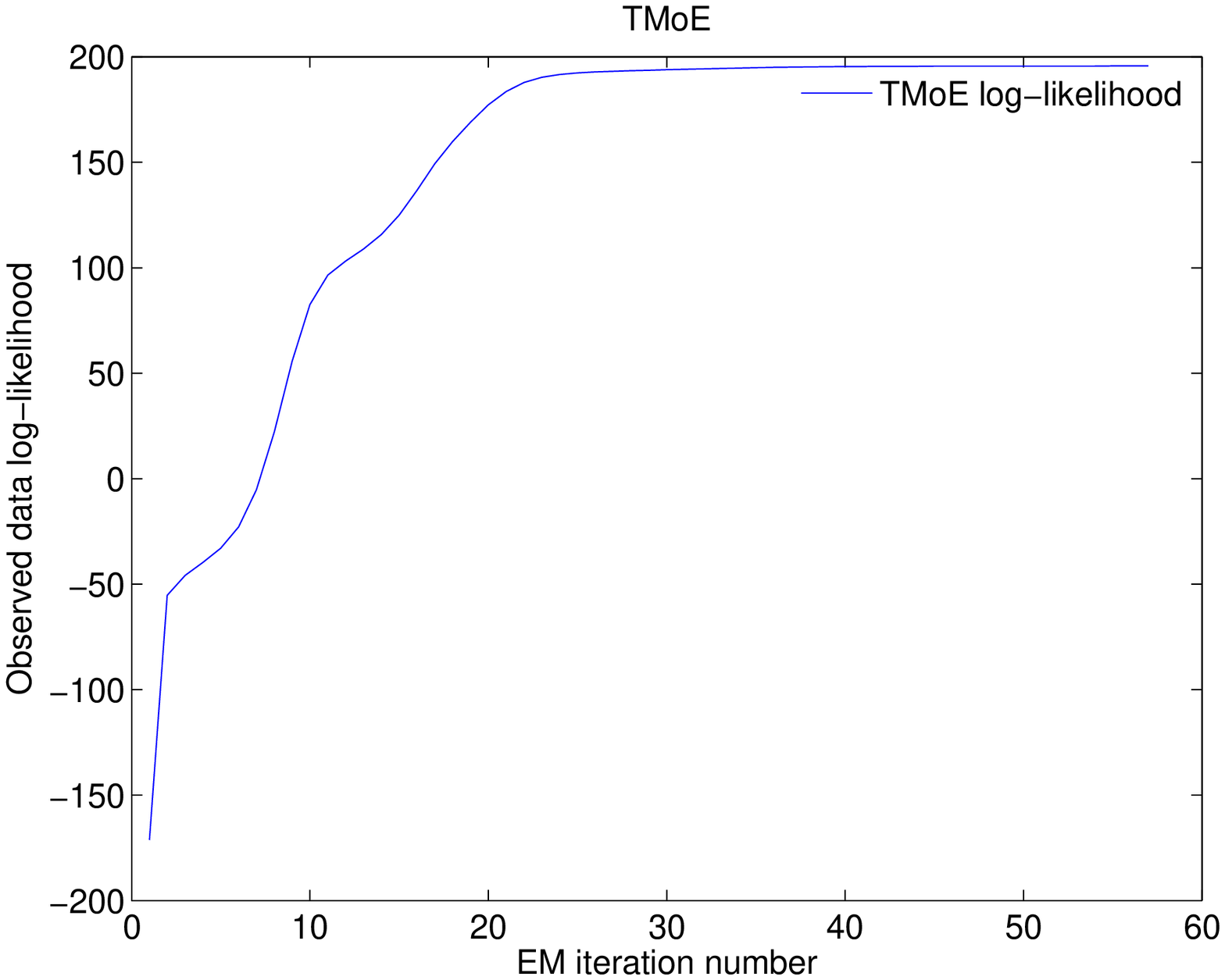} 
   \end{tabular}
      \caption{\label{fig. Tone data with outliers and loglik}The log-likelihood during the EM iterations when fitting the MoLE models to the tone data set with ten added outliers $(0,4)$. Left: NMoE model, Middle LMoE, and Right:  TMoE model.}
\end{figure}
The values of estimated MoE parameters in this case with outliers are given in Table \ref{tab. Estimated parameters for the tone perception data set with outliers}.
The regression coefficients for the second expert component are very similar for the three models. For the first component, the TMoE model retrieved a more heavy tailed component. Finally, for this data set, we can conclude that the TMoE provides the best solution. 
{\setlength{\tabcolsep}{4pt
\begin{table}[H]
\centering
{\small \begin{tabular}{l c c  c c c c c c c c c c}
\hline
param. & $\alpha_{10}$ & $\alpha_{11}$ & $\beta_{10}$ & $\beta_{11}$ & $\beta_{20}$ & $\beta_{21}$ & $\sigma_{1}$& $\sigma_{2}$ & $\lambda_{1}$ & $\lambda_{2}$ & $\nu_{1}$ & $\nu_{2}$ \\ 
model		& & & & & & & &  & &  & & \\
 \hline
 \hline
NMoE  	&  0.811	&  0.150	 & 3.117 & -0.285	&	1.907 & 0.046  & 0.700 & 0.050  & - & - & - & - \\
LMoE	&  -0.557 & -0.232  & 1.606 & 0.3047 & 1.9524 & 0.027 & - & - & 0.546 & 0.038 & - & - \\
TMoE   &  0.888   &	-0.236 & 0.002  & 0.999   & 1.971 & 0.020	& 0.002    & 0.024  & - & -  & 0.682 & 0.812\\
 \hline
\end{tabular}}
\caption{\label{tab. Estimated parameters for the tone perception data set with outliers} Values of the estimated MoE parameters  for the  tone perception data set with added outliers.}
\end{table}
}

\subsubsection{Temperature anomalies data set}
In this experiment, we examine another real-world data set related to climate change analysis. 
The NASA GISS Surface Temperature (GISTEMP) analysis provides a measure of the changing global surface temperature with monthly resolution for the period since 1880, when a reasonably global distribution of meteorological stations was established. 
The GISS analysis is updated monthly, however the data presented here\footnote{from \citet{TemperatureAnomalyData}, \url{http://cdiac.ornl.gov/ftp/trends/temp/hansen/gl_land.txt}} are updated annually as issued from the Carbon Dioxide Information Analysis Center (CDIAC), which has served as the primary climate-change data and information analysis center of the U.S. Department of Energy since 1982.
The data consist of $n = 135$ yearly measurements of the global annual temperature anomalies (in degrees C) computed using data from land meteorological stations for the period of $1882-2012$. 
These data have been analyzed earlier by \citet{Hansen1999,Hansen2001} and recently by \citet{Nguyen2014-MoLE} by using the Laplace mixture of linear experts (LMoLE). 

To apply the proposed $t$ mixture of expert model, we consider a mixture of two experts as in \citet{Nguyen2014-MoLE}. 
This number of components is also the one provided by the model selection criteria as shown later in Table \ref{tab. Model selection temperature anomalies data}. 
Indeed, as mentioned by \citet{Nguyen2014-MoLE},  \citet{Hansen2001} found that the data could be segmented into two periods of global warming (before 1940 and after 1965), separated by a transition period where there was a slight global cooling (i.e. 1940 to 1965). Documentation of the basic analysis method is provided by \citet{Hansen1999,Hansen2001}. 
We set the response $y_i (i=1,\ldots,135)$ as the temperature anomalies and the covariates $\bsx_i = \bsr_i = (1,x_i)^T$ where $x_i$ is the year of the $i$th observation.

%
%
Figures \ref{fig. temperature anomalies data and models experts}, 
\ref{fig. temperature anomalies data and models means}, and 
\ref{fig. temperature anomalies data and models loglik}
respectively show, for each of the three compared models, the fitted linear expert components, 
the corresponding means and confidence regions computed as plus and minus twice the estimated (pointwise) standard deviation as presented in Section \ref{sec: Prediction using the NNMoE},
and the log-likelihood profiles.
One can observe that the three models are successfully applied on the data set and provide very similar results. 
%
\begin{figure}[H]
   \centering 
   \begin{tabular}{ccc}
   \includegraphics[width=5cm]{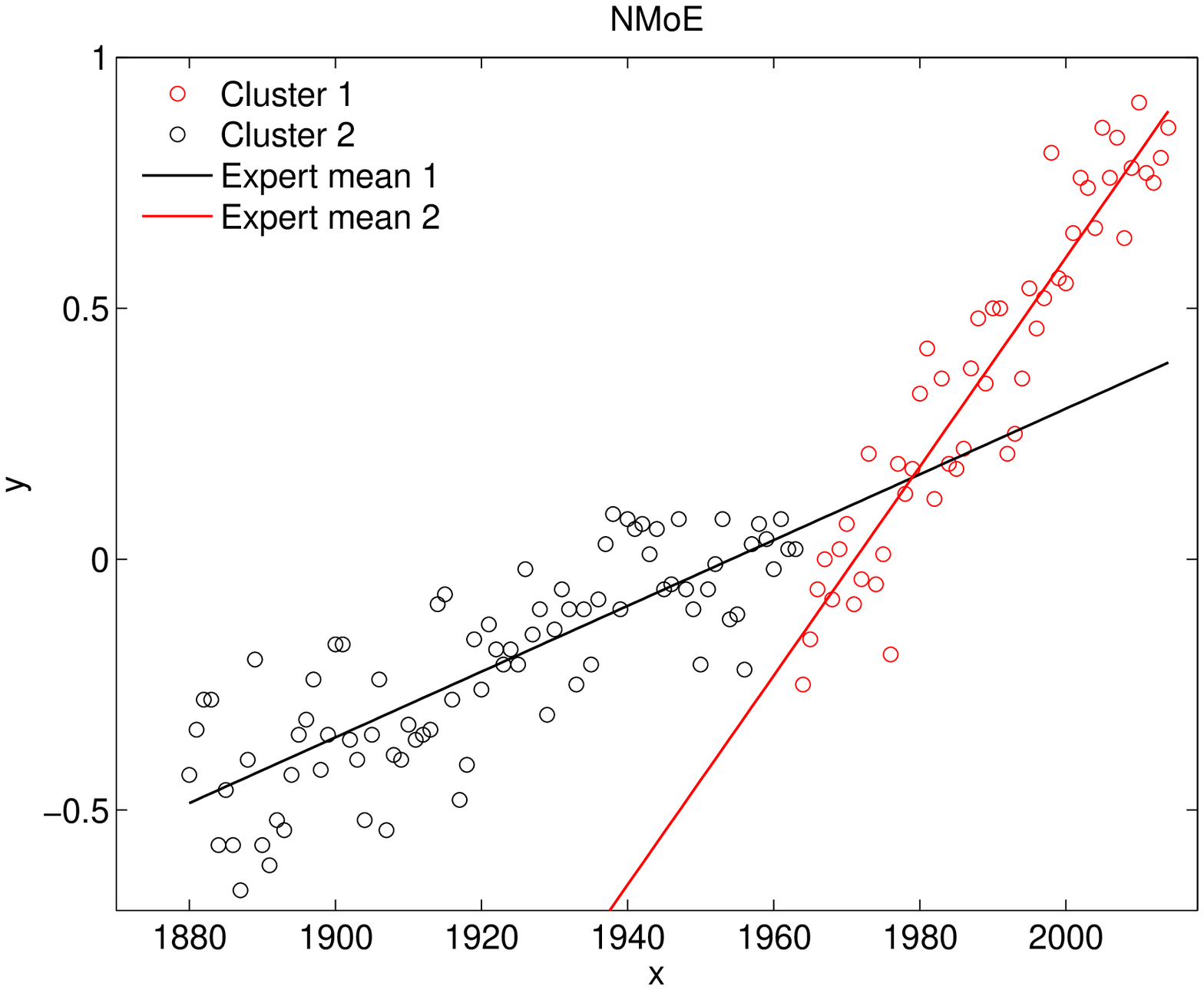} & 
\includegraphics[width=5cm]{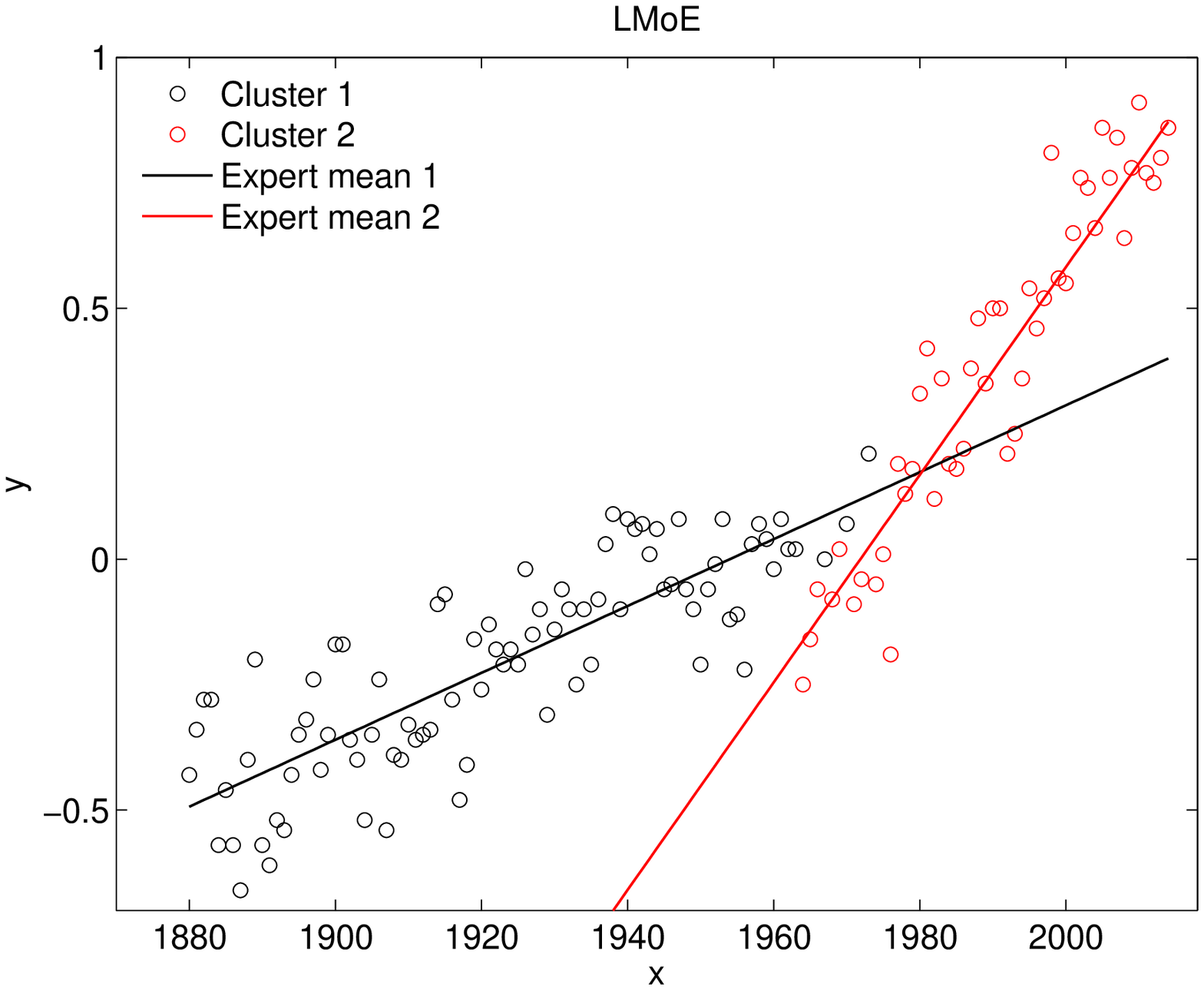} & 
\includegraphics[width=5cm]{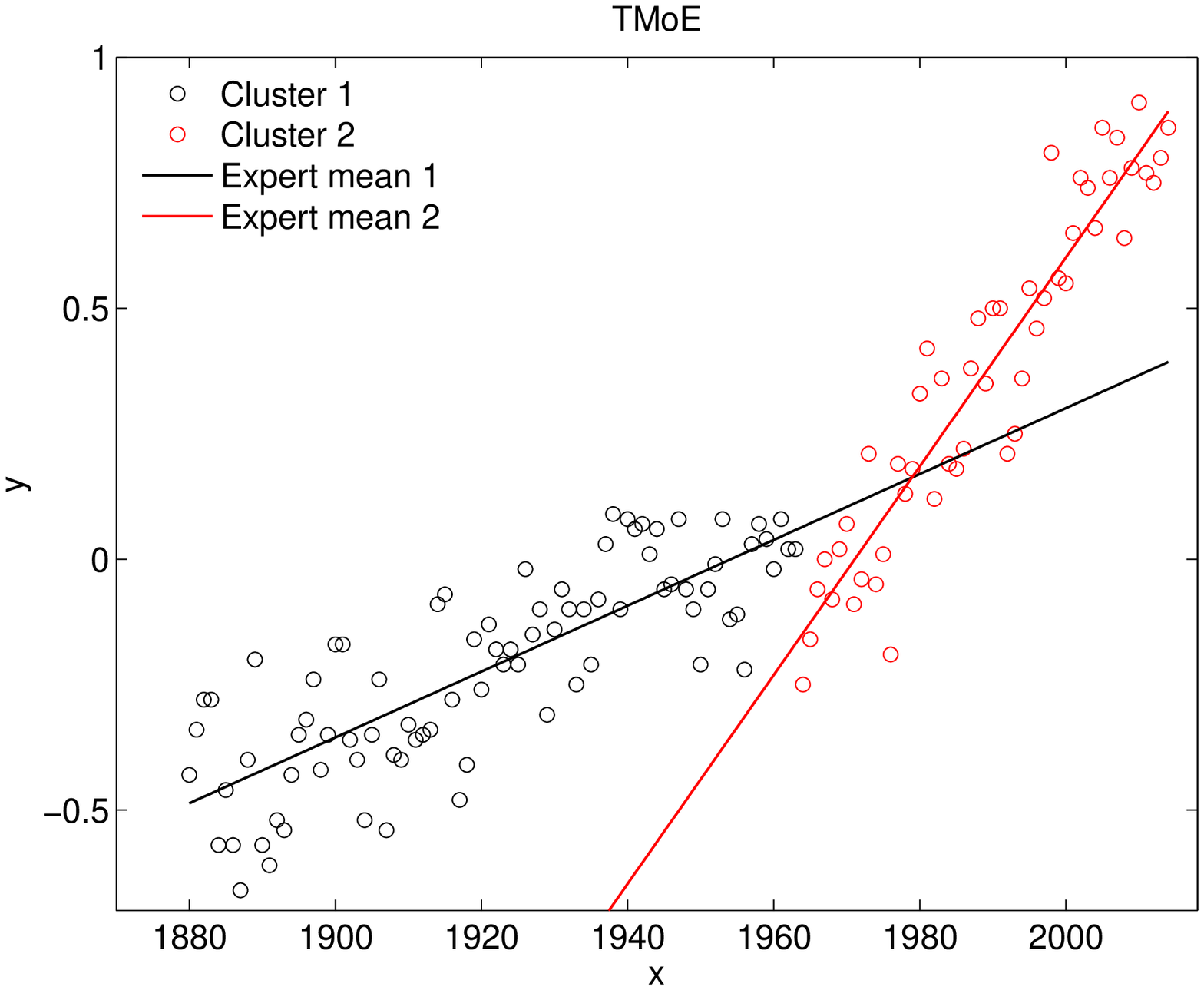} \\
\includegraphics[width=5cm]{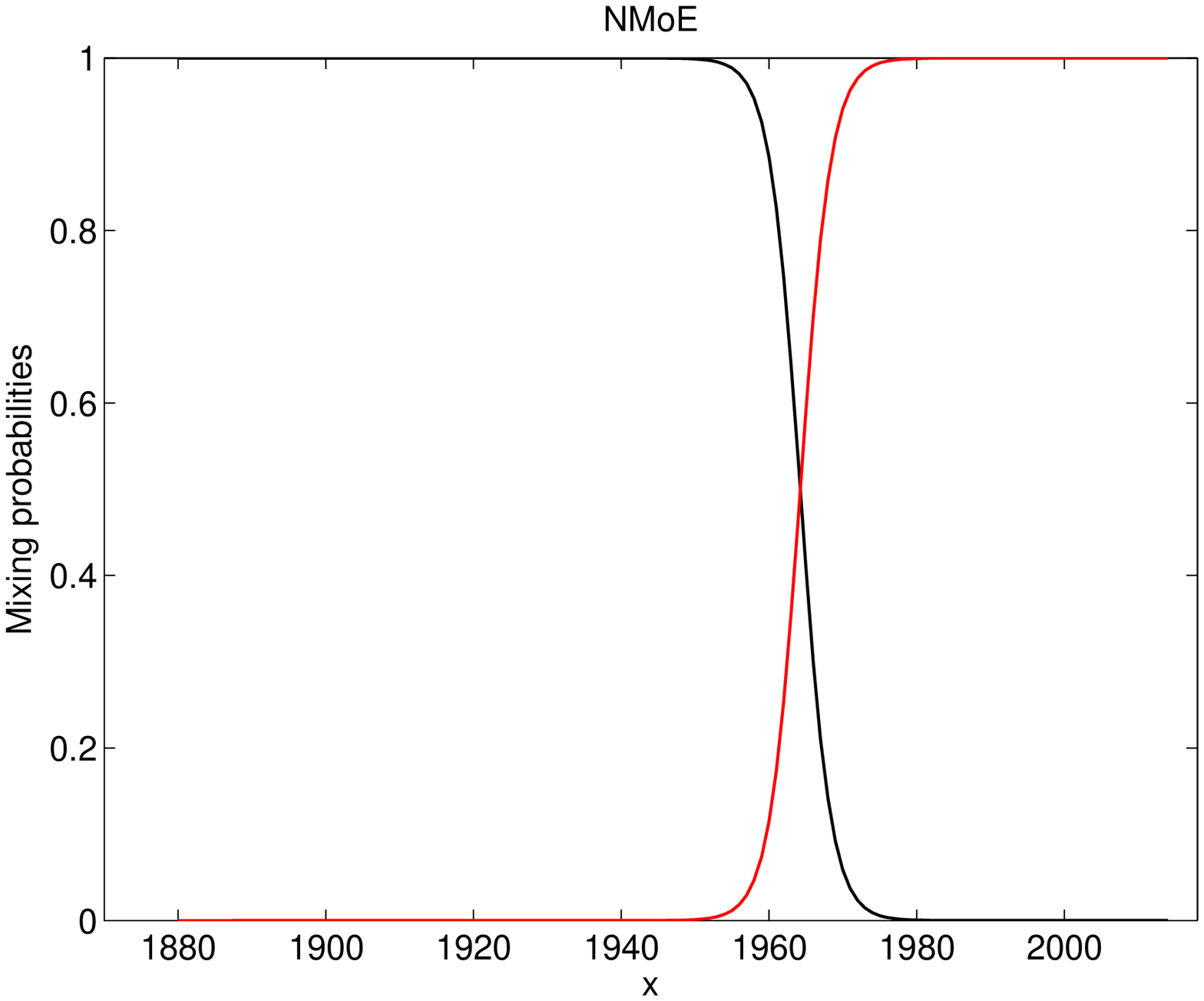} & 
\includegraphics[width=5cm]{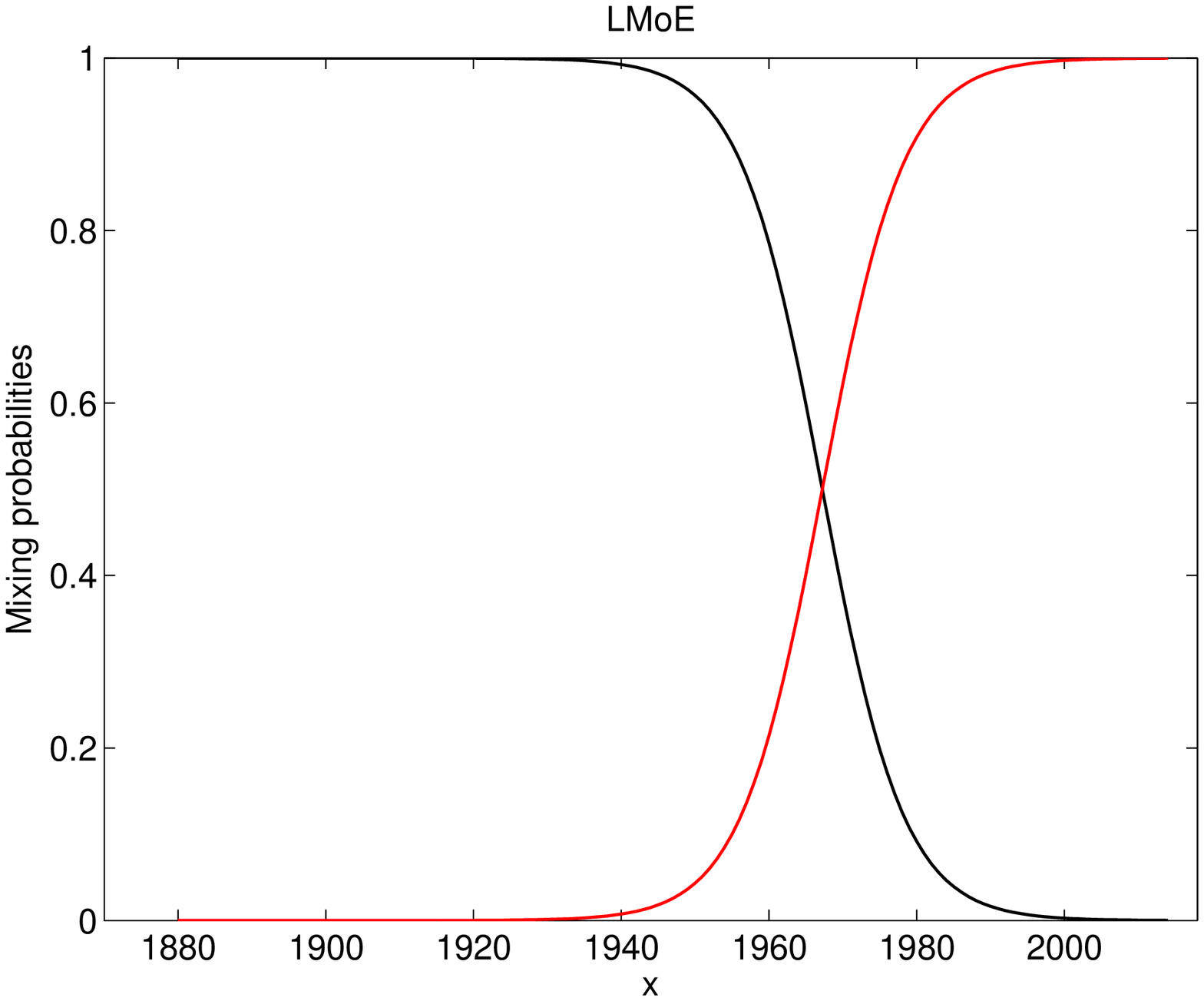} &
\includegraphics[width=5cm]{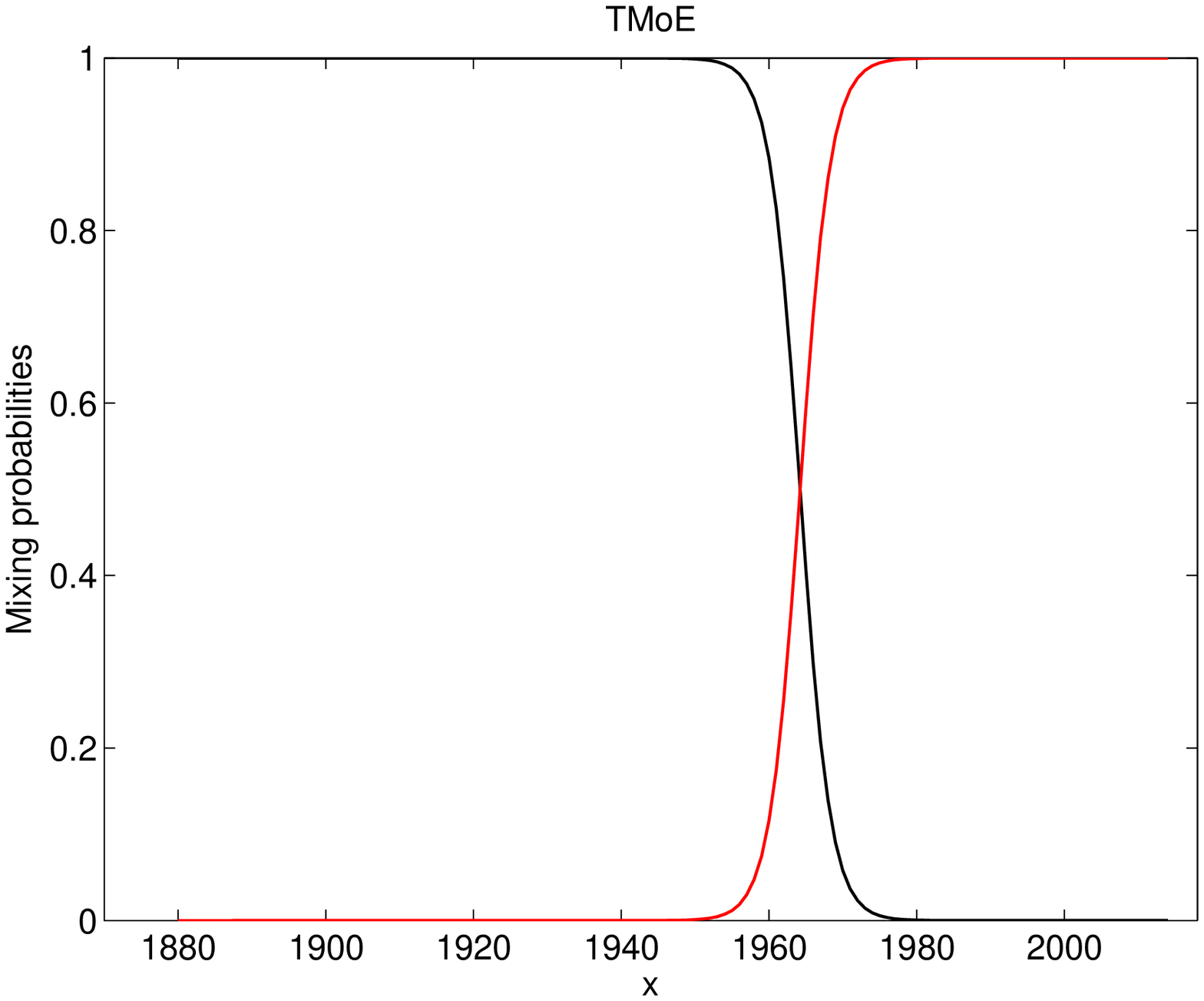}
   \end{tabular}
      \caption{\label{fig. temperature anomalies data and models experts}Fitting the MoLE models to the temperature anomalies data set. Left: NMoE model fit; Middle: LMoE model; Right: TMoE model. The predictor $x$ is the year and the response $y$ is the temperature anomaly.}
\end{figure}
\begin{figure}[H]
   \centering 
   \begin{tabular}{ccc}
   \includegraphics[width=5cm]{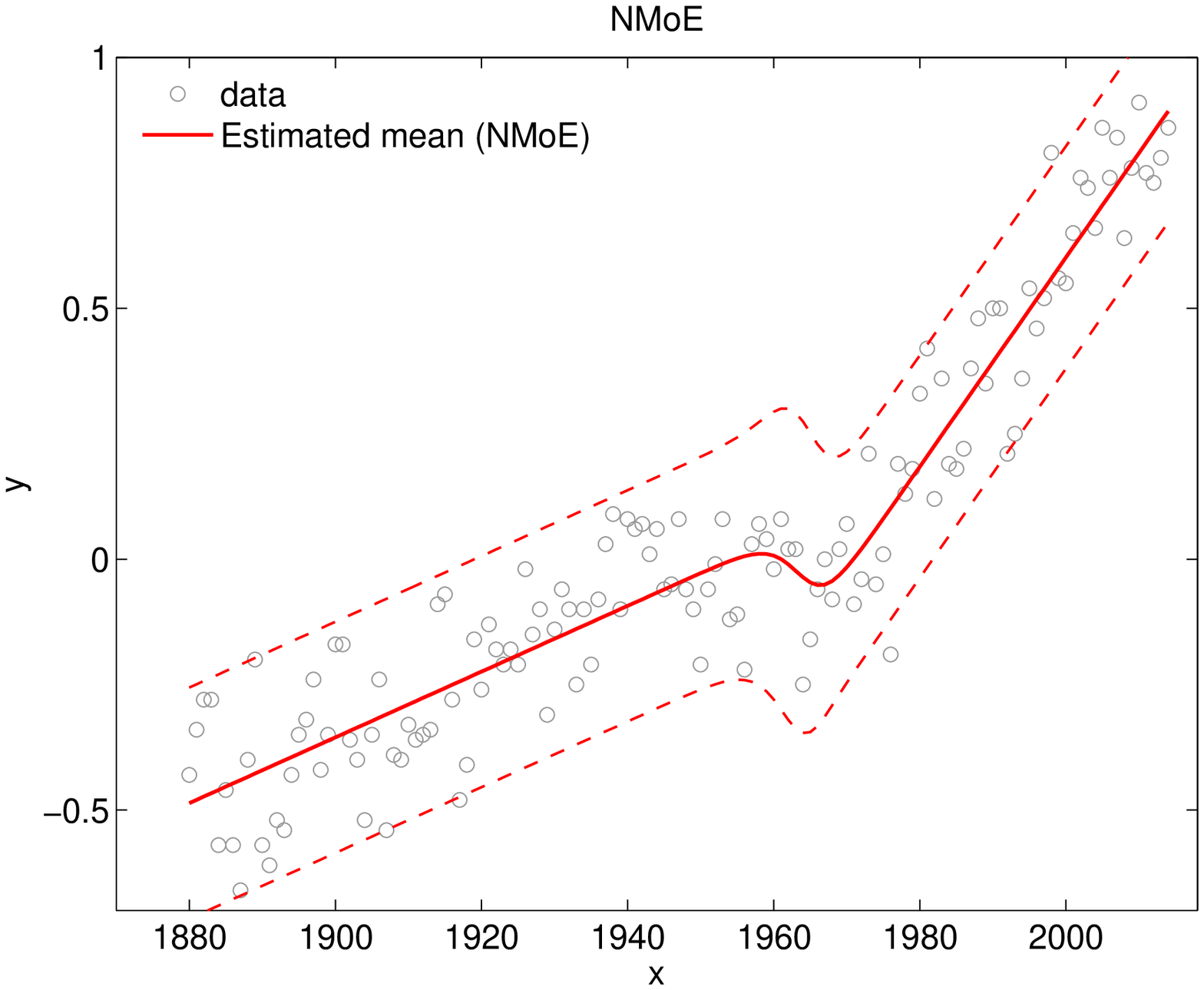} & 
\includegraphics[width=5cm]{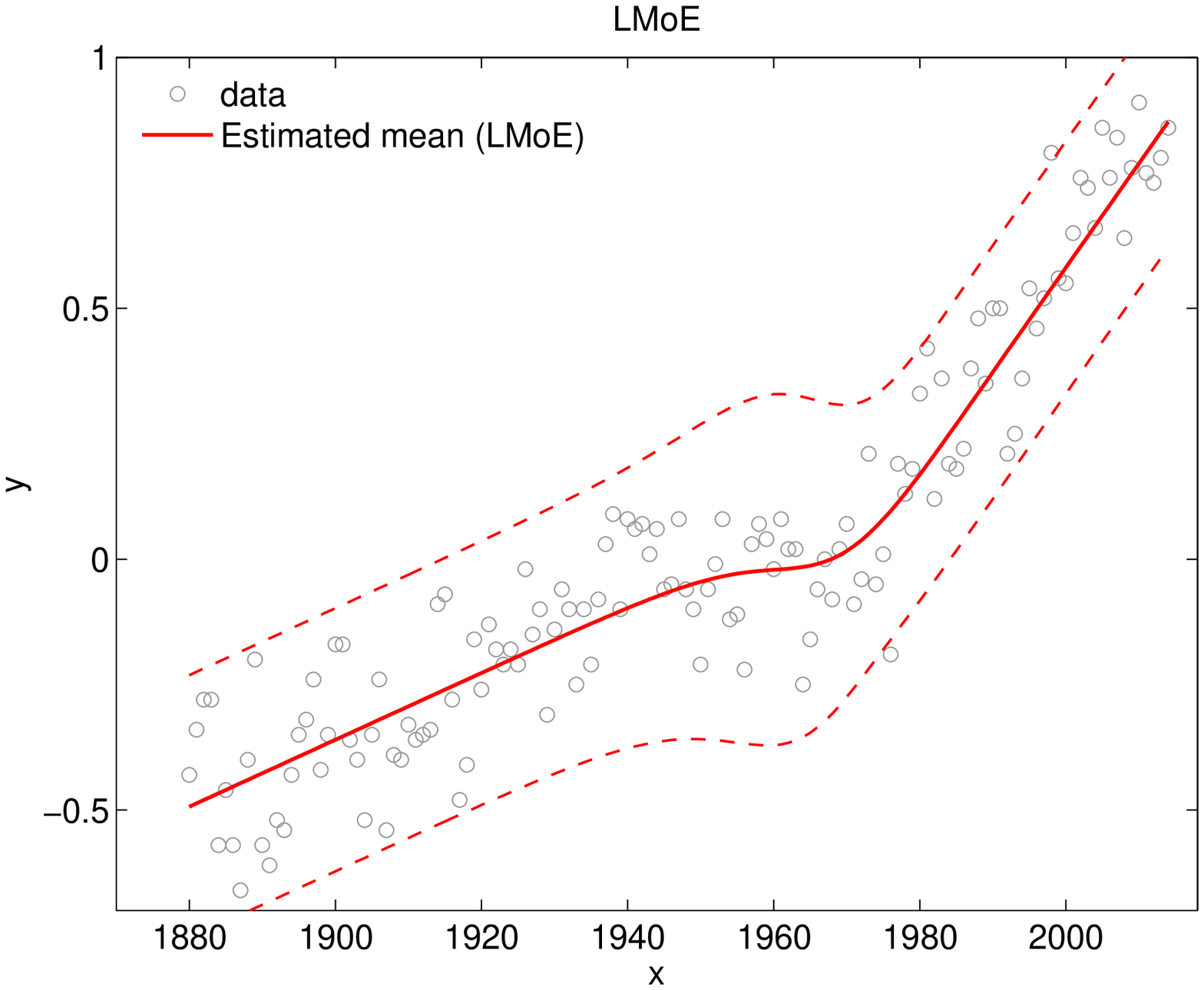} & 
\includegraphics[width=5cm]{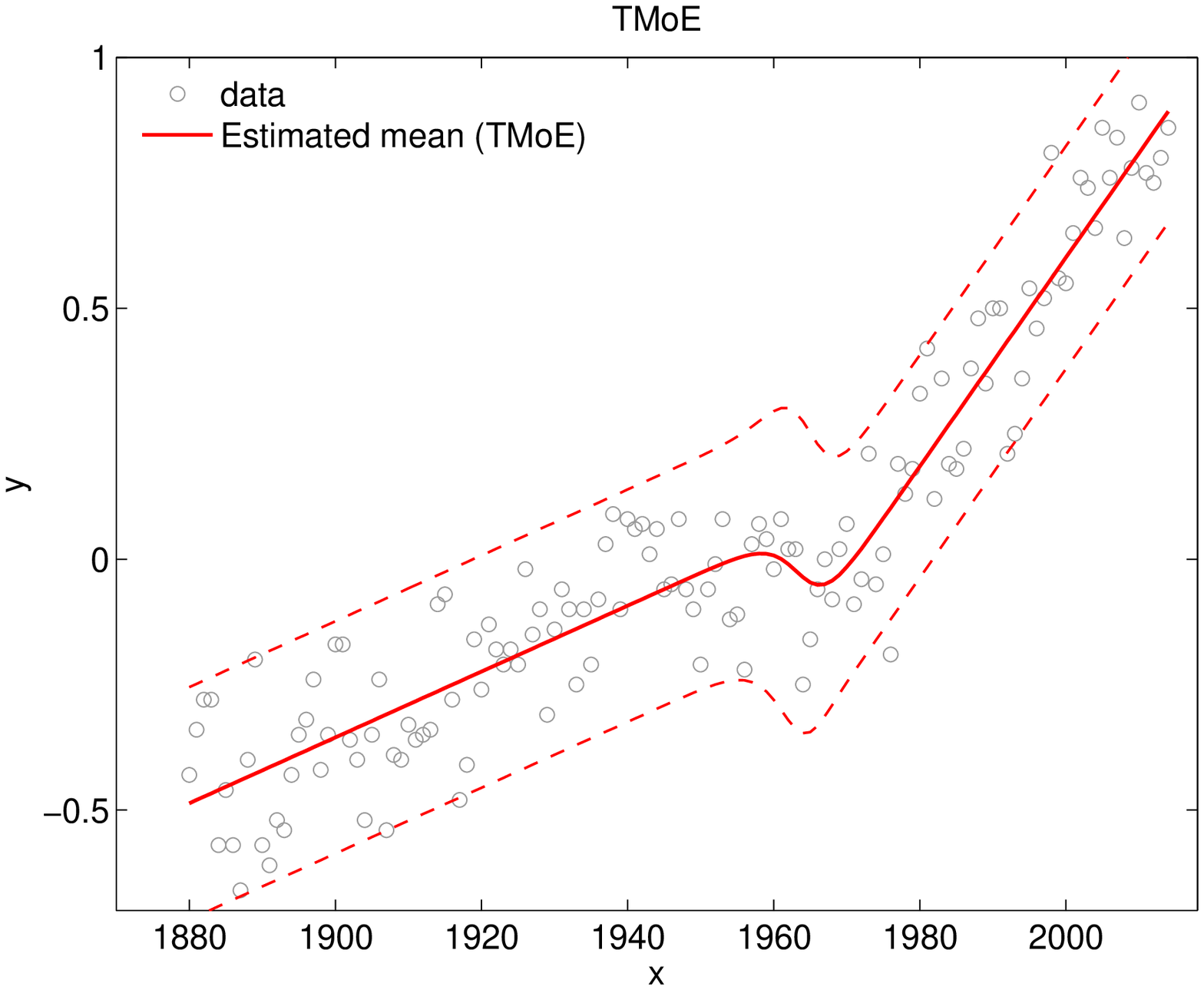} 
   \end{tabular}
      \caption{\label{fig. temperature anomalies data and models means}The fitted MoLE models to the temperature anomalies data set. Left: NMoE model; Middle: LMoE; Right: TMoE model. The predictor $x$ is the year and the response $y$ is the temperature anomaly.  The shaded region represents plus and minus twice the estimated (pointwise) standard deviation as presented in Section \ref{sec: Prediction using the NNMoE}.}
\end{figure}
\begin{figure}[H]
   \centering 
   \begin{tabular}{ccc}
   \includegraphics[width=5cm]{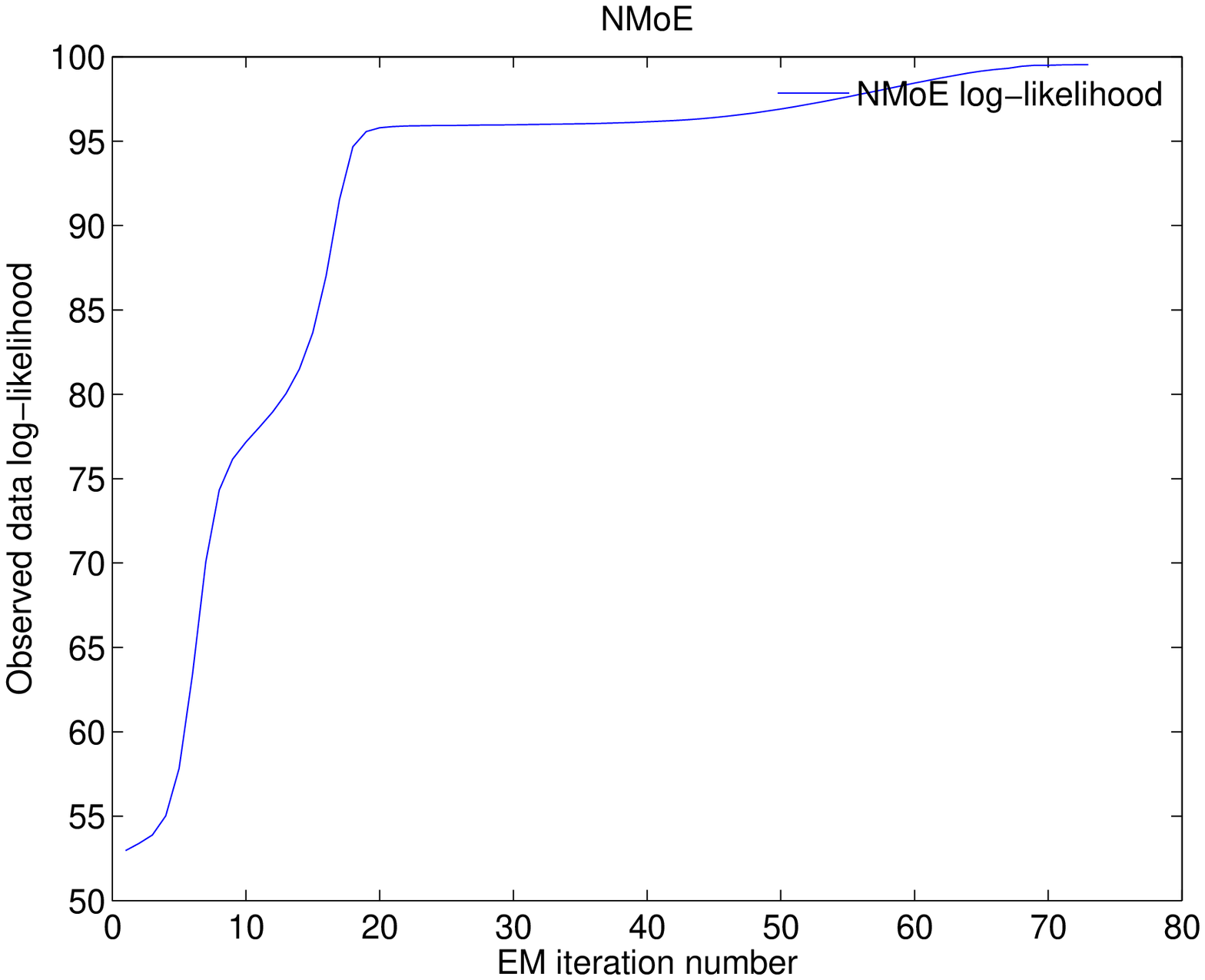} & 
  \includegraphics[width=5cm]{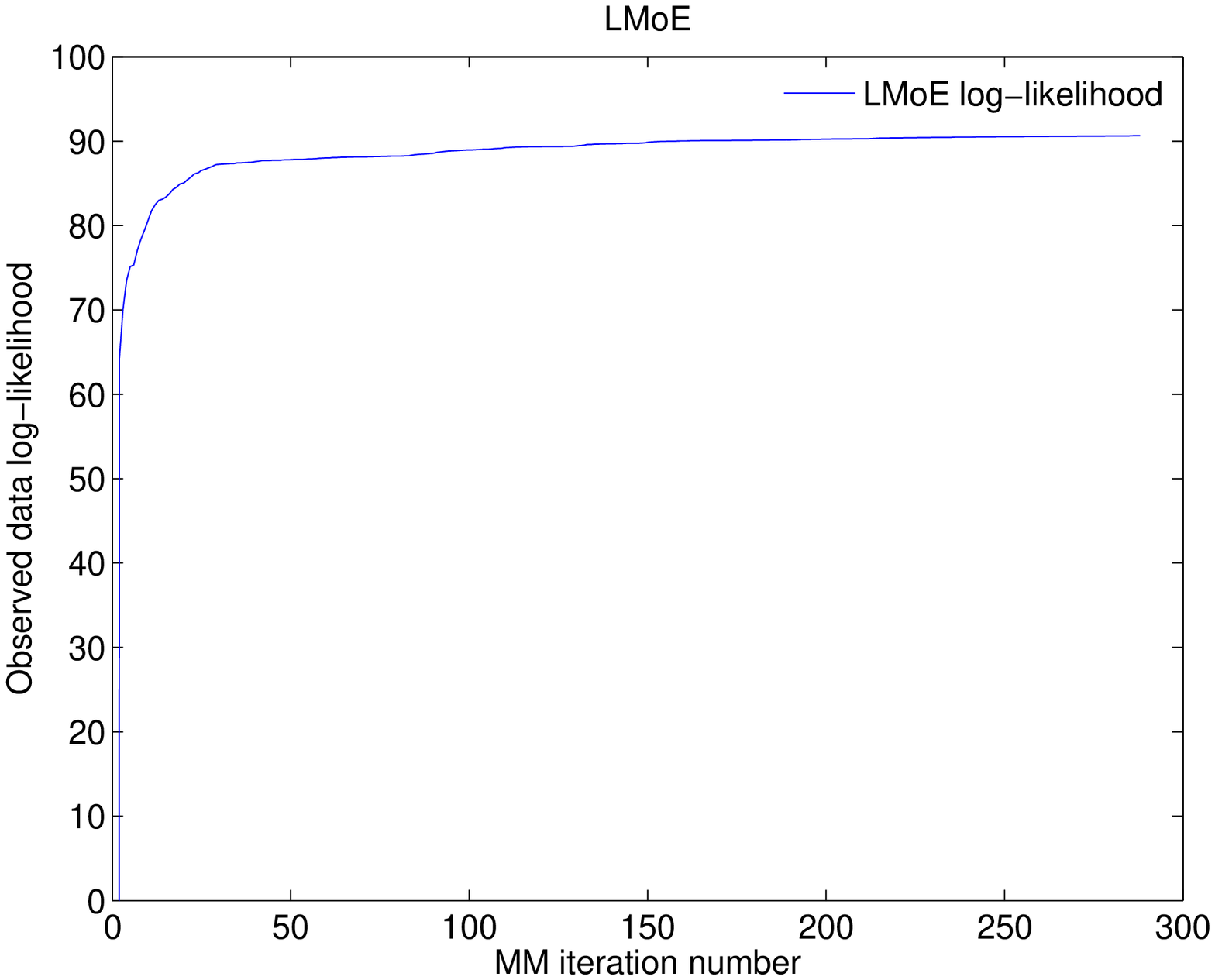} & 
  \includegraphics[width=5cm]{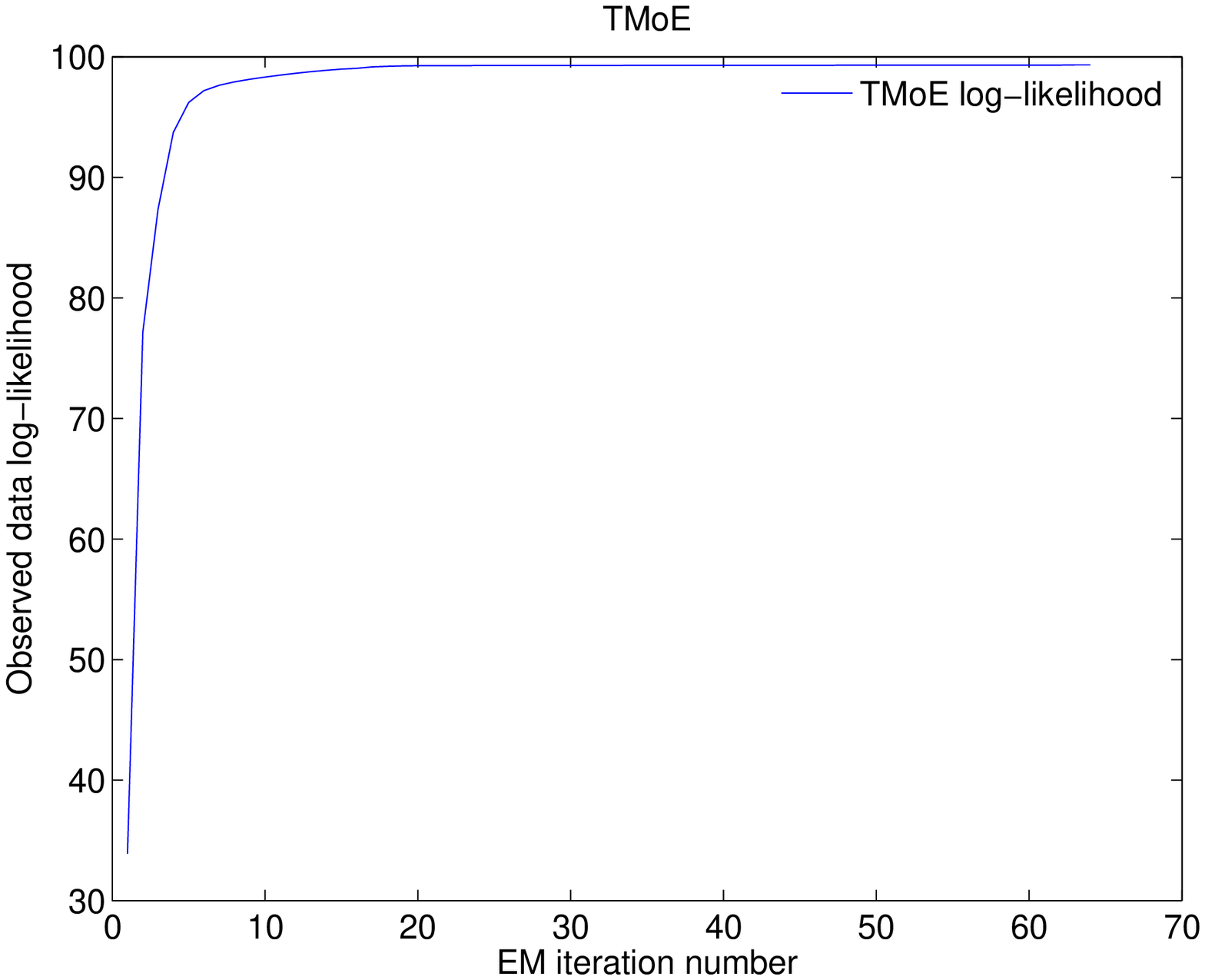} 
   \end{tabular}
      \caption{\label{fig. temperature anomalies data and models loglik}The log-likelihood during the EM iterations when fitting the MoLE models to the temperature anomalies data set. Left: NMoE model; Middle: LMoE; Right: TMoE model.}
\end{figure}
The values of estimated MoE parameters for the temperature anomalies data set are given in Table \ref{tab. estimated parameters for the temperature anomalies data set}.
%
One can see that the parameters common for the three models are quasi-identical, with a slight difference for the gating network parameters provided by the LMoE model. This slight difference results in the slight difference in the shape of the estimated mean curve. 
%
The TMoE  provides high degrees of freedom, which tends to approach a normal distribution.
This can also be seen on the log-likelihood profiles, which converges to almost the same value, meaning that the hypothesis of normality may be likely for this data set.
On the other hand, the regression coefficients are also similar to those found by \citet{Nguyen2014-MoLE} who used LMoE. 
{\setlength{\tabcolsep}{4pt
\begin{table}[H]
\centering
{\small
\begin{tabular}{l c c  c c c c c c c c c c}
\hline
param. & $\alpha_{10}$ & $\alpha_{11}$ & $\beta_{10}$ & $\beta_{11}$ & $\beta_{20}$ & $\beta_{21}$ & $\sigma_{1}$& $\sigma_{2}$ & $\lambda_{1}$ & $\lambda_{2}$ & $\nu_{1}$ & $\nu_{2}$ \\ 
model		& & & & & & & & & & \\
 \hline
 \hline
NMoE  	& 946.483 & -0.481  & -12.805 & 0.006 & -41.073   & 0.020 & 0.115 & 0.110  & -  & - & -  & -  \\
LMoE   & 354.076  &  -0.180 & -13.026  & 0.006 & -40.796  & 0.020 & - & - & 0.092  &  0.088 & - & - \\
TMoE  & 947.225 & -0.482 & -12.825 &  0.006 & -41.008     & 0.020 & 0.114 & 0.108 & -  & - & 70.82 & 54.38\\ 
 \hline
\end{tabular}}
\caption{\label{tab. estimated parameters for the temperature anomalies data set}Values of the estimated MoE parameters for the temperature anomalies data set.}
\end{table}
}
 
We performed a model selection procedure on the temperature anomalies data set to choose the best number of MoE components from values between 1 and 5. Table  \ref{tab. Model selection temperature anomalies data} gives the obtained values of the used model selection criteria, that is BIC, AIC, and ICL. 
One can see that, except the result provided by AIC for the NMoE model which provide a high number of components, and the one provided by ICL of the LMoE model, which underestimate the number of components, all the others results provide evidence for two components in the data.  
{\setlength{\tabcolsep}{4pt
\begin{table}[H]
\centering
{\small 
\begin{tabular}{l |ccc | ccc | ccc}
\hline
	& \multicolumn{3}{c|}{NMoE}	&	\multicolumn{3}{c}{LMoE} &  \multicolumn{3}{c}{TMoE}\\
\cline{2-10}
K			 &   BIC 	 & 	AIC 	&    ICL		 &   BIC 	 & 	AIC 	&    ICL		&   BIC 	 & 	AIC 	&    ICL\\
\hline
\hline
1 & 46.0623 &  50.4202 &  46.0623 & 39.2617 &  43.6196 &  \underline{-7.3579} &	43.5521 &  49.3627 &  43.5521 \\
2 & \underline{79.9163} &  91.5374 &  \underline{79.6241} & \underline{71.0153} &  \underline{82.6364} & -19.6211 & 	\underline{74.7960} &  \underline{89.3224} &  \underline{74.5279} \\
3 & 71.3963 &  90.2806 &  58.4874 &	61.9639 &  80.8482  &-31.8843 & 63.9709 &  87.2131 &  47.3643 \\
4 & 66.7276 &  92.8751 &  54.7524 & 49.9480  & 76.0955  &-44.1475 &56.8410  & 88.7990  & 45.1251\\
5 & 59.5100 &  \underline{92.9206} &  51.2429 & 40.3062 &  73.7169 &  -56.4107 & 43.7767  & 84.4505  & 29.3881\\ 
\hline 
\end{tabular}
}
\caption{\label{tab. Model selection temperature anomalies data}Choosing the number of expert components $K$ for the temperature anomalies data by using the information criteria BIC, AIC, and ICL. Underlined value indicates the highest value for each criterion.}
\end{table}}


\section{Conclusion and future work}
\label{sec: Conclusion}

 In this paper, we proposed a new robust non-normal MoE model, which generalizes the standard normal MoE. It is based on the $t$ distribution and named TMoE.
The TMoE model is suggested  for data with possibly outliers and heavy tail.
 We developed an EM algorithm and ECM extension to infer the proposed model  and  described its use in non-linear regression and prediction, as well as in model-based clustering.
The developed model is successfully applied and validated on simulated and real data sets.
The results obtained on simulated data confirm the good performance of the model in terms of density estimation, non-linear regression function approximation and clustering.
 In addition, the simulation results provide evidence of the robustness of the TMoE model to outliers, compared to the normal alternative model. 
The proposed model is  also successfully applied to two different real data sets, including a situation with outliers. 
The model selection using information criteria tends to promote using BIC and also ICL against AIC which performed poorly in the analyzed data. 
The obtained results support  the benefit of the proposed approach  for practical applications. Furthermore, compared to the LMoE model, the TMoE has been revealed to be more adapted in several situations.

In this paper, we only considered the MoE in their standard (non-hierarchical) version. One interesting future direction is therefore to extend the proposed models to the hierarchical MoE framework \citep{jordanHME}. Furthermore, a  natural  future extension of this work is to consider the case of   MoE for multiple regression on multivariate data rather than simple regression on univariate data.

\section*{References}
\bibliographystyle{elsarticle-harv}
\bibliography{TMoE-NN}
\end{document}